\PassOptionsToPackage{unicode}{hyperref}
\PassOptionsToPackage{hyphens}{url}
\PassOptionsToPackage{dvipsnames,svgnames,x11names}{xcolor}
\documentclass[
  12pt]{article}

\usepackage{amsmath,amssymb}
\usepackage{iftex}
\ifPDFTeX
  \usepackage[T1]{fontenc}
  \usepackage[utf8]{inputenc}
  \usepackage{textcomp} 
\else 
  \usepackage{unicode-math}
  \defaultfontfeatures{Scale=MatchLowercase}
  \defaultfontfeatures[\rmfamily]{Ligatures=TeX,Scale=1}
\fi
\usepackage{lmodern}
\ifPDFTeX\else  
\fi
\IfFileExists{upquote.sty}{\usepackage{upquote}}{}
\IfFileExists{microtype.sty}{
  \usepackage[]{microtype}
  \UseMicrotypeSet[protrusion]{basicmath} 
}{}
\makeatletter
\@ifundefined{KOMAClassName}{
  \IfFileExists{parskip.sty}{%
    \usepackage{parskip}
  }{
    \setlength{\parindent}{0pt}
    \setlength{\parskip}{6pt plus 2pt minus 1pt}}
}{
  \KOMAoptions{parskip=half}}
\makeatother
\usepackage{xcolor}
\makeatletter
\ifx\paragraph\undefined\else
  \let\oldparagraph\paragraph
  \renewcommand{\paragraph}{
    \@ifstar
      \xxxParagraphStar
      \xxxParagraphNoStar
  }
  \newcommand{\xxxParagraphStar}[1]{\oldparagraph*{#1}\mbox{}}
  \newcommand{\xxxParagraphNoStar}[1]{\oldparagraph{#1}\mbox{}}
\fi
\ifx\subparagraph\undefined\else
  \let\oldsubparagraph\subparagraph
  \renewcommand{\subparagraph}{
    \@ifstar
      \xxxSubParagraphStar
      \xxxSubParagraphNoStar
  }
  \newcommand{\xxxSubParagraphStar}[1]{\oldsubparagraph*{#1}\mbox{}}
  \newcommand{\xxxSubParagraphNoStar}[1]{\oldsubparagraph{#1}\mbox{}}
\fi
\makeatother

\usepackage{longtable,booktabs,array}
\usepackage{calc} 
\usepackage{etoolbox}
\makeatletter
\patchcmd\longtable{\par}{\if@noskipsec\mbox{}\fi\par}{}{}
\makeatother
\IfFileExists{footnotehyper.sty}{\usepackage{footnotehyper}}{\usepackage{footnote}}
\makesavenoteenv{longtable}
\usepackage{graphicx}
\makeatletter
\def\maxwidth{\ifdim\Gin@nat@width>\linewidth\linewidth\else\Gin@nat@width\fi}
\def\maxheight{\ifdim\Gin@nat@height>\textheight\textheight\else\Gin@nat@height\fi}
\makeatother
\setkeys{Gin}{width=\maxwidth,height=\maxheight,keepaspectratio}
\makeatletter
\def\fps@figure{htbp}
\makeatother

\makeatletter
\@ifpackageloaded{caption}{}{\usepackage{caption}}
\AtBeginDocument{%
\ifdefined\contentsname
  \renewcommand*\contentsname{Table of contents}
\else
  \newcommand\contentsname{Table of contents}
\fi
\ifdefined\listfigurename
  \renewcommand*\listfigurename{List of Figures}
\else
  \newcommand\listfigurename{List of Figures}
\fi
\ifdefined\listtablename
  \renewcommand*\listtablename{List of Tables}
\else
  \newcommand\listtablename{List of Tables}
\fi
\ifdefined\figurename
  \renewcommand*\figurename{Figure}
\else
  \newcommand\figurename{Figure}
\fi
\ifdefined\tablename
  \renewcommand*\tablename{Table}
\else
  \newcommand\tablename{Table}
\fi
}
\@ifpackageloaded{float}{}{\usepackage{float}}
\floatstyle{ruled}
\@ifundefined{c@chapter}{\newfloat{codelisting}{h}{lop}}{\newfloat{codelisting}{h}{lop}[chapter]}
\floatname{codelisting}{Listing}

\makeatother
\makeatletter
\@ifpackageloaded{caption}{}{\usepackage{caption}}
\@ifpackageloaded{subcaption}{}{\usepackage{subcaption}}
\makeatother

\ifLuaTeX
  \usepackage{selnolig}  
\fi
\usepackage[]{natbib}
\usepackage{bookmark}

\IfFileExists{xurl.sty}{\usepackage{xurl}}{} 
\hypersetup{
  pdftitle={Bayesian MI-LASSO for Variable Selection on Multiply-Imputed Data},
  pdfauthor={Jungang Zou; Sijian Wang; Qixuan Chen},
  pdfkeywords={Bayesian models, Group LASSO, Multiple imputation, Projection predictive variable selection},
  colorlinks=true,
  linkcolor={blue},
  filecolor={Maroon},
  citecolor={Blue},
  urlcolor={Blue}}

\usepackage{bbm}
\usepackage{multirow}
\usepackage{makecell}
\usepackage{enumitem}

\newcommand{\bX}{\mathbf{X}}
\newcommand{\bx}{\mathbf{x}}
\newcommand{\bY}{\mathbf{Y}}
\newcommand{\by}{\mathbf{y}}

\newcommand{\bb}{\boldsymbol{\beta}}

\newcommand{\E}{\mathrm{E}}

\newcommand{\cond}{{\, \vert \,}}
\DeclareMathOperator{\Tr}{tr}

\AtBeginEnvironment{table}{\spacingset{1}}

\newcommand{\anon}{1}

\begin{document}

\def\spacingset#1{\renewcommand{\baselinestretch}%
{#1}\small\normalsize} \spacingset{1}


\if1\anon
{
  \title{\vspace*{-1em}\bf Bayesian MI-LASSO for Variable Selection on Multiply-Imputed Data}
  \author{Jungang Zou\thanks{
    Corresponding author: Jungang Zou (\href{mailto:jz3183@cumc.columbia.edu}{jz3183@cumc.columbia.edu}).}\hspace{.2cm}\\
    Department of Biostatistics, Columbia University\\
    and \\
    Sijian Wang \\
    Department of Statistics, Rutgers University\\
    and \\
    Qixuan Chen \\
    Department of Biostatistics, Columbia University}
  \maketitle
} \fi

\if0\anon
{
  \bigskip
  \bigskip
  \bigskip
  \begin{center}
    {\LARGE\bf Bayesian MI-LASSO for Variable Selection on Multiply-Imputed Data}
\end{center}
  \medskip
} \fi

\bigskip
\begin{abstract}
Multiple imputation is widely used for handling missing data in real-world applications. For variable selection on multiply-imputed datasets, however, if selection is performed on each imputed dataset separately, it can result in different sets of selected variables across datasets. MI-LASSO, one of the most commonly used approaches to this problem, regards the same variable across all separate imputed datasets as a group variable and exploits the group LASSO to yield a consistent variable selection across all the multiply-imputed datasets. In this paper, we extend MI-LASSO into the Bayesian framework and propose four Bayesian MI-LASSO models for variable selection on multiply-imputed data, including three shrinkage prior-based and one Spike-and-Slab prior-based methods. To overcome the limitations of traditional, ad hoc threshold-based model selection, we develop a principled four-step projection predictive procedure that provides robust variable selection. This procedure also facilitates approximately calibrated post-selection inference and attains near-nominal coverage under a correct selection model. Simulation studies showed that the Bayesian MI-LASSO outperformed MI-LASSO and other alternative approaches, achieving higher specificity and lower mean squared error across a range of settings. We further demonstrated these methods via a case study using a multiply-imputed dataset from the University of Michigan Dioxin Exposure Study. The \texttt{R} package \href{https://cran.r-project.org/web/packages/BMIselect}{\texttt{BMIselect}} is available on \texttt{CRAN}.
\end{abstract}

\noindent%

\vfill

\newpage
\spacingset{1.8} 

\section{Introduction}
Variable selection is a classic but important area in statistical research. The goal of variable selection is to identify a subset of the most relevant and informative variables associated with the outcome variable. While most variable selection methods assume complete data, missing data are common in real-world applications. A variety of approaches have been developed to simultaneously address missingness and variable selection, including maximum likelihood estimation via the EM algorithm \citep{Garcia2010}, weighted estimating functions \citep{Johnson2008}, and fully Bayesian methods \citep{Yang2005,GarciaDonato2025}. In many healthcare studies, however, missing data have already been addressed upstream using multiple imputation (MI) \citep{Rubin1987} by separate data providers \citep{Schafer2001}, often using confidential information unavailable to analysts. In these settings, data providers generate multiply-imputed datasets, and analysts conduct data analysis using the already imputed datasets. Motivated by this real setting, we focus on variable selection using multiply-imputed data, assuming that the imputations are properly performed.

With multiply-imputed data, the common approach is to apply a complete-data method to each imputed data and then combine the results using Rubin's rules \citep{Rubin1987}. However, this approach may perform inadequately on variable selection, as different sets of relevant variables may be selected across the imputed datasets, leading to inconsistent variable selection results across multiple imputed data. In the literature, various solutions have been proposed to address this issue \citep{Heymans2007, Wood2008, Chen2013,Wan2015, Geronimi2017, Hu2021}. Among these, MI-LASSO \citep{Chen2013} has been widely adopted in applied research \citep{Weller2018, Becker2019, Anand2020}. This method treats the same variable across imputations as a group and applies a group LASSO penalty to either select or exclude the entire group, thereby promoting consistency in the selection across datasets. Empirical studies have shown strong performance of MI-LASSO for variable selection on multiply-imputed datasets.

While effective, the original MI-LASSO and its extensions \citep{Geronimi2017, Marino2017, Liu2016,Du2022} are frequentist approaches and, as such, lack a natural framework for incorporating prior information or providing coherent posterior uncertainty quantification. This motivates the development of a Bayesian counterpart. The Bayesian framework allows group-level information to be incorporated through shared penalized priors, provides posterior uncertainty estimates, and accommodates different forms of shrinkage and variable selection through prior specification. To bridge this gap, we propose four Bayesian MI-LASSO models and systematically evaluate their performance for variable selection with multiply-imputed data. These include three shrinkage prior-based models and one Spike-and-Slab model.

The posterior distributions from these models provide uncertainty information that can be used for variable selection and coefficient inference. In many applied settings, however, practitioners ultimately require a single selected model together with uncertainty quantification for its coefficients. These goals are not addressed directly by the fitted fully Bayesian models: choosing a final subset of variables generally relies on ad hoc thresholding rules \citep{Li2010,Carvalho2010,Barbieri2004}, while obtaining calibrated coefficient distributions for the resulting model remains difficult.

To address these challenges, we further develop a flexible four-step projection predictive variable selection procedure. Briefly, the procedure (i) generates candidate subsets over a grid of selection thresholds, (ii) projects the full-model posterior mean onto each candidate to obtain submodel point estimates, (iii) selects the subset that minimizes a modified version of the Bayesian information criterion, and (iv) projects the posterior draws onto the selected submodel and calibrates their variability to provide approximately calibrated inference conditional on the selected model. This procedure avoids reliance on a single ad hoc selection threshold, requires minimal tuning, and provides approximately calibrated post-selection inference for the coefficients of the selected model. 

The remainder of the paper is organized as follows. Section~\ref{sec:methods} presents notations, reviews the MI-LASSO method, and proposes the four Bayesian MI-LASSO models as well as the four-step projection predictive variable selection procedure. Section~\ref{sec:simulation} evaluates the performance of our proposed methods through extensive simulation studies. In Section~\ref{sec:case}, we apply the Bayesian MI-LASSO methods to existing multiply-imputed data from the University of Michigan Dioxin Exposure Study to identify important covariates of serum dioxin concentrations and estimate their associations. We close our paper with a discussion in Section~\ref{sec:discussion}.

\section{Methods}
\label{sec:methods}

\subsection{Notations}
Let $\bX$ and $\bY$ represent a $p$-dimensional covariate vector and outcome variable, respectively. We denote $\bX_{j}$ as the $j$-th covariate. We assume that each covariate $\bX_{j}$ is scaled to have mean zero and unit variance, and that the outcome $\bY$ is centered. Let $\bx = (\bx_1^\top, \ldots, \bx_n^\top)^\top \in \mathbb{R}^{n \times p}$ and $\by = (y_1, \ldots, y_n)^\top \in \mathbb{R}^n$ denote the data from $n$ independent and identically distributed units, with $\bx_i = (x_{i1}, \ldots, x_{ip})$ for unit $i$. We are interested in the Gaussian linear regression model of $\by \sim \mathcal{MVN}(\bx\bb, \sigma^2\mathbf{I})$, where $\mathbf{I}$ is the identity matrix, $\bb=(\beta_1,\beta_2,\ldots,\beta_p)^\top$ is a vector of coefficients and $\sigma^2 $ is the error variance. Assume data are incomplete and suppose $D$ imputations are generated, with $\bx^d = ({\bx_1^d}^\top, \ldots, {\bx_n^d}^\top)^\top$ and $\by^d = (y_1^d, \ldots, y_n^d)^\top$ denoting values of $\bX$ and $\bY$ in the $d$-th imputed dataset, respectively. We normalize each column of $\bx^d$ and center $\by^d$ for all $d = 1, \ldots, D$. For the linear regression on each imputed dataset, let $\bb^{d}=(\beta^d_{1},\beta^d_{2},\ldots,\beta^d_{p})^\top$ denote the regression coefficient vector on the $d$-th imputed dataset. We also define $\bb_{j}=(\beta^1_{j},\beta^2_{j},\ldots,\beta^D_{j})$ as the group of coefficients for the $j$-th covariate across all imputed datasets. 

\subsection{Overview of MI-LASSO}
\label{sec:MI_LASSO}
Although various models have been proposed to address the inconsistency of variable selection in MI settings, MI-LASSO \citep{Chen2013}  was the first to introduce the group LASSO \citep{Yuan2006} penalty to this problem. Different from stacking methods \citep{Wood2008, Wan2015,Du2022}, MI-LASSO regards the same variable across all imputed sets as a group of variables, and adopts group LASSO to either include or exclude the group of variables together. Specifically, the loss function for MI-LASSO is written as follows:
\begin{equation}
\label{eq:mi_lasso}
\operatorname*{min}_{\beta_{j}^d}\sum_{d=1}^D\|\by^{d}-\bx^{d}\bb^{d}\|_2^2+\xi \sum_{j=1}^{p}\|\bb_{j}\|_2. 
\end{equation}
where $\|\cdot\|_2$ is the $\ell_2$-norm, and $\xi$ is pre-specified to control the shrinkage of group coefficients. To select the optimal parameter $\xi$, \citet{Chen2013} developed a modified version of the Bayesian information criterion ($\text{BIC}_{m}$) for model selection:
\begin{equation} \label{eq:bic}
\begin{split}
    & \text{BIC}_{m} = \log\left(\sum_{d=1}^D\|\by^{d}-\bx^{d}\hat{\bb}^{d}\|_2^2/(Dn)\right)+df*\log(Dn)/(Dn), \\
   &  df = \sum_{d=1}^D\sum_{i=1}^n\frac{\text{cov}(\bx_i^d\hat{\bb}^d, y_i^d)}{\sigma^2},
   \end{split}
\end{equation}
where $\hat{\bb}^d$ is the MI-LASSO estimate for $\bb^d$, $df$ is the degree of freedom defined in \citet{Hastie1990}, and $\text{cov}(\cdot, \cdot)$ denotes the covariance. Let $\hat{\bb}_j$ be the MI-LASSO estimate for $\bb_j$, and $\hat{\bb}^{OLS}_j$ be the ordinary least square estimate of the full model for $\bb_j$. \citet{Chen2013} estimates $df$ by
\begin{equation}\label{eq:df_mi_lasso}
    \hat{df}_{MI-LASSO} = \sum_{j=1}^p\mathbbm{1}\left(\|\hat{\bb}_j\|_2>0\right) + \sum_{j=1}^p\frac{\|\hat{\bb}_j\|_2}{\|\hat{\bb}^{OLS}_j\|_2}(D-1),
\end{equation}
where $\mathbbm{1}(\cdot)$ is the indicator function. $\text{BIC}_{m}$ is evaluated over a sequence of $\xi$ values, and the value of $\xi$ that minimizes $\text{BIC}_m$ is selected.

\subsection{Bayesian MI-LASSO}
\label{sec:BMI_LASSO}
With the development of Bayesian group LASSO in recent years \citep{Raman2009, Carvalho2009, HernandezLobato2013,Xu2015}, it is appealing to extend the MI-LASSO into the Bayesian framework. In this section, we adapt some established Bayesian group LASSO approaches and propose four Bayesian MI-LASSO models. The corresponding variable selection strategy for these Bayesian MI-LASSO models is discussed in Section~\ref{sec:pvs}.

The outcome model for each Bayesian MI-LASSO is specified as:
\begin{equation}
    \by^{d} \sim \mathcal{MVN}(\bx^d\bb^d, \sigma^2\mathbf{I}), \ d=1,2,\ldots,D
\end{equation}
This model can be regarded as $D$ regressions with a common variance $\sigma^2$ to control overall variability. We further set non-informative priors $\pi(\sigma^2) \propto \frac{1}{\sigma^2}$ for variance $\sigma^2$, following \citet{Gelman2006}. The key idea of Bayesian MI-LASSO is to place a shared, penalized prior $\pi_j^{\beta}(\cdot)$ on $j$-th grouped coefficient vector $\bb_j$, i.e. $\beta_j^1,\beta_j^2,\ldots,\beta_j^D \stackrel{i.i.d}{\sim} \pi^{\beta}_j(\cdot)$, for $j=1,\ldots,p$. Similar to the original MI-LASSO, each covariate is treated as a grouped variable across imputations, and the group is either retained or excluded based on the posterior distribution of $\bb_j$. This structure enables consistent variable selection across all imputations. 

By specifying different priors $\pi^{\beta}_j(\cdot)$, we introduce four Bayesian MI-LASSO models: three shrinkage-based models and one Spike-and-Slab-type model. 

\subsubsection{Shrinkage Bayesian MI-LASSO Models}
\label{sec:shrinkage}
Shrinkage-based Bayesian MI-LASSO models apply continuous priors that concentrate mass near zero, encouraging coefficients to be small but not exactly zero. As a result, these models produce smooth posterior estimates of the coefficients, offering stable and robust inference in practice.

The first Bayesian MI-LASSO model specifies \textbf{Multi-Laplace} priors for grouped regression coefficients, inspired by the original Bayesian group LASSO proposed in \citet{Raman2009}. In essence, the posterior mode under this prior is the direct Bayesian analogue of the group LASSO penalty. To facilitate more efficient posterior computation, we follow the hierarchical formulation proposed in the original work, replacing the Multi-Laplace prior with a scale mixture of normals through a Gamma hyperprior on the group-specific variances. Throughout this paper, we use the shape-scale form for all Gamma distributions. Let $\boldsymbol{\Lambda}=diag(\lambda_1^2,\lambda_2^2,\ldots,\lambda_p^2)$, where $\lambda_j^2$ denotes the local shrinkage parameter for $j$-th group coefficients $\bb_j$. The Multi-Laplace Bayesian MI-LASSO is specified as follows:
\begin{equation*}
\operatorname{} \left \{
\begin{array}{c}
\bb^d \stackrel{i.i.d}{\sim} \mathcal{MVN}(0, \sigma^2 \boldsymbol{\Lambda}), \ d = 1, 2, \ldots, D \\
\lambda_j^2 \stackrel{i.i.d}{\sim} \text{Gamma}(\frac{D + 1}{2}, \frac{2}{D\rho}), \ j = 1, 2, \ldots, p \\
\rho \sim \text{Gamma}(h, v) \\
\end{array}
\right.
\end{equation*}
where $h$ and $v$ are positive hyperparameters to be specified in advance. \citet{Raman2009} showed $\bb_j \cond \rho \sim \text{Multi-Laplace}(0, \sqrt{\frac{\sigma^2}{D\rho}})$. Compared to a simple normal prior, the Multi-Laplace prior has heavier tails, preserving large values of $\bb$ in non-sparse regression.

The second shrinkage Bayesian MI-LASSO model leverages \textbf{Horseshoe} priors \citep{Carvalho2009}. Compared with the Multi-Laplace prior, the Horseshoe prior introduces an additional global parameter $\tau^2$ to control the overall shrinkage of all coefficients. By leveraging both global and local shrinkage effects, the Horseshoe prior adaptively facilitates strong shrinkage of noise coefficients while keeping large signals untouched, thereby achieving a highly desirable property for sparse modeling. The Horseshoe Bayesian MI-LASSO model is given by:
\begin{equation*}
\operatorname{} \left \{
\begin{array}{c}
\bb^d \stackrel{i.i.d}{\sim} \mathcal{MVN}(0, \sigma^2\tau^2\boldsymbol{\Lambda}), \ d = 1, 2, \ldots, D \\
\lambda_j \stackrel{i.i.d}{\sim} C^+(0, 1), \ j = 1, 2, \ldots, p \\
\tau \sim C^+(0, 1) \\
\end{array}
\right.
\end{equation*}
where $C^+(0, 1)$ denotes the standard half-Cauchy distribution.

The last shrinkage Bayesian MI-LASSO model we propose is a simple yet effective extension of the grouped Automatic Relevance Determination (\textbf{ARD}) model. The ARD prior, a classic model in the machine learning literature, effectively learns the relevance of each covariate by assigning it a unique precision parameter that is estimated from the data:
\begin{equation*}
\operatorname{} \left \{
\begin{array}{c}
\bb^d \stackrel{i.i.d}{\sim} \mathcal{MVN}(0, \sigma^2\boldsymbol{\Psi}^{-1}), \ d = 1, 2, \ldots, D \\
\pi(\psi^2_j) \stackrel{i.i.d}{\propto}  \frac{1}{\psi^2_j}, \ j = 1, 2, \ldots, p \\
\end{array}
\right.
\end{equation*}
where $\boldsymbol{\Psi}=diag(\psi^{2}_1,\psi^{2}_2,\ldots,\psi^{2}_p)$. Unlike Multi-Laplace and Horseshoe priors, which specify priors on variance, the ARD prior places non-informative priors directly on each group-specific precision parameter \citep{HernandezLobato2013}. When marginalizing over $\psi_j^2$, this specification induces an improper prior  $\pi(\beta_j^d \mid \sigma^2) \propto \int_0^\infty \mathcal{N}(0, \sigma^2/\psi_j^2) \frac{1}{\psi_j^2} d\psi_j^2 \propto \frac{1}{|\beta_j^d|}$, which has an infinite spike at zero and heavy tails. It therefore strongly shrinks small coefficients toward zero while allowing large coefficients to remain relatively unshrunk.

\subsubsection{Spike-and-Laplace Bayesian MI-LASSO Model}
While shrinkage-based Bayesian MI-LASSO methods can perform variable selection by shrinking coefficients toward zero, they do not produce exact zeros. In contrast, Spike-and-Slab priors \citep{Mitchell1988} perform explicit variable selection by modeling each coefficient as a mixture: a ``Spike'' at exactly zero for irrelevant variables and a continuous ``Slab'' for relevant ones. This formulation enables direct estimation of posterior inclusion probabilities for each variable.

A classic choice uses a normal ``Slab'' in Spike-and-Slab priors. In high-dimensional settings, however, heavier-tailed slabs provide better variable selection performance. As an alternative, the \textbf{Spike-and-Laplace} prior \citep{Xu2015} simply replaces the normal components with Multi-Laplace distributions for stronger shrinkage:
\begin{equation*}
\operatorname{} \left \{
\begin{array}{c}
\bb^d \stackrel{i.i.d}{\sim} \mathcal{MVN}(0, \sigma^2\boldsymbol{\Lambda})\boldsymbol{\gamma}+\boldsymbol{\delta}(\bb^d)(1 - \boldsymbol{\gamma}), \ d = 1, 2, \ldots, D \\
\lambda_j^2 \stackrel{i.i.d}{\sim} \text{Gamma}(\frac{D + 1}{2}, \frac{2}{D\rho}), \ j = 1, 2, \ldots, p \\
\gamma_j \stackrel{i.i.d}{\sim} \text{Bernoulli}(\theta_j), \ j = 1, 2, \ldots, p \\
\theta_j \stackrel{i.i.d}{\sim} \text{Beta}(1, 1), \ j = 1, 2, \ldots, p \\
\rho \sim \text{Gamma}(a, b)\\
\end{array}
\right.
\end{equation*}
where $\boldsymbol{\gamma}=(\gamma_1,\gamma_2,\ldots,\gamma_p)^\top$ is a binary inclusion vector, and $\boldsymbol{\delta}(\cdot)$ is a Dirac function placing point mass at zero. The hyperparameters $a$ and $b$ are positive, similar to those in the Multi-Laplace prior. This approach combines two selection mechanisms: the selection probability $\theta_j$ controls inclusion, while the Multi-Laplace shrinks the size of nonzero coefficients.

\subsubsection{Prior Specification and Posterior Computation}
\label{sec:prior}
Since the Horseshoe and ARD priors have no additional hyperparameters, no hyperparameter tuning is required for those models. For the Multi-Laplace and Spike-and-Laplace models, we first fix $h = 2$ and $a = 2$ to obtain weakly informative hyperpriors, and then tune $v$ and $b$. We adopt a strategy based on matching the prior variance of the coefficients $\beta_j^d$ to the error variance $\sigma^2$. As shown in Table~\ref{tab:hyperparameters}, $\text{Var}(\beta_j^d \cond \sigma^2)$ is an explicit function of $\sigma^2$ and the respective hyperparameters. Setting $Var(\beta_j^d|\sigma^2)=\sigma^2$ yields the default choices $v = \frac{D+1}{D}$ for the Multi-Laplace model and $b = \frac{D+1}{2D}$ for the Spike-and-Laplace model. For the Spike-and-Laplace model, the marginal variance additionally incorporates the prior inclusion probability $\mathrm{E}[\theta_j]=\tfrac12$; consequently, $b$ is one-half of the corresponding Multi-Laplace value so that the marginal prior variance remains equal to $\sigma^2$. This matching approach has two advantages: (i) it yields a default hyperparameter specification that adapts to the number of imputations $D$, and (ii) it ensures that the prior variability is on a comparable scale with the variability of the outcome, which allows these prior distributions to adapt to the variability in the data with consistent scaling across the two Bayesian MI-LASSO models. Table~\ref{tab:hyperparameters} shows the specifications of hyperparameters when $D=5$.

We estimate the four Bayesian MI-LASSO models using Markov chain Monte Carlo (MCMC). The sampling algorithms for all four models are provided as Web Algorithms 1--4 in Supplementary Materials S1.

\begin{table}[!htbp]
\centering
\begin{tabular}{cccc}
\toprule
\textbf{Model} & $\text{Var}(\beta_j^d \cond \sigma^2)$ & \textbf{Hyperparameters}  & \textbf{Default Value} \\
\midrule
 \multirow{2}{*}{Multi-Laplace} & \multirow{2}{*}{$\sigma^2 \frac{D+1}{Dv(h-1)}$} & $h$ & 2 \\ 
 & & $v$ & $\frac{D+1}{D}=1.2$ \\
\midrule
 \multirow{1}{*}{Horseshoe}  & \multicolumn{3}{c}{No hyperparameter} \\ 
\midrule
 \multirow{1}{*}{ARD} & \multicolumn{3}{c}{No hyperparameter}  \\ 
\midrule
\multirow{2}{*}{Spike-and-Laplace} & \multirow{2}{*}{$\frac{1}{2} \sigma^2 \frac{D+1}{Db(a-1)}$} & a & 2 \\ 
& & b & $\frac{D+1}{2D}=0.6$ \\
\bottomrule
\end{tabular}
\caption{Default values of hyperparameters for Bayesian MI-LASSO models, when $D=5$.}
\label{tab:hyperparameters}
\end{table}

\subsection{A Four-Step Projection Predictive Variable Selection Procedure}\label{sec:pvs}
Compared with the classic MI-LASSO, Bayesian MI-LASSO provides a Bayesian framework for combining inference across multiple imputations. Frequentist multiple-imputation analyses commonly combine estimates and their variances across imputations using Rubin's rules \citep{Rubin1987}. In a Bayesian analysis, posterior draws obtained from separate imputations can instead be combined directly, so that the resulting mixture posterior incorporates both within- and between-imputation uncertainty \citep{Gelman2013}. In our Bayesian MI-LASSO framework, we adopt this strategy by combining posterior draws across the $D$ imputations to obtain a pooled distribution for $\bb^{pool}$.  

However, selecting a final set of variables from the posterior of $\beta_j^{pool}$ remains challenging. Continuous shrinkage priors concentrate posterior draws around zero without producing exact zeros, while the Spike-and-Laplace prior allows variable inclusion to vary across MCMC draws. Common selection rules rely on fixed thresholds. For shrinkage models, the $j$-th covariate is included if the fixed-level credible interval (e.g., 95\%) of $\beta^{pool}_j$ excludes zero. For the Spike-and-Laplace model, the $j$-th covariate is selected if the posterior mean of its inclusion indicator $\gamma_j$ exceeds a specified cutoff. However, these thresholds are typically chosen in an ad hoc manner and the resulting selection can be sensitive to these thresholds. \citet{Li2010} found that selection based on 95\% credible intervals could retain too few important variables under the shrinkage models they considered. 

Obtaining coefficient estimates for a reduced model presents an additional challenge. Simply dropping excluded covariates while retaining the remaining coefficients does not generally yield estimates appropriate for the reduced model. Projection predictive inference \citep{Piironen2017, Piironen2020,McLatchie2025} provides an approach to obtain coefficient estimates for the reduced models while propagating uncertainty from the full model. It projects posterior draws from the full model onto a candidate submodel, adjusting the retained coefficients to preserve predictive information. Implementing this approach requires a strategy for generating candidate subsets and a criterion for selecting among them. However, projection predictive inference has its own potential limitation. When projecting the full model into submodels, the residual variance may absorb some of the posterior variance of projected regression coefficients, making them over-confident \citep{McLatchie2025}. This motivates the adjustment to the posterior variances of the projected coefficients for interval estimation.

To support variable selection and inference in reduced models, we propose a four-step procedure. First, we generate a candidate collection of variable subsets across a prespecified grid of selection thresholds, avoiding reliance on a single fixed threshold. Second, we project posterior draws from the full model onto each candidate subset to obtain reduced models that approximate the full model's predictions. Third, we evaluate each projected model using BIC$_m$, defined in~\eqref{eq:bic} and~\eqref{eq:bic_mi_B}, and select the subset of variables that minimizes BIC$_m$. Finally, we adjust the posterior variances of the projected coefficients to provide interval estimates for the selected submodel. This procedure applies to both continuous-shrinkage and Spike-and-Laplace Bayesian MI-LASSO models, using a grid of selection thresholds between 0 and 1.

\subsubsection*{Step 1. Candidate Set Generation via a Grid of Selection Thresholds}

Rather than using a single fixed threshold to determine the final selected model, we sweep the threshold over a grid to generate a collection of candidate subsets. We pre-specify a grid $\{\phi_k \in [0,1] : k = 1, \ldots, K\}$, whose default choice runs from $0$ to $1$ in increments of $0.01$.

For the shrinkage models (Multi-Laplace, Horseshoe, and ARD), at each level $\phi_k$, we retain the covariates whose central $\phi_k$-level pooled credible interval excludes zero,
\[ s_k = \left\{ j : 0 \notin \left[ q_{\frac{1-\phi_k}{2}}\!\left(\beta_j^{pool}\cond \text{data}\right),\; q_{\frac{1+\phi_k}{2}}\!\left(\beta_j^{pool}\cond \text{data}\right) \right] \right\}, \]
where $q_\alpha(\beta_j^{pool}\cond \text{data})$ is the $\alpha$-quantile of the pooled posterior of $\beta_j^{pool}$. A larger $\phi_k$ widens the interval and hence retains fewer covariates. For the Spike-and-Laplace model, at each level $\phi_k$, we retain the covariates whose posterior inclusion probability is at least $\phi_k$,
\[ s_k = \left\{ j : \pi\!\left(\gamma_j = 1 \cond \text{data}\right) \geq \phi_k \right\}. \]
When $\phi_k = 0.5$, $s_k$ recovers the median probability model \citep{Barbieri2004}. For both criteria, $\phi_k = 0$ retains all covariates, recovering the full model, whereas $\phi_k = 1$ retains at most the most strongly supported covariates and is typically empty.

We collect all unique subsets into the candidate collection $S = \{ s_k: k = 1, \ldots, K \}$. We discard any candidate whose design submatrix is rank-deficient, as well as any candidate with more than $n-2$ variables, which would over-saturate the projection and yield a degenerate $\mathrm{BIC}_m$. The next two steps evaluate a projected model for each unique $s \in S$.

\subsubsection*{Step 2. Posterior Projection onto Candidate Subsets}
For each candidate subset $s \in S$, we construct a reduced model by projecting the posterior draws from the Bayesian MI-LASSO full model onto the subspace spanned by the selected covariates. This projection step addresses the key challenge in submodels' coefficients estimation. A common but problematic practice for post-selection inference is to either (i) discard the coefficients of excluded variables or (ii) refit a new model using only the selected covariates. Both approaches ignore the posterior structure learned from the full model. In contrast, the projection predictive approach retains this information and yields a reduced model that preserves its predictive distribution \citep{Piironen2017, Piironen2020}.

We begin by introducing the notation used in the projection. Let $|s|$ denote the number of selected covariates for the subset $s$. For $d$-th imputed dataset, let $\bx^d_{[s]}$ and $\bb^d_{[s]}$ denote the submatrix and subvector of $\bx^d$ and $\bb^d$ corresponding to the selected subset $s$, with complements $\bx^d_{[-s]}$ and $\bb^d_{[-s]}$, respectively. We use superscripts $(t)$ to indicate the $t$-th posterior draw of a quantity; for example, $\bb^{d,(t)}$ denotes the $t$-th draw of $\bb^d$. We let $T$ denote the total number of posterior draws. We adopt the projection predictive approach \citep{Piironen2017, Piironen2020}, obtaining the projected coefficients $\bb^{(t)}_s \in \mathbb{R}^{|s| \times D}$ and residual variance ${\sigma_s^2}^{(t)}$ for the $t$-th posterior draw by minimizing the Kullback–Leibler divergence between the full and reduced models' predictive distributions:
\begin{equation}
\label{eq:KL}
    \min_{\bb^{d,(t)}_{s}, {\sigma^2_s}^{(t)}}KL\left(\prod_{d=1}^D\mathcal{MVN}(\bx^d\bb^{d,(t)}, {\sigma^2}^{(t)}\mathbf{I}), \prod_{d=1}^D\mathcal{MVN}(\bx_{[s]}^{d}\bb_{s}^{d,(t)}, {\sigma^2_s}^{(t)}\mathbf{I})\right)
\end{equation}
The closed-form solution of the optimization problem~\eqref{eq:KL} is 
\begin{align}
   & \bb^{d,(t)}_{s}=({\bx^d}^\top_{[s]}\bx^d_{[s]})^{-1}{\bx^d}^\top_{[s]}\bx^d\bb^{d,(t)} \label{eq:solution_beta} \\
   & {\sigma^2_s}^{(t)}={\sigma^2}^{(t)} + \frac{1}{nD}\sum_{d=1}^D\|\bx^d\bb^{d,(t)}-\bx_{[s]}^{d}\bb_{s}^{d,(t)}\|^2_2 \label{eq:solution_sigma2}
\end{align}
Since subsets with non-full-rank design submatrices are excluded, $({\bx^d}^\top_{[s]}\bx^d_{[s]})^{-1}$ exists for all $d=1,\ldots,D$, leading to a unique projection solution. This projection ensures that the predictive distribution of the reduced model remains as close as possible to that of the full model. Since the projected parameters are determined entirely by the fitted values of the full model, not by re-fitting to the data, the projection preserves the structural information of the full model. To see this explicitly, we can rewrite~\eqref{eq:solution_beta} as $\bb^{d,(t)}_{s}=\bb^{d,(t)}_{[s]}+({\bx^d}^\top_{[s]}\bx^d_{[s]})^{-1}{\bx^d}^\top_{[s]}\bx^d_{[-s]}\bb_{[-s]}^{d,(t)}$. This expression shows that the projected coefficients $\bb_s^{d,(t)}$ equal the original coefficients in the selected subset, $\bb^{d,(t)}_{[s]}$, plus a correction term that accounts for the contribution of excluded coefficients $\bb^{d,(t)}_{[-s]}$. Furthermore, if $\bx_{[s]}^d$ and $\bx_{[-s]}^d$ are orthogonal, or if $\bb_{[-s]}^{d,(t)} = 0$, then the correction term vanishes and the projection reduces to retaining only the selected subset coefficients, i.e., $\bb_s^{d,(t)} = \bb_{[s]}^{d,(t)}$. In contrast, the projected residual variance ${\sigma_s^2}^{(t)}$ is always greater than or equal to the original variance ${\sigma^2}^{(t)}$, with the difference reflecting the loss in predictive accuracy incurred by projecting onto a lower-dimensional subspace, as shown in~\eqref{eq:solution_sigma2}. When the column space of $\bx^d_{[s]}$ spans that of $\bx^d$, the two variances are equal, indicating that the excluded variables are redundant and the reduced model retains the full predictive capacity of the original model.

When posterior means of the projected coefficients are calculated, computational burden can be further reduced. The projection of the posterior mean satisfies $\E[\bb_s^{d}\cond \text{data}]=({\bx^d}^\top_{[s]}\bx^d_{[s]})^{-1}{\bx^d}^\top_{[s]}\bx^d\E[\bb^{d}\cond \text{data}]$, which is a linear function of $\E[\bb^{d}\cond \text{data}]$. So we only need to project once for each candidate subset $s \in S$, using the posterior mean from the full model.

\subsubsection*{Step 3. Model Selection via Modified BIC}
\label{sec:df}
After the first two steps, we have obtained the projected estimates for each candidate subset $s \in S$. We now evaluate model performance using the modified Bayesian information criterion (BIC$_m$) defined in~\eqref{eq:bic}. Since BIC is a point-estimation criterion, we compute it based on the posterior mean of the projected coefficients. Specifically, for each subset $s$, BIC$_m(s)$ is given by:
\begin{equation} \label{eq:bic_mi_B}
\text{BIC}_{m}(s) = \log\left(\sum_{d=1}^D\|\by^d - \bx_{[s]}^d\E[\bb_s^d\cond \text{data}]\|_2^2 / (Dn)\right)+df(s)\cdot\log(Dn)/(Dn)
\end{equation}
We then select the optimal subset $s_m$ that minimizes BIC$_m(s)$ over all $s \in S$.

A key component in computing BIC$_m(s)$ is the degrees of freedom ${df}(s)$. The existing formula~\eqref{eq:df_mi_lasso} applies specifically to the original MI-LASSO, and cannot be used directly for our projected Bayesian MI-LASSO models. Therefore, we derive separate expressions $\hat{df}(s)$ for estimating ${df}(s)$ tailored to each Bayesian MI-LASSO variant. These expressions are based on MCMC draws from the full model and take the following forms:
\begin{equation}
\resizebox{\textwidth}{!}{$
\begin{aligned}
\label{eq:df_B_MI_LASSO}
\hat{df}(s)_{\text{Multi-Laplace}} &= \frac{1}{T} \sum_{t=1}^T\sum_{d=1}^D \Tr\left(\bx^d_{[s]}({\bx_{[s]}^d}^\top\bx_{[s]}^d)^{-1}{\bx_{[s]}^d}^\top\bx^d [{\boldsymbol{\Lambda}^{(t)}}^{-1} + {\bx^d}^\top\bx^d]^{-1}{\bx^d}^\top\right) \\[0.5em]
\hat{df}(s)_{\text{Horseshoe}} &= \frac{1}{T}\sum_{t=1}^T\sum_{d=1}^D \Tr\left(\bx^d_{[s]}({\bx_{[s]}^d}^\top\bx_{[s]}^d)^{-1}{\bx_{[s]}^d}^\top\bx^d [({\tau^2}^{(t)}{\boldsymbol{\Lambda}}^{(t)})^{-1} + {\bx^d}^\top\bx^d]^{-1}{\bx^d}^\top\right) \\[0.5em]
\hat{df}(s)_{\text{ARD}} &= \frac{1}{T} \sum_{t=1}^T\sum_{d=1}^D \Tr\left(\bx^d_{[s]}({\bx_{[s]}^d}^\top\bx_{[s]}^d)^{-1}{\bx_{[s]}^d}^\top\bx^d [{\boldsymbol{\Psi}}^{(t)} + {\bx^d}^\top\bx^d]^{-1}{\bx^d}^\top\right) \\[0.5em]
\hat{df}(s)_{\text{Spike-and-Laplace}} &= \frac{1}{T} \sum_{t=1}^T\sum_{d=1}^D \Tr\left(\bx^d_{[s]}({\bx_{[s]}^d}^\top\bx_{[s]}^d)^{-1}{\bx_{[s]}^d}^\top\bx^d_{[\boldsymbol{\gamma}^{(t)}]} \left[({\boldsymbol{\Lambda}}^{(t)}_{[\boldsymbol{\gamma}^{(t)}]})^{-1}+{\bx^d}_{[\boldsymbol{\gamma}^{(t)}]}^\top \bx^d_{[\boldsymbol{\gamma}^{(t)}]}\right]^{-1}{\bx^d}_{[\boldsymbol{\gamma}^{(t)}]}^\top\right)
\end{aligned}
$}
\end{equation}

\noindent Here, $\Tr(.)$ is the trace of a matrix. $\bx^d_{[\boldsymbol{\gamma}^{(t)}]}$ denotes the submatrix of $\bx^d$ corresponding to the active covariates with $\gamma_j^{(t)} = 1$. The submatrix $\boldsymbol{\Lambda}^{(t)}_{[\boldsymbol{\gamma}^{(t)}]}$ of $\boldsymbol{\Lambda}^{(t)}$ includes only those diagonal entries for which $\gamma_j^{(t)} = 1$. We further define \[\bx^d_{[\boldsymbol{\gamma}^{(t)}]} \left[({\boldsymbol{\Lambda}}^{(t)}_{[\boldsymbol{\gamma}^{(t)}]})^{-1}+{\bx^d}_{[\boldsymbol{\gamma}^{(t)}]}^\top \bx^d_{[\boldsymbol{\gamma}^{(t)}]}\right]^{-1}{\bx^d}_{[\boldsymbol{\gamma}^{(t)}]}^\top\equiv0, \ \text{if }\gamma^{(t)} = 0.\] The four expressions share a common form. Each estimates the effective number of parameters used by the submodel, obtained by combining the submodel projection with the penalized projection of the full model. Although the submodel has $D|s|$ coefficients across the $D$ imputations, the priors shrink these coefficients toward zero, so the effective number is generally smaller and $\hat{df}(s)\leq D|s|$. Equality holds only without penalization, i.e., as the precision parameters $\boldsymbol{\Lambda}^{(t)-1}$, $({\tau^2}^{(t)}\boldsymbol{\Lambda}^{(t)})^{-1}$, $\boldsymbol{\Psi}^{(t)}$, or the active-set $\boldsymbol{\Lambda}^{(t)-1}_{[\boldsymbol{\gamma}^{(t)}]}$ approach the zero matrix. The derivations of both $\text{BIC}_{m}(s)$ and ${df}(s)$ are provided in Supplementary Materials S2.
 
\subsubsection*{Step 4. Approximately Calibrated Post-Selection Inference}
\label{sec:ps_infer}
After selecting the best subset $s_m$, we obtain projected estimates $\bb_{s_m}^{d,(t)}$ and ${\sigma_{s_m}^2}^{(t)}$ by projecting each posterior draw onto that subset, setting the coefficients of excluded variables to zero. As discussed earlier, the variability of the projected coefficient draws may understate uncertainty in the submodel coefficients. We therefore adjust the projected draws within each imputation while preserving their posterior means as point estimates. This adjustment makes their within-imputation posterior covariance match the marginal posterior covariance of the submodel coefficients under a flat prior, so that the interval widths are determined by the data, restoring approximately calibrated coverage. Let $\bb_{s_m,\text{calibrated}}^{d,(t)}$ be the calibrated value of $\bb_{s_m}^{d,(t)}$. The calibration step is given by:
\begin{equation}
\label{eq:calibration}
\bb_{s_m,\text{calibrated}}^{d,(t)} = (\mathbf{V}_{s_m}^d)^{1/2}(\mathbf{U}_{s_m}^d)^{-1/2}\left(\bb_{s_m}^{d,(t)} - \E[\bb_{s_m}^d \cond \text{data}]\right)+\E[\bb_{s_m}^d \cond \text{data}],
\end{equation}
where $\mathbf{U}_{s_m}^d = \mathrm{Cov}(\bb_{s_m}^d \cond \text{data})$ is the posterior covariance of projected coefficients $\bb_{s_m}^d$, and $\mathbf{V}_{s_m}^d = \mathrm{Cov}(\bb_{s_m,\text{calibrated}}^{d}\cond \text{data}) = \E[{\sigma}_{s_m}^2\cond \text{data}]({\bx^d_{[s_m]}}^\top\bx^d_{[s_m]})^{-1}$ is the target covariance. The calibration preserves the posterior mean within each imputation,  $\E[\bb_{s_m,\text{calibrated}}^{d}\cond \text{data}]=\E[\bb_{s_m}^{d}\cond \text{data}]$, so the posterior-mean point estimates remain unchanged after calibration (see Supplementary Materials S3 for derivations). After calibrating coefficients, posterior draws of ${\sigma_{s_m}^2}^{(t)}$ are discarded, and the calibrated residual variance is then updated by ${\sigma_{s_m, \text{calibrated}}^2}^{(t)} = {\sigma^2}^{(t)} + \frac{1}{nD}\sum_{d=1}^D\|\bx^d\bb^{d,(t)}-\bx_{[s_m]}^{d}\bb_{s_m, \text{calibrated}}^{d,(t)}\|^2_2$. Pooled estimates are then obtained by combining the calibrated posterior draws for each coefficient across $D$ imputed datasets.  In our simulations, the calibrated intervals attain approximately nominal coverage for the regression coefficients of the selected variables under correct variable selection.

\section{Simulation Study}
\label{sec:simulation}

\subsection{Design}
We conducted a series of simulation studies to evaluate the variable selection performance of different methods on multiply-imputed datasets. We considered three simulations: the first two were based on independent or correlated continuous covariates, and the third generated correlated binary covariates. In each simulation scenario, covariates $\bX$, true regression coefficients $\bb$, and the outcome $\bY$ were specified from a Gaussian linear regression model: \begin{equation}
\label{eq:sim_outcome}
    \bY \sim \mathcal{N}(\bX\bb, \sigma^2),
\end{equation}
where $\sigma^2$ was determined according to a pre-specified signal-to-noise ratio $\mathrm{SNR}=\bb^\top \Sigma \bb / \sigma^2$, and $\Sigma$ denotes the covariance matrix of $\bX$. 

For each scenario, missing values were imposed in $\bX$ under either an MCAR or MAR mechanism. After imposing missingness, multiple imputation was performed using the \texttt{mice} package in \texttt{R} \citep{van2011mice}. The \texttt{mice} algorithm employs chained equations to specify univariate conditional models for each variable with missing values, conditional on all other variables (including $\bY$).  The default predictive mean matching imputation method was used for continuous variables with five donors, and logistic regression imputation for binary variables. The number of imputations was set to $D=5$ with five iterations each, in all simulations.

We compared the four Bayesian MI-LASSO models with existing methods, including four models applied to full or incomplete datasets and three methods for multiply-imputed data:
\begin{itemize}
    \item {\bf LASSO} (full dataset): Applied to the full dataset before imposing missing values. The penalty parameter $\xi$ was set on a grid of $2^{l}$, with $l \in [-3, 7]$ in increments of 0.1. The optimal $\xi$ was selected by minimizing the classic BIC.
    
    \item {\bf BLASSO} (full dataset): The Bayesian LASSO \citep{Park2008}, implemented via the \texttt{blasso} function in the \texttt{R} package \texttt{monomvn}, which infers models by Reversible Jump MCMC. Variables with posterior median coefficients equal to zero were excluded.
    
    \item {\bf CC-LASSO} (complete cases): LASSO applied to complete cases only, with tuning performed as in LASSO.
    
    \item {\bf CC-BLASSO} (complete cases): Bayesian LASSO applied to complete cases only, with implementation identical to BLASSO.
    
\item {\bf MI-LASSO} (MI data): Used the same $\xi$ grid as LASSO, with the optimal parameter selected by minimizing $\mathrm{BIC}_m$ in~\eqref{eq:bic}.
    
    \item {\bf SaENET} (MI data): Stacks multiply-imputed datasets into a single dataset and applies an adaptive elastic net for variable selection \citep{Du2022}. Tuning parameters were selected via 5-fold cross-validation using the \texttt{R} package \texttt{miselect}.
    
    \item {\bf GaLASSO} (MI data): A variant of MI-LASSO that replaces the group LASSO with adaptive group LASSO \citep{Du2022}. Tuning was performed identically to SaENET using 5-fold cross-validation in \texttt{miselect}.
\end{itemize}
BLASSO, CC-BLASSO and the four Bayesian MI-LASSO models were estimated using MCMC with 5,000 burn-in iterations followed by 5,000 sampling iterations. Hyperparameters for the Bayesian MI-LASSO models were set to default values (see Table~\ref{tab:hyperparameters}). We normalized each $\bx^d$ and centered $\by^d$ before fitting models. Each replicate used a distinct random seed for the data generation, the imposed missingness, and the sampling. Using a single chain per model, we monitored the rank-normalized split-$\hat R$ and confirmed its maximum over the full model, calibrated coefficients, and the residual variance remained below 1.01 \citep{Vehtari2021}.

Five metrics were used to evaluate model performance. The first three assess variable selection accuracy, treating selection as a binary classification task; the last two measure post-selection estimation and sign recovery of the estimated coefficients:
\begin{itemize}[
  itemsep=0.3em,
  parsep=0em
]
  \item Sensitivity (\textbf{SEN}): the proportion of truly important variables that are correctly selected;

  \item Specificity (\textbf{SPE}): the proportion of truly unimportant variables that are correctly removed;

  \item F$_1$ score (\textbf{F}$_1$): $\frac{2\,\text{PRE}\times\text{SEN}}{\text{PRE} + \text{SEN}}$, where PRE = the proportion of selected variables that are truly important;

  \item Mean squared error (\textbf{MSE}): $\E[(\tilde{\boldsymbol\beta} - \boldsymbol\beta)^\top\bX^\top\bX(\tilde{\boldsymbol\beta} - \boldsymbol\beta)]$, where $\tilde{\boldsymbol\beta}$ is the estimated coefficient vector;

  \item  Sign accuracy (\textbf{SIGN}): $\frac{1}{p}\sum_{j=1}^p\mathbbm{1}(\mathrm{sign}(\tilde{\beta}_j)=\mathrm{sign}(\beta_j))$, where $\mathrm{sign}(\cdot)$ is the sign function returning $1$, $-1$ and $0$ for positive, negative, and zero inputs, respectively.
\end{itemize}
For Bayesian MI-LASSO models, $\tilde{\boldsymbol\beta}$ denotes the pooled posterior mean of the calibrated coefficients for selected variables, with excluded coefficients set to 0. For BLASSO and CC-BLASSO, we used the posterior medians; for LASSO, CC-LASSO, and SaENET, we used the point estimates directly. For MI-LASSO and GaLASSO, we averaged the coefficient estimates across the $D$ imputed datasets, using refitted estimates for MI-LASSO \citep{Chen2013} and penalized estimates for GaLASSO \citep{Du2022}. We summarized the results by reporting the mean of each metric across 200 simulation replicates.

\subsection{Simulation Scenario 1: Independent Continuous Covariates}
In the first simulation, each covariate $\bX_j$ was independently generated from a standard normal distribution. We used a baseline configuration with sample size $n=100$, number of covariates $p=20$, and an SNR of 1.5, meaning that approximately 60\% of the variation in $\bY$ is explained by the linear terms $\bX\bb$, and the remaining 40\% is due to noise. We then varied one factor at a time across eight additional settings: higher missingness rate, increased sample size ($n=200$), increased dimensionality ($p=40$), higher SNR (2.5), lower SNR (0.5), weaker signal (SNR=$0.2$), relatively higher dimensionality ($p=80$), and high dimensionality ($p=200$).

\begin{table}[!htbp]
\centering
\resizebox{\textwidth}{!}{
\begin{tabular}{cccccccccccccccccccccccc}
\toprule
 & \multicolumn{10}{c}{{\normalsize \textbf{n=100, p=20, SNR=1.5, moderate missing}}} & & & & \multicolumn{10}{c}{\normalsize \textbf{n=100, p=20, SNR=1.5, high missing}} \\ 
 \midrule
\textbf{Full Data} &  \multicolumn{2}{c}{\textbf{SEN}} & \multicolumn{2}{c}{\textbf{SPE}} & \multicolumn{2}{c}{\textbf{F}$_1$} & \multicolumn{2}{c}{\textbf{MSE}} & \multicolumn{2}{c}{\textbf{SIGN}} & & & & \multicolumn{2}{c}{\textbf{SEN}} & \multicolumn{2}{c}{\textbf{SPE}} & \multicolumn{2}{c}{\textbf{F}$_1$} & \multicolumn{2}{c}{\textbf{MSE}} & \multicolumn{2}{c}{\textbf{SIGN}} \\
LASSO & \multicolumn{2}{c}{100} & \multicolumn{2}{c}{79} & \multicolumn{2}{c}{82} & \multicolumn{2}{c}{82} & \multicolumn{2}{c}{85} & & & & \multicolumn{2}{c}{100} & \multicolumn{2}{c}{79} & \multicolumn{2}{c}{82} & \multicolumn{2}{c}{82} & \multicolumn{2}{c}{85}\\  
 BLASSO & \multicolumn{2}{c}{100} & \multicolumn{2}{c}{91} & \multicolumn{2}{c}{91} & \multicolumn{2}{c}{52} & \multicolumn{2}{c}{93} & & & & \multicolumn{2}{c}{100} & \multicolumn{2}{c}{91} & \multicolumn{2}{c}{91} & \multicolumn{2}{c}{52} & \multicolumn{2}{c}{93}\\
 \midrule
\multirow{2}*{\textbf{Models}} & \multicolumn{5}{c}{MCAR} & & \multicolumn{5}{c}{MAR} & & \multicolumn{5}{c}{MCAR} & & \multicolumn{5}{c}{MAR} \\ 
\cmidrule{2-6} \cmidrule{8-12} \cmidrule{14-18} \cmidrule{20-24}
 & \textbf{SEN} & \textbf{SPE} & \textbf{F}$_1$ & \textbf{MSE} & \textbf{SIGN} & & \textbf{SEN} & \textbf{SPE} & \textbf{F}$_1$ & \textbf{MSE} & \textbf{SIGN} & & \textbf{SEN} & \textbf{SPE} & \textbf{F}$_1$ & \textbf{MSE} & \textbf{SIGN} & & \textbf{SEN} & \textbf{SPE} & \textbf{F}$_1$ & \textbf{MSE} & \textbf{SIGN} \\
  CC-LASSO & 91 & 75 & 73 & 164 & 80 & & 68 & 81 & 57 & 312 & 77 & & 72 & 67 & 54 & 462 & 68 & & 39 & 79 & 31 & 516 & 67\\ 
  CC-BLASSO & 86 & 93 & 84 & 230 & 91 & & 54 & \textbf{96} & 60 & 447 & 83 & & 36 & \textbf{97} & 44 & 522 & 79 & & 11 & \textbf{99} & 16 & 585 & 72\\ 
  SaENET & \textbf{100} & 76 & 80 & 69 & 83 & & \textbf{100} & 74 & 78 & 77 & 82 & & 99 & 73 & 78 & 83 & 81 & & 99 & 67 & 74 & 104 & 77\\ 
  GaLASSO & 99 & 94 & 93 & 54 & 95 & & 98 & 93 & 92 & 58 & 95 & & 98 & 93 & 92 & 62 & 95 & & 98 & 91 & 90 & 77 & 93\\ 
  MI-LASSO & \textbf{100} & 78 & 80 & 86 & 85 & & \textbf{100} & 76 & 79 & 92 & 83 & & \textbf{100} & 76 & 79 & 91 & 83 & & \textbf{100} & 70 & 75 & 108 & 79\\ 
    \cmidrule{1-1}
  Multi-Laplace & 99 & \textbf{97} & \textbf{96} & 48 & \textbf{97} & & 98 & \textbf{96} & \textbf{95} & 55 & \textbf{97} & & 98 & 95 & \textbf{94} & 60 & \textbf{96} & & 97 & 93 & \textbf{92} & 77 & \textbf{94}\\ 
  Horseshoe & 99 & 95 & 95 & 50 & 96 & & 99 & 95 & 94 & 56 & 96 & & 98 & 94 & 93 & 61 & 95 & & 98 & 91 & 90 & 78 & 93\\ 
  ARD & 99 & 95 & 95 & 50 & 96 & & 99 & 95 & 94 & 55 & 96 & & 98 & 94 & 93 & 60 & 95 & & 98 & 91 & 90 & 78 & 93\\ 
  Spike-and-Laplace & 99 & \textbf{97} & \textbf{96} & \textbf{46} & \textbf{97} & & 98 & \textbf{96} & \textbf{95} & \textbf{54} & \textbf{97} & & 98 & 96 & \textbf{94} & \textbf{59} & \textbf{96} & & 98 & 93 & \textbf{92} & \textbf{75} & \textbf{94}\\ 
\bottomrule
\toprule
 & \multicolumn{10}{c}{{\normalsize \textbf{n=200, p=20, SNR=1.5, moderate missing}}} & & & & \multicolumn{10}{c}{\normalsize \textbf{n=100, p=40, SNR=1.5, moderate missing}} \\ 
 \midrule
\textbf{Full Data} &  \multicolumn{2}{c}{\textbf{SEN}} & \multicolumn{2}{c}{\textbf{SPE}} & \multicolumn{2}{c}{\textbf{F}$_1$} & \multicolumn{2}{c}{\textbf{MSE}} & \multicolumn{2}{c}{\textbf{SIGN}} & & & & \multicolumn{2}{c}{\textbf{SEN}} & \multicolumn{2}{c}{\textbf{SPE}} & \multicolumn{2}{c}{\textbf{F}$_1$} & \multicolumn{2}{c}{\textbf{MSE}} & \multicolumn{2}{c}{\textbf{SIGN}} \\
LASSO & \multicolumn{2}{c}{100} & \multicolumn{2}{c}{85} & \multicolumn{2}{c}{86} & \multicolumn{2}{c}{39} & \multicolumn{2}{c}{89} & & & & \multicolumn{2}{c}{79} & \multicolumn{2}{c}{83} & \multicolumn{2}{c}{69} & \multicolumn{2}{c}{478} & \multicolumn{2}{c}{82}\\  
 BLASSO & \multicolumn{2}{c}{100} & \multicolumn{2}{c}{93} & \multicolumn{2}{c}{93} & \multicolumn{2}{c}{19} & \multicolumn{2}{c}{95} & & & & \multicolumn{2}{c}{86} & \multicolumn{2}{c}{92} & \multicolumn{2}{c}{83} & \multicolumn{2}{c}{426} & \multicolumn{2}{c}{90}\\
 \midrule
\multirow{2}*{\textbf{Models}} & \multicolumn{5}{c}{MCAR} & & \multicolumn{5}{c}{MAR} & & \multicolumn{5}{c}{MCAR} & & \multicolumn{5}{c}{MAR} \\ 
\cmidrule{2-6} \cmidrule{8-12} \cmidrule{14-18} \cmidrule{20-24}
 & \textbf{SEN} & \textbf{SPE} & \textbf{F}$_1$ & \textbf{MSE} & \textbf{SIGN} & & \textbf{SEN} & \textbf{SPE} & \textbf{F}$_1$ & \textbf{MSE} & \textbf{SIGN} & & \textbf{SEN} & \textbf{SPE} & \textbf{F}$_1$ & \textbf{MSE} & \textbf{SIGN} & & \textbf{SEN} & \textbf{SPE} & \textbf{F}$_1$ & \textbf{MSE} & \textbf{SIGN} \\
  CC-LASSO & \textbf{100} & 81 & 83 & 63 & 87 & & 97 & 78 & 79 & 159 & 84 & & 45 & 85 & 40 & 969 & 72 & & 12 & 96 & 12 & 1135 & 71\\ 
  CC-BLASSO & \textbf{100} & 92 & 92 & 36 & 94 & & 97 & 91 & 90 & 129 & 93 & & 43 & 96 & 52 & 952 & 80 & & 14 & \textbf{98} & 22 & 1157 & 73\\ 
  SaENET & \textbf{100} & 82 & 84 & 27 & 87 & & \textbf{100} & 77 & 81 & 34 & 84 & & \textbf{94} & 74 & 74 & \textbf{363} & 80 & & \textbf{93} & 70 & 71 & 432 & 77\\ 
  GaLASSO & \textbf{100} & 96 & 96 & 19 & 98 & & \textbf{100} & 95 & 95 & 23 & 96 & & 62 & \textbf{98} & 72 & 578 & 87 & & 59 & 97 & 69 & 630 & 86\\ 
  MI-LASSO & \textbf{100} & 82 & 83 & 39 & 87 & & \textbf{100} & 79 & 81 & 42 & 85 & & 85 & 86 & 76 & 465 & 85 & & 81 & 83 & 72 & 524 & 83\\ 
    \cmidrule{1-1}
  Multi-Laplace & \textbf{100} & 98 & \textbf{98} & \textbf{18} & \textbf{99} & & \textbf{100} & \textbf{98} & \textbf{98} & 21 & 98 & & 78 & 95 & 82 & 392 & \textbf{90} & & 77 & 93 & 79 & 442 & 88\\ 
  Horseshoe & \textbf{100} & 98 & \textbf{98} & 19 & 98 & & \textbf{100} & 97 & 97 & 21 & 98 & & 82 & 94 & \textbf{83} & 374 & \textbf{90} & & 81 & 92 & \textbf{80} & \textbf{427} & 88\\ 
  ARD & \textbf{100} & 98 & \textbf{98} & \textbf{18} & 98 & & \textbf{100} & 97 & 97 & 21 & 98 & & 82 & 94 & \textbf{83} & 376 & \textbf{90} & & 80 & 92 & \textbf{80} & 438 & 88\\ 
  Spike-and-Laplace & \textbf{100} & \textbf{99} & \textbf{98} & \textbf{18} & \textbf{99} & & \textbf{100} & \textbf{98} & \textbf{98} & \textbf{20} & \textbf{99} & & 80 & 95 & \textbf{83} & 387 & \textbf{90} & & 79 & 93 & \textbf{80} & 440 & \textbf{89}\\ 
\bottomrule
\toprule
 & \multicolumn{10}{c}{{\normalsize \textbf{n=100, p=20, SNR=2.5, moderate missing}}} & & & & \multicolumn{10}{c}{\normalsize \textbf{n=100, p=20, SNR=0.5, moderate missing}} \\ 
 \midrule
\textbf{Full Data} &  \multicolumn{2}{c}{\textbf{SEN}} & \multicolumn{2}{c}{\textbf{SPE}} & \multicolumn{2}{c}{\textbf{F}$_1$} & \multicolumn{2}{c}{\textbf{MSE}} & \multicolumn{2}{c}{\textbf{SIGN}} & & & & \multicolumn{2}{c}{\textbf{SEN}} & \multicolumn{2}{c}{\textbf{SPE}} & \multicolumn{2}{c}{\textbf{F}$_1$} & \multicolumn{2}{c}{\textbf{MSE}} & \multicolumn{2}{c}{\textbf{SIGN}} \\
LASSO & \multicolumn{2}{c}{100} & \multicolumn{2}{c}{79} & \multicolumn{2}{c}{82} & \multicolumn{2}{c}{48} & \multicolumn{2}{c}{85} & & & & \multicolumn{2}{c}{63} & \multicolumn{2}{c}{90} & \multicolumn{2}{c}{60} & \multicolumn{2}{c}{318} & \multicolumn{2}{c}{82}\\  
 BLASSO & \multicolumn{2}{c}{100} & \multicolumn{2}{c}{92} & \multicolumn{2}{c}{92} & \multicolumn{2}{c}{25} & \multicolumn{2}{c}{94} & & & & \multicolumn{2}{c}{70} & \multicolumn{2}{c}{94} & \multicolumn{2}{c}{74} & \multicolumn{2}{c}{351} & \multicolumn{2}{c}{87}\\
 \midrule
\multirow{2}*{\textbf{Models}} & \multicolumn{5}{c}{MCAR} & & \multicolumn{5}{c}{MAR} & & \multicolumn{5}{c}{MCAR} & & \multicolumn{5}{c}{MAR} \\ 
\cmidrule{2-6} \cmidrule{8-12} \cmidrule{14-18} \cmidrule{20-24}
 & \textbf{SEN} & \textbf{SPE} & \textbf{F}$_1$ & \textbf{MSE} & \textbf{SIGN} & & \textbf{SEN} & \textbf{SPE} & \textbf{F}$_1$ & \textbf{MSE} & \textbf{SIGN} & & \textbf{SEN} & \textbf{SPE} & \textbf{F}$_1$ & \textbf{MSE} & \textbf{SIGN} & & \textbf{SEN} & \textbf{SPE} & \textbf{F}$_1$ & \textbf{MSE} & \textbf{SIGN} \\
  CC-LASSO & \textbf{100} & 71 & 77 & 84 & 80 & & 90 & 71 & 69 & 176 & 77 & & 33 & 93 & 33 & 470 & 75 & & 13 & \textbf{98} & 15 & 550 & 72\\ 
  CC-BLASSO & 99 & 92 & 91 & 81 & 94 & & 85 & 93 & 83 & 235 & 91 & & 30 & \textbf{98} & 39 & 532 & 78 & & 13 & \textbf{98} & 18 & 583 & 73\\ 
  SaENET & \textbf{100} & 78 & 81 & 41 & 84 & & \textbf{100} & 77 & 80 & 43 & 84 & & \textbf{91} & 71 & 72 & \textbf{227} & 77 & & \textbf{90} & 66 & 68 & \textbf{261} & 73\\ 
  GaLASSO & \textbf{100} & \textbf{96} & \textbf{96} & \textbf{27} & \textbf{97} & & \textbf{100} & \textbf{96} & \textbf{96} & \textbf{28} & \textbf{97} & & 71 & 92 & 74 & 282 & 86 & & 69 & 91 & 71 & 307 & 84\\ 
  MI-LASSO & \textbf{100} & 75 & 78 & 50 & 83 & & \textbf{100} & 75 & 79 & 54 & 83 & & 77 & 88 & 74 & 273 & 85 & & 74 & 86 & 69 & 299 & 82\\ 
    \cmidrule{1-1}
  Multi-Laplace & \textbf{100} & \textbf{96} & \textbf{96} & 28 & \textbf{97} & & \textbf{100} & \textbf{96} & \textbf{96} & 29 & \textbf{97} & & 65 & 96 & 73 & 273 & \textbf{87} & & 63 & 95 & 71 & 299 & \textbf{86}\\ 
  Horseshoe & \textbf{100} & 95 & 95 & 28 & 96 & & \textbf{100} & 95 & 94 & 31 & 96 & & 69 & 95 & \textbf{75} & 263 & \textbf{87} & & 67 & 94 & \textbf{73} & 293 & \textbf{86}\\ 
  ARD & \textbf{100} & 95 & 95 & 28 & 96 & & \textbf{100} & 95 & 94 & 30 & 96 & & 67 & 95 & 74 & 273 & \textbf{87} & & 65 & 94 & 71 & 303 & 85\\ 
  Spike-and-Laplace & \textbf{100} & \textbf{96} & \textbf{96} & \textbf{27} & \textbf{97} & & \textbf{100} & \textbf{96} & \textbf{96} & 29 & \textbf{97} & & 64 & 96 & 72 & 292 & 86 & & 62 & 95 & 69 & 323 & 85\\ 
\bottomrule
\end{tabular}
}
\caption{Results of sensitivity (×100), specificity (×100), $\text{F}_1$ score (×100), MSE (×100) and sign accuracy (×100) in Scenario 1. The best value per metric within each setting is in bold.}
\label{tab:sim_A}
\end{table}

For settings with $p = 20$, we set the true regression coefficients in model~\eqref{eq:sim_outcome} as $\beta_j = 1$ for $j = 1, 2, 5, 11, 12, 15$ and $\beta_j = 0$ otherwise. Missing values were introduced in  $\bX_j,\ j=11,12,\ldots,20$ under either MCAR or MAR. For MCAR, 5\% of entries were randomly removed from each $\bX_j$, yielding approximately 60\% complete cases. For MAR, missingness was generated via a logistic model: $\text{logit}(\Pr(x_{ij} \text{ is missing} \cond x_{i(j-10)}, y_i)) = -3.4 + 0.5 x_{i(j-10)} + 0.5 y_i$. In the high-missingness setting, we increased the missing rate in each covariate to 10\% for MCAR, and for MAR used a higher baseline probability with $\text{logit}(\Pr(x_{ij} \text{ is missing} \cond x_{i(j-10)}, y_i)) = -2.1 + 0.5 x_{i(j-10)} + 0.5 y_i$, leading to roughly 35\% complete cases. For $p = 40$, $80$ or $200$, we set $\beta_j = 1$ for $j = 1, 2, 5, 11, 12, 15, 21, 22, 25, 31, 32, 35$, and zero otherwise. Missingness was applied to $\bX_j$, $j=11,12,\ldots,20,31,32,\ldots,40$. Under MCAR, we imposed a uniform 2.5\% missing rate; under MAR, missingness followed $\text{logit}(\Pr(x_{ij} \text{ is missing} \cond x_{i(j-10)}, y_i)) = -4.3 + 0.5 x_{i(j-10)} + 0.5 y_i$, again resulting in approximately 60\% complete cases.

The results of Scenario 1 are summarized in Table~\ref{tab:sim_A}, while results of weaker signal (SNR=$0.2$), relatively higher dimensionality ($p=80$), and high dimensionality ($p=200$) are reported in Supplementary Materials S4.1. The four proposed Bayesian MI-LASSO models demonstrated consistently strong and similar performance, outperforming the alternatives in variable selection and estimation accuracy across most of settings. Multi-Laplace and Spike-and-Laplace tended to yield slightly higher specificity, while Horseshoe and ARD achieved slightly better sensitivity, especially under $p=40$ or $80$, SNR = $0.5$ or $0.2$. Under the baseline setting, all four models attained near-perfect sensitivity comparable to LASSO, BLASSO, MI-LASSO, SaENET, and GaLASSO, but with higher specificity, higher F$_1$ scores, higher sign accuracy, and lower MSE. Their performance remained stable under MAR and high-missingness conditions, with smaller increases in MSE than other methods. In contrast, MI-LASSO and SaENET showed reduced specificity, indicating a tendency to select more irrelevant variables, while GaLASSO performed similarly to Bayesian MI-LASSO but slightly worse. 

When the sample size increased to $n = 200$ or the SNR increased to 2.5, signal strength improved, and our Bayesian MI-LASSO models remained comparable to GaLASSO while outperforming all the other methods, even slightly better than BLASSO on full data. MI-LASSO, SaENET, and LASSO continued to suffer from low specificity. In the more challenging scenarios with higher dimensionality ($p=40$ or $80$), the Bayesian MI‐LASSO models continued to perform reasonably well, achieving the highest F$_1$ score and sign accuracy. Although sensitivity decreased slightly, specificity remained high, indicating conservative selection with few false positives. However, when the SNR reduced to $0.5$ or $0.2$, sensitivity declined sharply, while specificity remained high. In contrast, GaLASSO's sensitivity dropped sharply as dimensionality increased to $p=40$ or $80$, accompanied by low F$_1$ score and high MSE.

The setting with $p=200 > n$ was the most challenging. The four Bayesian MI-LASSO models retained high sensitivity but had substantially lower specificity, selecting overly large submodels with many unimportant covariates and consequently lowered F$_1$ scores. Their MSE was also large, reflecting the difficulty of estimating coefficients in these oversized submodels. The comparison methods tended to select too few covariates, with GaLASSO and CC-BLASSO retaining almost no variables. SaENET achieved the highest sensitivity and F$_1$ score, because stacking the $D$ imputed datasets created a larger design, which may have improved numerical stability.

Lastly, BLASSO outperformed LASSO using the full data but performed poorly in the complete cases analysis, suggesting potential gains of using Bayesian LASSO over frequentist LASSO models while highlighting the importance of handling missing covariates when performing variable selection.

The median running time of each model, measured on a Mac Studio equipped with an Apple M2 Max chip, is reported in Supplementary Materials S4.2. Runtime increased moderately with the sample size $n$ and more steeply with the number of covariates $p$. Among the Bayesian MI-LASSO models, Spike-and-Laplace was consistently the most computationally expensive, particularly at $p=200$, whereas the other three shrinkage models remained computationally competitive with the frequentist comparators. The additional four-step selection procedure required at most approximately $0.6$ minutes in the settings considered. It increased with $n$ and $p$ more slowly than the MCMC and so became a small share of the total computational cost as the MCMC fit became more expensive. Although the number of imputations $D$ was not varied in our simulations, increasing $D$ primarily increases the number of imputation-specific regression coefficients, while several shrinkage parameters are shared across imputations; thus, we expect the computational burden associated with increasing $D$ to be less pronounced than that associated with increasing $p$ (such as $\lambda_j^2$, $\gamma_j$) shared across imputations.

\subsection{Simulation Scenario 2: Correlated Continuous Covariates}
In the second simulation, covariates were generated with a first-order autoregressive (AR(1)) covariance structure, defined by $\Sigma_{jl} = \varrho^{|j - l|}$. We fixed the sample size at $n=100$, the number of covariates at $p=20$, and SNR at 1.5, while varying the correlation strength and the proportion of missingness. Specifically, we considered $\varrho = 0.3, 0.5$ under both moderate and high levels of missingness, and $\varrho = 0.7, 0.9$ under moderate missingness. For each setting, the regression coefficients and MCAR mechanisms were the same as in Simulation 1. For MAR settings, we adjusted the intercepts of the logistic missingness models to $-3.6$ for moderate missingness and $-1.9$ for high missingness, resulting in approximately 60\% and 35\% complete cases, respectively.

\begin{table}[!htbp]
\centering
\resizebox{\textwidth}{!}{
\begin{tabular}{cccccccccccccccccccccccc}
\toprule
 & \multicolumn{10}{c}{{\normalsize \textbf{n=100, p=20, SNR=1.5, moderate missing, $\boldsymbol{\varrho=0.3}$}}} & & & & \multicolumn{10}{c}{\normalsize \textbf{n=100, p=20, SNR=1.5, high missing, $\boldsymbol{\varrho=0.3}$}} \\ 
 \midrule
\textbf{Full Data} &  \multicolumn{2}{c}{\textbf{SEN}} & \multicolumn{2}{c}{\textbf{SPE}} & \multicolumn{2}{c}{\textbf{F}$_1$} & \multicolumn{2}{c}{\textbf{MSE}} & \multicolumn{2}{c}{\textbf{SIGN}} & & & & \multicolumn{2}{c}{\textbf{SEN}} & \multicolumn{2}{c}{\textbf{SPE}} & \multicolumn{2}{c}{\textbf{F}$_1$} & \multicolumn{2}{c}{\textbf{MSE}} & \multicolumn{2}{c}{\textbf{SIGN}} \\
LASSO & \multicolumn{2}{c}{99} & \multicolumn{2}{c}{82} & \multicolumn{2}{c}{83} & \multicolumn{2}{c}{91} & \multicolumn{2}{c}{87} & & & & \multicolumn{2}{c}{99} & \multicolumn{2}{c}{82} & \multicolumn{2}{c}{83} & \multicolumn{2}{c}{91} & \multicolumn{2}{c}{87}\\  
 BLASSO & \multicolumn{2}{c}{99} & \multicolumn{2}{c}{91} & \multicolumn{2}{c}{90} & \multicolumn{2}{c}{70} & \multicolumn{2}{c}{93} & & & & \multicolumn{2}{c}{99} & \multicolumn{2}{c}{91} & \multicolumn{2}{c}{90} & \multicolumn{2}{c}{70} & \multicolumn{2}{c}{93}\\
 \midrule
\multirow{2}*{\textbf{Models}} & \multicolumn{5}{c}{MCAR} & & \multicolumn{5}{c}{MAR} & & \multicolumn{5}{c}{MCAR} & & \multicolumn{5}{c}{MAR} \\ 
\cmidrule{2-6} \cmidrule{8-12} \cmidrule{14-18} \cmidrule{20-24}
 & \textbf{SEN} & \textbf{SPE} & \textbf{F}$_1$ & \textbf{MSE} & \textbf{SIGN} & & \textbf{SEN} & \textbf{SPE} & \textbf{F}$_1$ & \textbf{MSE} & \textbf{SIGN} & & \textbf{SEN} & \textbf{SPE} & \textbf{F}$_1$ & \textbf{MSE} & \textbf{SIGN} & & \textbf{SEN} & \textbf{SPE} & \textbf{F}$_1$ & \textbf{MSE} & \textbf{SIGN} \\
  CC-LASSO & 93 & 79 & 78 & 167 & 83 & & 70 & 82 & 61 & 347 & 79 & & 78 & 69 & 61 & 451 & 71 & & 41 & 80 & 35 & 593 & 68\\ 
  CC-BLASSO & 88 & 92 & 85 & 218 & 91 & & 65 & 93 & 68 & 460 & 85 & & 47 & \textbf{96} & 55 & 548 & 81 & & 11 & \textbf{98} & 16 & 719 & 72\\ 
  SaENET & \textbf{99} & 75 & 78 & 84 & 82 & & 98 & 72 & 76 & 97 & 80 & & \textbf{99} & 72 & 76 & 97 & 80 & & \textbf{98} & 62 & 69 & 142 & 72\\ 
  GaLASSO & 98 & 90 & 89 & 73 & 92 & & 98 & 87 & 87 & 86 & 90 & & 97 & 87 & 87 & 88 & 90 & & 97 & 76 & 78 & 136 & 82\\ 
  MI-LASSO & \textbf{99} & 81 & 83 & 95 & 87 & & \textbf{99} & 78 & 80 & 104 & 85 & & \textbf{99} & 79 & 81 & 104 & 85 & & \textbf{98} & 71 & 75 & 128 & 79\\ 
    \cmidrule{1-1}
  Multi-Laplace & 95 & \textbf{96} & 93 & 71 & \textbf{96} & & 94 & \textbf{95} & \textbf{92} & 81 & \textbf{95} & & 94 & 95 & \textbf{91} & 85 & \textbf{94} & & 91 & 90 & 85 & 122 & 90\\ 
  Horseshoe & 96 & 95 & 93 & \textbf{68} & \textbf{96} & & 96 & 94 & \textbf{92} & \textbf{78} & 94 & & 95 & 94 & \textbf{91} & \textbf{84} & \textbf{94} & & 93 & 89 & \textbf{86} & \textbf{119} & 90\\ 
  ARD & 96 & \textbf{96} & 93 & \textbf{68} & \textbf{96} & & 95 & 94 & 91 & 80 & 94 & & 95 & 93 & \textbf{91} & \textbf{84} & \textbf{94} & & 92 & 89 & 85 & 121 & 90\\ 
  Spike-and-Laplace & 95 & \textbf{96} & \textbf{94} & 69 & \textbf{96} & & 95 & \textbf{95} & \textbf{92} & \textbf{78} & \textbf{95} & & 93 & 94 & \textbf{91} & 87 & \textbf{94} & & 92 & 90 & \textbf{86} & 121 & \textbf{91}\\ 
\bottomrule
\toprule
 & \multicolumn{10}{c}{{\normalsize \textbf{n=100, p=20, SNR=1.5, moderate missing, $\boldsymbol{\varrho=0.5}$}}} & & & & \multicolumn{10}{c}{\normalsize \textbf{n=100, p=20, SNR=1.5, high missing, $\boldsymbol{\varrho=0.5}$}} \\ 
 \midrule
\textbf{Full Data} &  \multicolumn{2}{c}{\textbf{SEN}} & \multicolumn{2}{c}{\textbf{SPE}} & \multicolumn{2}{c}{\textbf{F}$_1$} & \multicolumn{2}{c}{\textbf{MSE}} & \multicolumn{2}{c}{\textbf{SIGN}} & & & & \multicolumn{2}{c}{\textbf{SEN}} & \multicolumn{2}{c}{\textbf{SPE}} & \multicolumn{2}{c}{\textbf{F}$_1$} & \multicolumn{2}{c}{\textbf{MSE}} & \multicolumn{2}{c}{\textbf{SIGN}} \\
LASSO & \multicolumn{2}{c}{98} & \multicolumn{2}{c}{80} & \multicolumn{2}{c}{82} & \multicolumn{2}{c}{98} & \multicolumn{2}{c}{86} & & & & \multicolumn{2}{c}{98} & \multicolumn{2}{c}{80} & \multicolumn{2}{c}{82} & \multicolumn{2}{c}{98} & \multicolumn{2}{c}{86}\\  
 BLASSO & \multicolumn{2}{c}{97} & \multicolumn{2}{c}{90} & \multicolumn{2}{c}{89} & \multicolumn{2}{c}{93} & \multicolumn{2}{c}{92} & & & & \multicolumn{2}{c}{97} & \multicolumn{2}{c}{90} & \multicolumn{2}{c}{89} & \multicolumn{2}{c}{93} & \multicolumn{2}{c}{92}\\
 \midrule
\multirow{2}*{\textbf{Models}} & \multicolumn{5}{c}{MCAR} & & \multicolumn{5}{c}{MAR} & & \multicolumn{5}{c}{MCAR} & & \multicolumn{5}{c}{MAR} \\ 
\cmidrule{2-6} \cmidrule{8-12} \cmidrule{14-18} \cmidrule{20-24}
 & \textbf{SEN} & \textbf{SPE} & \textbf{F}$_1$ & \textbf{MSE} & \textbf{SIGN} & & \textbf{SEN} & \textbf{SPE} & \textbf{F}$_1$ & \textbf{MSE} & \textbf{SIGN} & & \textbf{SEN} & \textbf{SPE} & \textbf{F}$_1$ & \textbf{MSE} & \textbf{SIGN} & & \textbf{SEN} & \textbf{SPE} & \textbf{F}$_1$ & \textbf{MSE} & \textbf{SIGN} \\
  CC-LASSO & 90 & 78 & 75 & 174 & 81 & & 64 & 84 & 59 & 409 & 78 & & 73 & 70 & 59 & 515 & 71 & & 34 & 83 & 31 & 735 & 68\\ 
  CC-BLASSO & 85 & 90 & 81 & 245 & 88 & & 61 & 93 & 65 & 549 & 83 & & 51 & \textbf{94} & 58 & 609 & 81 & & 14 & \textbf{98} & 19 & 845 & 73\\ 
  SaENET & 96 & 72 & 75 & 100 & 80 & & 95 & 69 & 73 & 114 & 77 & & 96 & 70 & 74 & 112 & 78 & & 94 & 60 & 67 & 167 & 70\\ 
  GaLASSO & 94 & 90 & 87 & \textbf{93} & 91 & & 94 & 88 & 85 & \textbf{109} & 90 & & 95 & 86 & 84 & \textbf{104} & 89 & & 91 & 76 & 76 & 167 & 81\\ 
  MI-LASSO & \textbf{99} & 81 & 82 & 103 & 86 & & \textbf{97} & 79 & 80 & 111 & 84 & & \textbf{98} & 78 & 80 & \textbf{104} & 84 & & \textbf{96} & 70 & 73 & \textbf{136} & 77\\ 
    \cmidrule{1-1}
  Multi-Laplace & 87 & \textbf{96} & \textbf{88} & 103 & \textbf{93} & & 86 & \textbf{94} & \textbf{86} & 115 & \textbf{92} & & 87 & \textbf{94} & \textbf{86} & 112 & \textbf{92} & & 83 & 88 & 79 & 159 & \textbf{87}\\ 
  Horseshoe & 88 & 95 & \textbf{88} & 104 & \textbf{93} & & 87 & 93 & 85 & 117 & 91 & & 87 & 93 & \textbf{86} & 114 & 91 & & 85 & 88 & \textbf{80} & 155 & \textbf{87}\\ 
  ARD & 87 & 95 & 87 & 107 & \textbf{93} & & 86 & 93 & 85 & 122 & 91 & & 87 & 93 & \textbf{86} & 116 & 91 & & 84 & 88 & 79 & 157 & 86\\ 
  Spike-and-Laplace & 87 & \textbf{96} & \textbf{88} & 104 & \textbf{93} & & 85 & \textbf{94} & 85 & 123 & 91 & & 86 & \textbf{94} & \textbf{86} & 117 & \textbf{92} & & 83 & 88 & 79 & 163 & \textbf{87}\\ 
\bottomrule
\toprule
 & \multicolumn{10}{c}{{\normalsize \textbf{n=100, p=20, SNR=1.5, moderate missing, $\boldsymbol{\varrho=0.7}$}}} & & & & \multicolumn{10}{c}{\normalsize \textbf{n=100, p=20, SNR=1.5, moderate missing, $\boldsymbol{\varrho=0.9}$}} \\ 
 \midrule
\textbf{Full Data} &  \multicolumn{2}{c}{\textbf{SEN}} & \multicolumn{2}{c}{\textbf{SPE}} & \multicolumn{2}{c}{\textbf{F}$_1$} & \multicolumn{2}{c}{\textbf{MSE}} & \multicolumn{2}{c}{\textbf{SIGN}} & & & & \multicolumn{2}{c}{\textbf{SEN}} & \multicolumn{2}{c}{\textbf{SPE}} & \multicolumn{2}{c}{\textbf{F}$_1$} & \multicolumn{2}{c}{\textbf{MSE}} & \multicolumn{2}{c}{\textbf{SIGN}} \\
LASSO & \multicolumn{2}{c}{92} & \multicolumn{2}{c}{78} & \multicolumn{2}{c}{76} & \multicolumn{2}{c}{127} & \multicolumn{2}{c}{82} & & & & \multicolumn{2}{c}{65} & \multicolumn{2}{c}{77} & \multicolumn{2}{c}{59} & \multicolumn{2}{c}{144} & \multicolumn{2}{c}{73}\\  
 BLASSO & \multicolumn{2}{c}{88} & \multicolumn{2}{c}{86} & \multicolumn{2}{c}{80} & \multicolumn{2}{c}{166} & \multicolumn{2}{c}{87} & & & & \multicolumn{2}{c}{61} & \multicolumn{2}{c}{84} & \multicolumn{2}{c}{61} & \multicolumn{2}{c}{632} & \multicolumn{2}{c}{77}\\
 \midrule
\multirow{2}*{\textbf{Models}} & \multicolumn{5}{c}{MCAR} & & \multicolumn{5}{c}{MAR} & & \multicolumn{5}{c}{MCAR} & & \multicolumn{5}{c}{MAR} \\ 
\cmidrule{2-6} \cmidrule{8-12} \cmidrule{14-18} \cmidrule{20-24}
 & \textbf{SEN} & \textbf{SPE} & \textbf{F}$_1$ & \textbf{MSE} & \textbf{SIGN} & & \textbf{SEN} & \textbf{SPE} & \textbf{F}$_1$ & \textbf{MSE} & \textbf{SIGN} & & \textbf{SEN} & \textbf{SPE} & \textbf{F}$_1$ & \textbf{MSE} & \textbf{SIGN} & & \textbf{SEN} & \textbf{SPE} & \textbf{F}$_1$ & \textbf{MSE} & \textbf{SIGN} \\
  CC-LASSO & 79 & 77 & 68 & 227 & 78 & & 60 & 83 & 56 & 463 & 76 & & 52 & 78 & 52 & 238 & 71 & & 40 & 83 & 43 & 626 & 70\\ 
  CC-BLASSO & 73 & 86 & 71 & 352 & 82 & & 57 & \textbf{90} & 60 & 728 & 80 & & 48 & 88 & 54 & 1040 & \textbf{76} & & 32 & \textbf{90} & 39 & 1684 & \textbf{73}\\ 
  SaENET & 88 & 63 & 69 & 130 & 70 & & 88 & 52 & 63 & 148 & 63 & & \textbf{78} & 50 & 55 & \textbf{138} & 58 & & \textbf{79} & 42 & 52 & \textbf{181} & 51\\ 
  GaLASSO & 79 & 87 & \textbf{76} & 144 & \textbf{85} & & 78 & 83 & \textbf{72} & 169 & 81 & & 44 & 86 & 49 & 209 & 73 & & 44 & 79 & 45 & 287 & 68\\ 
  MI-LASSO & \textbf{93} & 74 & 74 & \textbf{117} & 80 & & \textbf{91} & 72 & \textbf{72} & \textbf{137} & 78 & & 71 & 70 & \textbf{59} & 152 & 69 & & 68 & 69 & \textbf{57} & 192 & 67\\ 
    \cmidrule{1-1}
  Multi-Laplace & 69 & \textbf{92} & 73 & 163 & \textbf{85} & & 66 & 89 & 69 & 186 & \textbf{82} & & 35 & 90 & 44 & 218 & 74 & & 36 & 86 & 42 & 251 & 71\\ 
  Horseshoe & 69 & \textbf{92} & 73 & 160 & \textbf{85} & & 66 & 89 & 69 & 181 & \textbf{82} & & 35 & \textbf{91} & 45 & 204 & 74 & & 35 & 87 & 42 & 246 & 71\\ 
  ARD & 68 & 91 & 72 & 164 & 84 & & 66 & 89 & 69 & 182 & \textbf{82} & & 35 & \textbf{91} & 45 & 208 & 74 & & 35 & 87 & 42 & 257 & 71\\ 
  Spike-and-Laplace & 67 & \textbf{92} & 72 & 167 & 84 & & 65 & \textbf{90} & 69 & 188 & \textbf{82} & & 36 & 90 & 45 & 213 & 74 & & 36 & 86 & 42 & 271 & 71\\ 
\bottomrule
\end{tabular}
}
\caption{Results of Sensitivity (×100), Specificity (×100), $\text{F}_1$ score (×100), MSE (×100), and Sign accuracy (×100) in Scenario 2. The best value per metric within each setting is in bold.}
\label{tab:sim_B}
\end{table}

Table~\ref{tab:sim_B} presents the results of Scenario 2. The four Bayesian MI-LASSO models continued to perform well under moderately correlated covariates with $\varrho\leq 0.5$, achieving the best overall balance of sensitivity and specificity, highest F$_1$ score, near-lowest MSE, and highest sign accuracy. When $\varrho$ increased to $0.7$ or $0.9$, the Bayesian MI-LASSO models retained high specificity and sign accuracy, but their sensitivity decreased and MSE increased. GaLASSO exhibited higher sensitivity but lower specificity than the Bayesian MI-LASSO models when $\varrho\leq0.5$, with both sensitivity and specificity declining at higher correlations. MI-LASSO achieved higher sensitivity and lower MSE than the Bayesian MI-LASSO methods when $\varrho=0.7$ and $0.9$, showing an advantage not observed with independent or weakly correlated covariates.

\subsection{Simulation Scenario 3: Correlated Binary Covariates}

This simulation evaluates model performance under binary covariates, with $n = 100$, $p = 20$, and $\text{SNR} = 1.5$. Covariates $\bX$ were first generated from a multivariate normal distribution with an AR(1) covariance structure and autocorrelation parameter $\varrho = 0.5$, as in Simulation 2. The continuous covariates were then dichotomized into binary variables using $0$ as the cutoff.  
Since each binary $\bX_j$ had mean $0.5$, MSE is calculated as $\mathrm{MSE}=(\tilde{\boldsymbol\beta} - \boldsymbol\beta)^\top(\Sigma+0.25)(\tilde{\boldsymbol\beta} - \boldsymbol\beta)$, where $\Sigma_{jl} = \frac{1}{2\pi} \arcsin(0.5^{|j - l|})$. We excluded SaENET and GaLASSO from this comparison, as \citet{Du2022} developed and validated these methods only on continuous covariates through the simulation studies. Simulation results are shown in Table~\ref{tab:sim_C}. The findings are similar to those observed in Simulation 2, except that the Bayesian MI-LASSO models now yielded lower MSE compared to BLASSO applied to the full data.

\begin{table}[!htbp]
\centering
\resizebox{\textwidth}{!}{
\begin{tabular}{cccccccccccccccccccccccc}
\toprule
 & \multicolumn{10}{c}{{\normalsize \textbf{n=100, p=20, SNR=1.5, moderate missing, $\boldsymbol{\varrho=0.5}$}}} & & & & \multicolumn{10}{c}{\normalsize \textbf{n=100, p=20, SNR=1.5, high missing, $\boldsymbol{\varrho=0.5}$}} \\ 
 \midrule
\textbf{Full Data} &  \multicolumn{2}{c}{\textbf{SEN}} & \multicolumn{2}{c}{\textbf{SPE}} & \multicolumn{2}{c}{\textbf{F}$_1$} & \multicolumn{2}{c}{\textbf{MSE}} & \multicolumn{2}{c}{\textbf{SIGN}} & & & & \multicolumn{2}{c}{\textbf{SEN}} & \multicolumn{2}{c}{\textbf{SPE}} & \multicolumn{2}{c}{\textbf{F}$_1$} & \multicolumn{2}{c}{\textbf{MSE}} & \multicolumn{2}{c}{\textbf{SIGN}} \\
LASSO & \multicolumn{2}{c}{99} & \multicolumn{2}{c}{81} & \multicolumn{2}{c}{83} & \multicolumn{2}{c}{66} & \multicolumn{2}{c}{86} & & & & \multicolumn{2}{c}{99} & \multicolumn{2}{c}{81} & \multicolumn{2}{c}{83} & \multicolumn{2}{c}{66} & \multicolumn{2}{c}{86}\\  
 BLASSO & \multicolumn{2}{c}{99} & \multicolumn{2}{c}{90} & \multicolumn{2}{c}{90} & \multicolumn{2}{c}{45} & \multicolumn{2}{c}{93} & & & & \multicolumn{2}{c}{99} & \multicolumn{2}{c}{90} & \multicolumn{2}{c}{90} & \multicolumn{2}{c}{45} & \multicolumn{2}{c}{93}\\
 \midrule
\multirow{2}*{\textbf{Models}} & \multicolumn{5}{c}{MCAR} & & \multicolumn{5}{c}{MAR} & & \multicolumn{5}{c}{MCAR} & & \multicolumn{5}{c}{MAR} \\ 
\cmidrule{2-6} \cmidrule{8-12} \cmidrule{14-18} \cmidrule{20-24}
 & \textbf{SEN} & \textbf{SPE} & \textbf{F}$_1$ & \textbf{MSE} & \textbf{SIGN} & & \textbf{SEN} & \textbf{SPE} & \textbf{F}$_1$ & \textbf{MSE} & \textbf{SIGN} & & \textbf{SEN} & \textbf{SPE} & \textbf{F}$_1$ & \textbf{MSE} & \textbf{SIGN} & & \textbf{SEN} & \textbf{SPE} & \textbf{F}$_1$ & \textbf{MSE} & \textbf{SIGN} \\
  CC-LASSO & 94 & 79 & 79 & 130 & 84 &  & 84 & 80 & 70 & 281 & 81 &  & 74 & 70 & 59 & 312 & 71 &  & 50 & 79 & 44 & 577 & 70\\
  CC-BLASSO & 89 & 91 & 85 & 219 & 90 &  & 76 & 92 & 76 & 494 & 87 &  & 47 & \textbf{96} & 55 & 766 & 81 &  & 24 & \textbf{98} & 31 & 1000 & 76\\
  MI-LASSO & \textbf{99} & 81 & 82 & 72 & 86 &  & \textbf{99} & 80 & 82 & 75 & 86 &  & \textbf{99} & 80 & 82 & 76 & 86 &  & \textbf{99} & 78 & 80 & 76 & 84\\
    \cmidrule{1-1}
  Multi-Laplace & 96 & \textbf{96} & 93 & 33 & \textbf{96} &  & 95 & \textbf{96} & \textbf{93} & 34 & \textbf{95} &  & 95 & 95 & \textbf{92} & 36 & \textbf{95} &  & 94 & 95 & \textbf{92} & 39 & \textbf{95}\\
  Horseshoe & 97 & 95 & 93 & 32 & 95 &  & 96 & 94 & 92 & 33 & \textbf{95} &  & 96 & 94 & 91 & 34 & 94 &  & 95 & 94 & 91 & \textbf{37} & 94\\
  ARD & 96 & 95 & 92 & 32 & 95 &  & 96 & 94 & 92 & 32 & \textbf{95} &  & 95 & 93 & 91 & \textbf{33} & 94 &  & 95 & 94 & 91 & \textbf{37} & 94\\
  Spike-and-Laplace & 95 & \textbf{96} & \textbf{94} & \textbf{30} & \textbf{96} &  & 96 & 95 & \textbf{93} & \textbf{31} & \textbf{95} &  & 94 & 95 & 91 & 34 & 94 &  & 94 & 95 & 91 & \textbf{37} & \textbf{95}\\
\bottomrule
\end{tabular}
}
\caption{Results of Sensitivity (×100), Specificity (×100), $\text{F}_1$ score (×100), MSE (×100), and Sign accuracy (×100) in Scenario 3. The best value per metric within each setting is in bold.}
\label{tab:sim_C}
\end{table}

\subsection{Coverage Analysis}

To evaluate post-selection variance calibration, we assessed the coverage and width of the $95\%$ credible intervals for the coefficients in the selected model. Let $\text{CI}_j$ denote the interval for coefficient $j$, taken as the $95\%$ quantile interval of the pooled calibrated draws when $j\in s_m$ and as $\{0\}$ when $j\notin s_m$. A missed signal then never covers its nonzero $\beta_j$ and a correctly excluded unimportant variable always covers zero, so coverage reflects both selection and inference. We report three coverage quantities and the mean interval width:
\begin{itemize}[
  itemsep=0.3em,
  parsep=0em
]
  \item Marginal coverage of signals (\textbf{Cov}$_{\ne0}$): $\dfrac{\sum_{j=1}^p \mathbbm{1}(\beta_j\neq0)\,\mathbbm{1}(\beta_j\in\text{CI}_j)}{\sum_{j=1}^p \mathbbm{1}(\beta_j\neq0)}$, the proportion of truly nonzero coefficients whose interval covers $\beta_j$;

  \item Conditional coverage of selected signals (\textbf{Cov}$_{\ne0}^{\text{sel}}$): $\dfrac{\sum_{j=1}^p \mathbbm{1}(\beta_j\neq0,\,j\in s_m)\,\mathbbm{1}(\beta_j\in\text{CI}_j)}{\sum_{j=1}^p \mathbbm{1}(\beta_j\neq0,\,j\in s_m)}$, restricting $\text{Cov}_{\ne0}$ to the signals that were selected;

\item Coverage of unimportant variables (\textbf{Cov}$_{=0}$): $\dfrac{\sum_{j=1}^p \mathbbm{1}(\beta_j=0)\,\mathbbm{1}(0\in\text{CI}_j)}{\sum_{j=1}^p \mathbbm{1}(\beta_j=0)}$, the proportion of truly unimportant variables whose interval covers $0$;

  \item Mean interval width (\textbf{Width}): $\dfrac{1}{|s_m|}\sum_{j\in s_m}|\text{CI}_j|$, the average length of the $95\%$ intervals over the selected coefficients.
\end{itemize}
The marginal coverage $\text{Cov}_{\ne0}$ combines the selection of the signals with the calibration of their intervals, whereas $\text{Cov}_{\ne0}^{\text{sel}}$ isolates the calibration by conditioning on selection. We additionally included an oracle regime that fixes the selected set as the true signal set, to evaluate the calibrated coverage based on correct selection. We also included the coverage of projected coefficients before calibration, to assess the effectiveness of the calibration step. We present coverage results of six MAR settings in Table~\ref{tab:coverage}, including the baseline, high missingness, increased sample size ($n=200$) and high dimensionality ($p=200$) in Scenario 1, and $\varrho=0.5$ and $0.9$ in Scenario 2.

Under the oracle regime, the calibrated intervals yielded near-nominal coverage across all settings, whereas the uncalibrated projected intervals had undercoverage, especially when $p$ was large, $\varrho$ was high, or the percentage of complete cases was low. With best select, however, coverage depended on selection performance. In the baseline, high-missingness, $n=200$, and moderately correlated ($\varrho=0.5$) settings, the calibrated intervals achieved conditional coverage of $90$ -- $95\%$ for selected signals and coverage of $94$ -- $98\%$ for unimportant variables. The projected intervals were narrower, with coverage several points lower. As selection became harder, marginal coverage of signals declined quickly. Under $\varrho=0.5$, this decline was attributable to missed signals, as conditional coverage of the selected signals stayed high. Under strong correlation ($\varrho=0.9$) and high dimensionality ($p=200$), conditional coverage also declined, suggesting that selection errors biased inference even for retained variables. 

\begin{table}[!htbp]
\centering
\resizebox{\textwidth}{!}{
\begin{tabular}{l cccccccc c cccccccc}
\toprule
 & \multicolumn{8}{c}{\textbf{Scenario 1, n=100, p=20, SNR=1.5, moderate missing}} & & \multicolumn{8}{c}{\textbf{Scenario 1, n=100, p=20, SNR=1.5, high missing}} \\
\cmidrule(lr){2-9}\cmidrule(lr){11-18}
 & \multicolumn{4}{c}{Calibrated} & \multicolumn{4}{c}{Projected} & & \multicolumn{4}{c}{Calibrated} & \multicolumn{4}{c}{Projected} \\
\cmidrule(lr){2-5}\cmidrule(lr){6-9}\cmidrule(lr){11-14}\cmidrule(lr){15-18}
 & Cov$_{\ne0}$ & Cov$_{\ne0}^{\text{sel}}$ & Cov$_{=0}$ & Width & Cov$_{\ne0}$ & Cov$_{\ne0}^{\text{sel}}$ & Cov$_{=0}$ & Width & & Cov$_{\ne0}$ & Cov$_{\ne0}^{\text{sel}}$ & Cov$_{=0}$ & Width & Cov$_{\ne0}$ & Cov$_{\ne0}^{\text{sel}}$ & Cov$_{=0}$ & Width \\
\midrule
\multicolumn{18}{l}{\textit{Best select}}\\
\quad Multi-Laplace & 93 & 94 & 97 & 87 & 89 & 91 & 96 & 78 & & 92 & 95 & 96 & 91 & 88 & 90 & 94 & 81 \\
\quad Horseshoe & 91 & 92 & 97 & 84 & 89 & 90 & 96 & 80 & & 90 & 92 & 96 & 88 & 88 & 89 & 94 & 83 \\
\quad ARD & 91 & 93 & 97 & 83 & 90 & 91 & 96 & 80 & & 90 & 92 & 96 & 87 & 88 & 90 & 94 & 84 \\
\quad Spike-and-Laplace & 92 & 94 & 96 & 83 & 91 & 93 & 96 & 80 & & 92 & 94 & 95 & 88 & 91 & 93 & 94 & 84 \\
\cmidrule(l){1-18}
\multicolumn{18}{l}{\textit{True select (oracle)}}\\
\quad Multi-Laplace & 94 & 94 & 100 & 89 & 90 & 90 & 100 & 78 & & 93 & 93 & 100 & 92 & 88 & 88 & 100 & 81 \\
\quad Horseshoe & 92 & 92 & 100 & 85 & 90 & 90 & 100 & 80 & & 91 & 91 & 100 & 90 & 87 & 87 & 100 & 83 \\
\quad ARD & 92 & 92 & 100 & 85 & 91 & 91 & 100 & 80 & & 91 & 91 & 100 & 89 & 88 & 88 & 100 & 83 \\
\quad Spike-and-Laplace & 94 & 94 & 100 & 85 & 92 & 92 & 100 & 80 & & 93 & 93 & 100 & 90 & 90 & 90 & 100 & 83 \\
\bottomrule
\toprule
 & \multicolumn{8}{c}{\textbf{Scenario 1, n=200, p=20, SNR=1.5, moderate missing}} & & \multicolumn{8}{c}{\textbf{Scenario 1, n=100, p=200, SNR=1.5, moderate missing}} \\
\cmidrule(lr){2-9}\cmidrule(lr){11-18}
 & \multicolumn{4}{c}{Calibrated} & \multicolumn{4}{c}{Projected} & & \multicolumn{4}{c}{Calibrated} & \multicolumn{4}{c}{Projected} \\
\cmidrule(lr){2-5}\cmidrule(lr){6-9}\cmidrule(lr){11-14}\cmidrule(lr){15-18}
 & Cov$_{\ne0}$ & Cov$_{\ne0}^{\text{sel}}$ & Cov$_{=0}$ & Width & Cov$_{\ne0}$ & Cov$_{\ne0}^{\text{sel}}$ & Cov$_{=0}$ & Width & & Cov$_{\ne0}$ & Cov$_{\ne0}^{\text{sel}}$ & Cov$_{=0}$ & Width & Cov$_{\ne0}$ & Cov$_{\ne0}^{\text{sel}}$ & Cov$_{=0}$ & Width \\
\midrule
\multicolumn{18}{l}{\textit{Best select}}\\
\quad Multi-Laplace & 93 & 93 & 98 & 61 & 91 & 91 & 98 & 57 & & 50 & 61 & 86 & 209 & 46 & 54 & 84 & 190 \\
\quad Horseshoe & 92 & 92 & 98 & 60 & 91 & 91 & 97 & 58 & & 20 & 28 & 75 & 53 & 17 & 23 & 73 & 44 \\
\quad ARD & 93 & 93 & 98 & 59 & 91 & 91 & 97 & 58 & & 29 & 38 & 77 & 82 & 24 & 32 & 75 & 71 \\
\quad Spike-and-Laplace & 93 & 93 & 98 & 59 & 92 & 93 & 98 & 58 & & 22 & 31 & 75 & 55 & 16 & 22 & 73 & 33 \\
\cmidrule(l){1-18}
\multicolumn{18}{l}{\textit{True select (oracle)}}\\
\quad Multi-Laplace & 94 & 94 & 100 & 61 & 91 & 91 & 100 & 57 & & 96 & 96 & 100 & 133 & 32 & 32 & 100 & 27 \\
\quad Horseshoe & 93 & 93 & 100 & 60 & 91 & 91 & 100 & 58 & & 95 & 95 & 100 & 135 & 28 & 28 & 100 & 24 \\
\quad ARD & 93 & 93 & 100 & 60 & 92 & 92 & 100 & 58 & & 94 & 94 & 100 & 126 & 28 & 28 & 100 & 23 \\
\quad Spike-and-Laplace & 94 & 94 & 100 & 60 & 93 & 93 & 100 & 58 & & 94 & 94 & 100 & 125 & 38 & 38 & 100 & 33 \\
\bottomrule
\toprule
 & \multicolumn{8}{c}{\textbf{Scenario 2, n=100, p=20, SNR=1.5, moderate missing, $\boldsymbol{\varrho=0.5}$}} & & \multicolumn{8}{c}{\textbf{Scenario 2, n=100, p=20, SNR=1.5, moderate missing, $\boldsymbol{\varrho=0.9}$}} \\
\cmidrule(lr){2-9}\cmidrule(lr){11-18}
 & \multicolumn{4}{c}{Calibrated} & \multicolumn{4}{c}{Projected} & & \multicolumn{4}{c}{Calibrated} & \multicolumn{4}{c}{Projected} \\
\cmidrule(lr){2-5}\cmidrule(lr){6-9}\cmidrule(lr){11-14}\cmidrule(lr){15-18}
 & Cov$_{\ne0}$ & Cov$_{\ne0}^{\text{sel}}$ & Cov$_{=0}$ & Width & Cov$_{\ne0}$ & Cov$_{\ne0}^{\text{sel}}$ & Cov$_{=0}$ & Width & & Cov$_{\ne0}$ & Cov$_{\ne0}^{\text{sel}}$ & Cov$_{=0}$ & Width & Cov$_{\ne0}$ & Cov$_{\ne0}^{\text{sel}}$ & Cov$_{=0}$ & Width \\
\midrule
\multicolumn{18}{l}{\textit{Best select}}\\
\quad Multi-Laplace & 81 & 93 & 96 & 114 & 78 & 88 & 95 & 100 & & 21 & 51 & 93 & 249 & 19 & 45 & 91 & 211 \\
\quad Horseshoe & 79 & 90 & 96 & 110 & 77 & 87 & 95 & 103 & & 18 & 48 & 92 & 234 & 17 & 44 & 90 & 218 \\
\quad ARD & 78 & 90 & 95 & 109 & 77 & 88 & 95 & 104 & & 17 & 44 & 91 & 234 & 16 & 42 & 90 & 222 \\
\quad Spike-and-Laplace & 78 & 91 & 94 & 109 & 78 & 89 & 94 & 103 & & 17 & 43 & 90 & 242 & 16 & 40 & 89 & 226 \\
\cmidrule(l){1-18}
\multicolumn{18}{l}{\textit{True select (oracle)}}\\
\quad Multi-Laplace & 95 & 95 & 100 & 117 & 91 & 91 & 100 & 101 & & 97 & 97 & 100 & 336 & 93 & 93 & 100 & 251 \\
\quad Horseshoe & 91 & 91 & 100 & 114 & 85 & 85 & 100 & 102 & & 96 & 96 & 100 & 334 & 76 & 76 & 100 & 209 \\
\quad ARD & 90 & 90 & 100 & 113 & 85 & 85 & 100 & 103 & & 96 & 96 & 100 & 333 & 73 & 73 & 100 & 202 \\
\quad Spike-and-Laplace & 92 & 92 & 100 & 113 & 88 & 88 & 100 & 102 & & 95 & 95 & 100 & 326 & 71 & 71 & 100 & 197 \\
\bottomrule
\end{tabular}
}
\caption{Empirical coverage (×100) and mean interval width (×100) of the 95\% credible intervals under MAR, averaged over 200 replicates, for the calibrated and the uncalibrated projected intervals under the selected set (\textit{Best select}) and the true signal set (\textit{oracle}). }
\label{tab:coverage}
\end{table}

\subsection{Sensitivity Analysis}
\label{sec:SA}
We conducted a sensitivity analysis to assess robustness to hyperparameter choices in the Multi-Laplace and Spike-and-Laplace models. We evaluated both the fitted full models before variable selection and the submodels selected by the four-step projection predictive procedure. Before selection,  full-model estimates were moderately robust to hyperparameter choices, except when $v$ in the Multi-Laplace model or $b$ in the Spike-and-Laplace model was very small, conditions under which the shrinkage effect on the regression coefficients was substantially reduced. This weaker shrinkage increased MSE for the Multi-Laplace model but often reduced MSE for the Spike-and-Laplace model. This difference stems from the Spike-and-Laplace prior's ability to maintain variable selection through the binary indicators $\boldsymbol{\gamma}$ under weak shrinkage. However, the reduced variability of the Spike-and-Laplace model with small $b$ may suffer from a loss of exploration for enough candidate subsets by the grid of selection thresholds. After selection, robustness improved substantially: most settings produced similar selections and MSE. Smaller hyperparameter values of $v$ and $b$ yielded stable MSE performance for the Multi-Laplace model, whereas the Spike-and-Laplace model exhibited slightly increased sensitivity, with reduced specificity and sign accuracy. Overall, the models selected by our four-step procedure are highly robust across a range of hyperparameters, with the default settings performing consistently well. However, we recommend avoiding extremely small $v$ or $b$, as they may lead to instability. 

Another sensitivity analysis investigated the resolution of the threshold grid. We repeated the four-step selection using grid increments ranging from $0.1$ to $0.001$ for the Horseshoe and Spike-and-Laplace models. The coarsest increment of $0.1$ slightly worsened selection performance, whereas an increment of $0.02$ or finer yielded nearly identical results. Although a finer grid may be useful as $p$ increases, the procedure is generally robust to grid resolution, since there are usually only a few unique candidate sets generated by threshold-grid search. Given the larger number of covariates in the University of Michigan Dioxin Exposure Study than in the simulation settings, we used the $0.01$ increment and set it as the default.

Complete details are provided in Supplementary Materials S5, with results summarized in Web Tables 3--7. 

\section{Case Study}
\label{sec:case}
\subsection{Data Description}
\label{sec:motivating}
Dioxin is a chemical pollutant with high toxicity and a long environmental half-life. The University of Michigan Dioxin Exposure Study (UMDES) is a population-based study of dioxin exposure conducted among residents of Midland, Michigan. It was designed to investigate the association between environmental exposure factors and human serum dioxin concentrations, particularly in relation to discharges of dioxin compounds into the nearby river and air by the Dow Chemical Company. The UMDES dataset includes serum dioxin measurements from blood samples, dioxin levels from household dust and soil, and detailed participant-level information on demographics, health status, residential history, occupational and recreational activities, lifetime food consumption, and current diet. Further details of this study can be found elsewhere \citep{Garabrant2009, Garabrant2009a, Hedgeman2009}.

Our goal was to identify important circumstances and exposure pathways that predict serum concentration of the most toxic dioxin compound, 2,3,7,8-tetrachlorodibenzo-\textit{p}-dioxin (TCDD), in a log-10 scale. We focus on 448 individuals residing in or near the Tittabawassee River floodplain, between the Dow Chemical plant in Midland and the confluence with the Shiawassee River in Saginaw. The dataset includes 63 continuous and 68 binary covariates capturing demographic and exposure-related characteristics. Correlations among covariates are generally weak: 96\% of pairwise correlations fall within $[-0.2, 0.2]$, with a mean of 0.03. 

As is typical in environmental health research, the UMDES data contain missing values in many covariates and outcomes. Among the 448 individuals in our sample, only 190 (42\%) have complete data. To address missingness, \citet{Olson2006} used sequential regression imputation implemented in \textit{IVEware} software \citep{Raghunathan2001} to generate five multiply-imputed datasets. Our analysis was conducted using these five previously imputed datasets.

\subsection{Variable Selection Results}
Following the recommendation of \citet{Hippel2007}, we excluded 48 individuals with missing outcomes. The analysis was then conducted on the 400 individuals with observed outcomes, across all five imputed datasets. We estimated the SNR to be about 1.4 by fitting linear regressions on each imputed dataset and combining the results using Rubin’s rules, which aligns closely with the setting used in the simulation study (Section~\ref{sec:simulation}). We then applied the four Bayesian MI-LASSO models and reported the selected variables, along with their posterior means and 95\% equal-tailed credible intervals for the pooled calibrated regression coefficients. We ran four chains for each model and monitored the rank-normalized $\hat R$ to ensure it remained below 1.01 \citep{Vehtari2021}. We reported the maximum $\hat R$ and the results of the first chain, together with its minimum bulk effective sample size across the calibrated coefficients and the residual variance. We normalized each $\bx^d$ and centered each $\by^d$ prior to model fitting, and rescaled the estimated coefficients back to the original scale afterward. Hyperparameters were set the same as in Table~\ref{tab:hyperparameters}.

\begin{table}[!htbp]
\centering
\resizebox{\textwidth}{!}{
\begin{tabular}{l@{\hspace{2pt}}lc@{\hspace{1pt}}cc@{\hspace{1pt}}cc@{\hspace{1pt}}cc@{\hspace{1pt}}cc@{\hspace{1pt}}c}
\toprule
\multicolumn{2}{l}{\textbf{Variables}} & \multicolumn{2}{c}{\textbf{Multi-Laplace}} & \multicolumn{2}{c}{\textbf{Horseshoe}} & \multicolumn{2}{c}{\textbf{ARD}} & \multicolumn{2}{c}{\textbf{Spike-and-Laplace}} \\
\midrule
(0) & Intercept & 0.176 & (0.105, 0.247) & 0.163 & (0.097, 0.230) & 0.159 & (0.094, 0.225) & 0.161 & (0.096, 0.227) \\
(1) & Age & 0.014 & (0.011, 0.017) & 0.014 & (0.011, 0.017) & 0.014 & (0.011, 0.017) & 0.013 & (0.010, 0.016) \\
(2) & Female & 0.146 & (0.081, 0.212) & 0.172 & (0.109, 0.234) & 0.176 & (0.114, 0.237) & 0.174 & (0.113, 0.236) \\
(3) & Age $\times$ female & 0.005 & (0.001, 0.009) & 0.005 & (0.001, 0.008) & 0.005 & (0.001, 0.008) & 0.005 & (0.001, 0.009) \\
(4) & BMI & 0.011 & (0.006, 0.017) & 0.011 & (0.006, 0.016) & 0.011 & (0.006, 0.016) & 0.011 & (0.006, 0.016) \\
(5) & BMI change & -0.020 & (-0.034, -0.007) & -0.021 & (-0.033, -0.008) & -0.021 & (-0.034, -0.009) & -0.022 & (-0.035, -0.010) \\
(6) & Ever smoking & -0.083 & (-0.141, -0.026) & -0.087 & (-0.140, -0.034) & -0.088 & (-0.142, -0.036) & -0.090 & (-0.143, -0.038) \\
(7) & \makecell[lt]{Years of living in \\ Midland: 1960-79} & 0.009 & (0.005, 0.013) & 0.009 & (0.005, 0.012) & 0.009 & (0.005, 0.013) & 0.009 & (0.005, 0.013) \\
(8) & \makecell[lt]{Fishing in the \\ Tittabawassee \\ River below the \\ Tridge after 1980 \\ (>0, <1/month)} & -0.057 & (-0.126, 0.012) & -0.070 & (-0.136, -0.005) & -0.080 & (-0.146, -0.015) & -0.089 & (-0.155, -0.024) \\
(9) & \makecell[lt]{Years of working in \\  Dow Chemical \\ company after 1980} & 0.013 & (0.003, 0.024) & 0.013 & (0.002, 0.023) & 0.012 & (0.002, 0.023) & 0.013 & (0.003, 0.023) \\
(10) & \makecell[lt]{Years of working in \\  foundry: 1960-79} & -0.014 & (-0.026, -0.003) & -0.018 & (-0.029, -0.007) & -0.019 & (-0.030, -0.008) & -0.019 & (-0.030, -0.008) \\
(11) & \makecell[lt]{Years of living with \\ somebody working \\in Dow Chemical \\ company: 1940-59} & \multicolumn{2}{c}{---} & 0.018 & (0.002, 0.034) & 0.019 & (0.003, 0.034) & 0.020 & (0.005, 0.036) \\
(12) & \makecell[lt]{Months of \\ breast-feeding} & -0.005 & (-0.009, -0.001) & -0.005 & (-0.009, -0.001) & -0.005 & (-0.009, -0.002) & -0.005 & (-0.009, -0.002) \\
(13) & \makecell[lt]{Ever hunting around \\ Saginaw River and \\ Bay: 1960-79} & \multicolumn{2}{c}{---} & 0.105 & (0.013, 0.204) & 0.112 & (0.018, 0.210) & 0.123 & (0.028, 0.218) \\
(14) & \makecell[lt]{Years of spraying \\ chemicals to kill \\ plants: 1960-79} & \multicolumn{2}{c}{---} & \multicolumn{2}{c}{---} & 0.016 & (0.000, 0.034) & 0.019 & (0.001, 0.036) \\
\midrule
\multicolumn{2}{l}{BIC$_m$} & \multicolumn{2}{c}{-2.464} & \multicolumn{2}{c}{-2.467} & \multicolumn{2}{c}{-2.460} & \multicolumn{2}{c}{-2.452} \\
\multicolumn{2}{l}{$\hat{df}$} & \multicolumn{2}{c}{48.1} & \multicolumn{2}{c}{57.5} & \multicolumn{2}{c}{63.3} & \multicolumn{2}{c}{66.0} \\
\midrule
\multicolumn{2}{l}{Max $\hat{R}$} & \multicolumn{2}{c}{1.000} & \multicolumn{2}{c}{1.001} & \multicolumn{2}{c}{1.001} & \multicolumn{2}{c}{1.000} \\
\multicolumn{2}{l}{Min bulk-ESS} & \multicolumn{2}{c}{2678} & \multicolumn{2}{c}{2349} & \multicolumn{2}{c}{2749} & \multicolumn{2}{c}{4079} \\
\bottomrule
\end{tabular}
}
\caption{Comparison of the four Bayesian MI-LASSO models in selecting important factors that predict log-10 serum 2,3,7,8-tetrachlorodibenzo-\textit{p}-dioxin concentration in the flood plain and near-flood plain samples from UMDES data. The reported $\hat{df}$ is the effective number of parameters across the $D=5$ imputations, at most $D|s|$ for a submodel of $|s|$ selected variables and reduced further by the priors' penalization of coefficients. The last two rows report the maximum rank-normalized $\hat{R}$ and the minimum bulk effective sample size across the calibrated coefficients and the residual variance \citep{Vehtari2021}.}
\label{tab:real_data}
\end{table}

As shown in Table~\ref{tab:real_data}, all four Bayesian MI-LASSO models selected a common set of 11 variables along with the intercept. The Multi-Laplace model selected only these 11 variables and was therefore the most parsimonious. The Horseshoe model selected two additional variables, \textit{Years of living with somebody working in Dow Chemical company: 1940--59} and \textit{Ever hunting around Saginaw River and Bay: 1960--79}, for a total of 13 variables. The ARD and Spike-and-Laplace models further selected \textit{Years of spraying chemicals to kill plants: 1960--79}, for a total of 14 variables each.

Compared to the original MI-LASSO of \citet{Chen2013}, which selected 13 variables along with the intercept, the Bayesian MI-LASSO models, particularly the Multi-Laplace variant, tended to select fewer variables while consistently identifying important covariates such as \textit{Age}, \textit{Female}, \textit{Age $\times$ Female}, \textit{BMI change}, \textit{ever smoking}, and \textit{Years of living in Midland: 1960--79}. A clearer contrast is in the uncertainty of the estimated associations. MI-LASSO retained some variables whose 95\% confidence intervals included zero, whereas the 95\% credible intervals excluded zero for nearly all variables selected by the Bayesian MI-LASSO models. The main exceptions were \textit{Fishing in the Tittabawassee River below the Tridge after 1980} under the Multi-Laplace model and \textit{Years of spraying chemicals to kill plants: 1960-79} under the ARD model. Thus, the Bayesian MI-LASSO models generally provided stronger evidence of nonzero associations among the selected variables.

Another advantage of the Bayesian MI-LASSO models is that they yielded consistent coefficient directions across all four specifications, providing a coherent interpretation of covariate effects. For example, the association between age and serum TCDD concentration differs by gender: as individuals age, TCDD levels tend to increase more in females than in males. Higher BMI is also associated with elevated serum TCDD levels, while weight gain, months of breast-feeding, and ever smoking are negatively associated with TCDD concentration. These findings are supported by earlier studies \citep{Chen2013, Garabrant2009}. In addition to individual characteristics, our analysis identified several residential, occupational, recreational, and property-use factors that likely reflect major exposure pathways. Higher serum TCDD concentrations were observed among individuals who lived longer in Midland County during 1960–1979, worked at Dow Chemical after 1980, lived with someone employed at Dow between 1940 and 1959, sprayed herbicides during 1960–1979, or hunted near the Saginaw River and Bay during that period.

\section{Discussion}
\label{sec:discussion}
While multiple imputation is a common approach for handling missing data, it poses challenges for achieving consistent variable selection across imputed datasets. MI-LASSO, introduced by \citet{Chen2013}, addresses this issue by treating each variable across imputations as a group and applying group LASSO to enforce consistent selection. However, there has been little development of Bayesian counterparts to this approach. Compared to frequentist methods, Bayesian models offer greater flexibility by allowing the incorporation of various types of prior information. In this paper, we extend MI-LASSO to the Bayesian framework and introduce four Bayesian MI-LASSO models, including three with continuous shrinkage priors, and one with a Spike-and-Laplace prior. These priors enable significant shrinkage of weak signals while allowing larger coefficients to retain substantial magnitude.  

Our simulation studies showed that the Bayesian MI-LASSO models outperformed MI-LASSO and other existing methods in variable selection and post-selection coefficient estimation across most scenarios considered. Notably, they also outperformed standard LASSO and performed comparably to Bayesian LASSO, applied to fully observed data. With robust hyperparameter specifications and a consistent selection procedure, the four models produced highly similar results. The Horseshoe and ARD priors exhibited slightly higher sensitivity, while the Multi-Laplace and Spike-and-Laplace priors showed slightly higher specificity. 

MI-LASSO exhibits a higher sensitivity, favoring larger models by including any variables that show signals of importance in any imputation, whereas the Bayesian MI-LASSO models tend to have higher specificity, selecting only variables whose pooled posterior draws of the regression coefficients provide strong overall evidence of importance. These patterns were observed not only in the simulations but also in the UMDES data application. In the UMDES application, all covariates (except one variable from Multi-Laplace and one variable from ARD) selected by the Bayesian MI-LASSO models {had} 95\% probability intervals excluding zero. However, three covariates selected by MI-LASSO were insignificant \citep{Chen2013}. Additionally, the Multi-Laplace prior yielded more parsimonious selections.

Another contribution is that we propose a novel and robust four-step projection predictive variable selection procedure for Bayesian MI-LASSO. Existing Bayesian methods often rely on ad hoc thresholds, and the selected variables can be sensitive to these choices. Our approach avoids relying on a single arbitrary threshold by generating candidate subsets across a prespecified grid and selecting among them using the complexity-penalized BIC$_m$. Projection maps the full-model posterior onto each reduced model to retain its predictive information. A subsequent calibration step adjusts the posterior variances of the projected coefficients to improve interval estimation for the selected model. This flexible framework applies to both shrinkage and Spike-and-Laplace models and requires only a grid of pre-specified thresholds between 0 and 1 to generate candidate subsets, making it easy to tune in practice.

Additional simulations further compare our four-step procedure with several alternatives. First, we compared it with fixed-threshold selection. For the shrinkage models, the four-step procedure performed similarly to the fixed threshold when the covariates were independent or weakly correlated. As the correlation increased, the $95\%$ credible-interval rule retained fewer variables, attaining higher specificity at the expense of sensitivity. So the advantage of the four-step procedure over selection using 95\% credible intervals increased as correlation among covariates increased. For the Spike-and-Laplace model, in contrast, the four-step procedure performed better than selection using a posterior inclusion probability threshold of 0.5 when covariates were independent, but worse under strong correlation (Supplementary Materials S6.1). Second, we examined alternative approaches for generating candidate subsets. The scaled neighborhood criterion \citep{Li2010} produced nearly identical results to our thresholds-based selection for the shrinkage models but higher MSE for the Spike-and-Laplace model (Supplementary Materials S6.2). The forward stepwise search, commonly used in the projection predictive inference \citep{Piironen2020,McLatchie2025}, performed similarly to threshold-grid search but required more computation time, especially in higher dimensions (Supplementary Materials S6.3). Finally, we compared BIC$_m$ with selection criteria based on PSIS-LOO expected log predictive density (ELPD) \citep{Piironen2020,McLatchie2025,Vehtari2017}. Selecting the model with the highest PSIS-LOO ELPD favored larger models and yielded poorer variable selection performance than $\mathrm{BIC}_m$ at $p=20$. The PSIS-LOO one standard error (1-SE) rule produced broadly comparable selection performance to $\mathrm{BIC}_m$ but required more computation time (Supplementary Materials S6.3).

Some practical considerations arise when applying our methods. For users seeking a robust, high-performance method with minimal tuning, we recommend starting with the Horseshoe or ARD priors. These models have no hyperparameters, deliver strong performance, and tend to favor higher sensitivity. If a more parsimonious selection is desired and the user is comfortable with hyperparameter specification, the Multi-Laplace and Spike-and-Laplace models offer slightly higher specificity. For these, we recommend our default strategy of matching the marginal prior variance to the error variance, which has proven effective and robust, particularly when coupled with our four-step selection procedure.

Our methods have {three} major limitations. First, our approaches were primarily developed for the setting with $n>p$, as in the UMDES application. In our simulation with $p>n$, they tended to select many unimportant covariates and yield large MSEs. This high-dimensional regime is indeed challenging for group-based methods in general, as MI-LASSO cannot be fit because of a non-full-rank design, and GaLASSO selected very few variables. The stacking-based method SaENET performed somewhat better in this setting, since stacking the imputations into one dataset creates a larger design matrix, which may improve numerical stability. However, stacking methods nevertheless tended to over-select variables and inadequately account for between-imputation uncertainty in variance estimation (Supplementary Materials S6.4). Second, when covariates were highly correlated, the Bayesian MI-LASSO models retained the highest specificity among the comparison methods but had lower sensitivity and larger MSEs than MI-LASSO. Third, our variance calibration is an approximate, empirically validated adjustment without a formal theoretical guarantee. Its coverage depends on selection accuracy, with near-nominal coverage observed only when the selected set is correct. Further, the reported intervals treat the selected model as fixed and do not account for selection uncertainty. A related limitation arises in frequentist multiple-imputation analyses: \citet{Liu2019} showed that applying Rubin's rules after variable selection without accounting for selection uncertainty can lead to substantial undercoverage.

There are several directions for future work. One natural extension is to explore alternative Bayesian shrinkage or Spike-and-Slab priors beyond those studied in this paper, such as the regularized Horseshoe prior \citep{piironen2017sparsity}. The framework could also incorporate prior information, for example by setting the Beta hyperparameters on the inclusion probabilities of the Spike-and-Laplace model to reflect the prior expected number of included covariates, or by assigning higher prior inclusion probabilities to important covariates identified in previous studies. Another direction is to extend the current framework to accommodate different outcome types, such as binary or ordinal responses, through latent variable models (e.g., probit models). These extensions may require numerical optimization, efficient approximation strategies, and appropriate degrees of freedom development.

\section*{Disclosure statement}\label{disclosure-statement}

The authors declare no conflicts of interest.

\phantomsection\label{supplementary-material}
\bigskip

\begin{center}

{\large\bf SUPPLEMENTARY MATERIAL}

\end{center}

\begin{description}
\item[Supplementary Material:]
Detailed Gibbs sampling algorithms of four Bayesian MI-LASSO models, derivations of estimated degrees of freedom and modified BIC, sensitivity analysis and additional simulations.
\item[R package BMIselect:]
Implementation of the proposed methods, available on \texttt{CRAN}.
\end{description}

\bibliography{bibliography}

\end{document}


\def\spacingset#1{\renewcommand{\baselinestretch}%
{#1}\small\normalsize} \spacingset{1}


\if1\anon
{
  \title{\bf Supporting Information for \\ 
"Bayesian MI-LASSO for Variable Selection on Multiply-Imputed Data"}
  \author{Jungang Zou\thanks{
    Corresponding author: Jungang Zou (\href{mailto:jz3183@cumc.columbia.edu}{jz3183@cumc.columbia.edu}).}\hspace{.2cm}\\
    Department of Biostatistics, Columbia University\\
    and \\
    Sijian Wang \\
    Department of Statistics, Rutgers University\\
    and \\
    Qixuan Chen \\
    Department of Biostatistics, Columbia University}
  \maketitle
} \fi

\if0\anon
{
  \bigskip
  \bigskip
  \bigskip
  \begin{center}
    {\LARGE\bf Supporting Information for \\ 
"Bayesian MI-LASSO for Variable Selection on Multiply-Imputed Data"}
\end{center}
  \medskip
} \fi

\bigskip
\begin{abstract}
This document provides additional supporting information for "Bayesian MI-LASSO for Variable Selection on Multiply-Imputed Data", including detailed Gibbs sampling algorithms of four Bayesian MI-LASSO models, derivations of estimated degrees of freedom and modified BIC, sensitivity analysis and additional simulations.
\end{abstract}
\vfill

\newpage
\spacingset{1.8} 

\section{Gibbs Samplers of the Four Bayesian MI-LASSO Models}\label{su_sec:gibbs}
We present the algorithms of Gibbs samplers to update the four Bayesian MI-LASSO models from Section S1.1 to S1.4. Throughout this manuscript, we use shape-scale form for all Gamma distributions. 

\subsection{Gibbs Sampler of the Multi-Laplace Bayesian MI-LASSO model}
Let $\mathbf{I}$ denote the identity matrix. The posterior update of Gibbs sampler for the Multi-Laplace model is listed in Web Algorithm~\ref{alg:ml}. To ensure conjugacy, we leverage the Generalized inverse Gaussian (GIG) distribution $\text{GIG}(g_1,g_2,g_3)$ for posterior distributions of $\lambda^2_j$, with parameters $g_1\in \mathbb{R}$, $g_2>0$, and $g_3>0$. Its density function is 
\[f(x; g_1,g_2,g_3)=\frac{(g_2/g_3)^{g_1/2}}{2K_{g_1}(\sqrt{g_2g_3})}x^{g_1-1}\exp(-(g_2x+g_3/x)/2),\]
where $K_{g_1}$ is a modified Bessel function of the second kind. In $\texttt{R}$, we use $\texttt{rgig}$ function in $\texttt{GIGrvg}$ package to sample from the GIG distribution. In Web Algorithm~\ref{alg:ml}, we update parameters related to Multi-Laplace prior at steps 1--2, and update regression parameters at steps 3--5.

\begin{algorithm}[ht]
\caption{The Gibbs sampler to update Multi-Laplace Bayesian MI-LASSO}
\label{alg:ml}
\begin{algorithmic}[1]
\Require MI datasets $\{(\bx^d,\by^d),\ d=1,2,\ldots,D\}$, hyperparameters $h$ and $v$.
\State Sample $\lambda_j^2$ from $\text{GIG}\left(0.5, D\rho,  \frac{\bb_j^\top\bb_j}{\sigma^2}\right)$, $j=1,2,\ldots,p$. Set $\boldsymbol{\Lambda}=diag(\lambda_1^2,\lambda_2^2,\ldots, \lambda_p^2)$.
\State Sample $\rho$ from $\text{Gamma}\left(h + \frac{p(D+1)}{2}, \frac{1}{\frac{1}{v} + \frac{D}{2}\sum_{j=1}^p\lambda_j^2} \right)$.
\State Sample $\bb^d$ from $\mathcal{MVN}\left([\boldsymbol{\Lambda}^{-1} + {\bx^d}^\top\bx^d]^{-1}{\bx^d}^\top\by^d,\sigma^2[\boldsymbol{\Lambda}^{-1} + {\bx^d}^\top\bx^d]^{-1}\right)$, $d=1,2,\ldots,D$.
\State Calculate $SSE = \sum_{d=1}^D\sum_{i=1}^n\left(y_i^d-{\bx_i^d}\bb^d\right)^2$.
\State Sample $\sigma^2$ from $\text{InvGamma}\left(\frac{D(n+p)}{2}, \frac{1}{2}(SSE + \sum_{j=1}^p\frac{\bb_j^\top\bb_j}{\lambda_j^2})\right)$.
\end{algorithmic}
\end{algorithm}

\subsection{Gibbs Sampler of the Horseshoe Bayesian MI-LASSO model}
The Gibbs sampler of the Horseshoe prior is based on the auxiliary inverse Gamma variables $(\eta, \kappa_1, \kappa_2,\ldots,\kappa_p)$, introduced in \citet{Makalic2016}. The half-Cauchy prior $\tau \sim C^+(0, 1)$ is equivalent to a scale-mixture representation $\tau^2 \cond \eta \sim \text{InvGamma}(0.5, \frac{1}{\eta})$, $\eta \sim \text{InvGamma}(0.5, 1)$. This representation is the same for all $\lambda_j$ with auxiliary variables $\kappa_j$, respectively. The collapsed Gibbs sampler for the Horseshoe model is listed in Web Algorithm~\ref{alg:horseshoe}. To break the strong coupling between each local scale $\lambda_j^2$ and its coefficients, we sample $\lambda_j^2$ from its marginal full conditional in Web Formula~\eqref{eq:hs_lambda_marg}, with $\beta_j^1,\ldots,\beta_j^D$ integrated out, by slice sampling \citep{Neal2003}, and then sample the coefficients from their closed-form full conditional.

\begin{equation}\label{eq:hs_lambda_marg}
\resizebox{0.9\textwidth}{!}{$
\pi(\lambda_j^2\mid\cdot)\ \propto\ (\lambda_j^2)^{-\frac{D+3}{2}}\exp\!\left(-\frac{1}{\kappa_j\lambda_j^2}\right)\prod_{d=1}^D\left({\bx^d_{[j]}}^\top\bx^d_{[j]}+\frac{1}{\tau^2\lambda_j^2}\right)^{-\frac{1}{2}}\exp\!\left(\frac{\left({\bx^d_{[j]}}^\top(\by^d-\bx^d_{[-j]}\bb^d_{[-j]})\right)^2}{2\sigma^2\left({\bx^d_{[j]}}^\top\bx^d_{[j]}+\frac{1}{\tau^2\lambda_j^2}\right)}\right).
$}
\end{equation}

\begin{algorithm}[ht]
\caption{The collapsed Gibbs sampler to update Horseshoe Bayesian MI-LASSO}
\label{alg:horseshoe}
\begin{algorithmic}[1]
\Require MI datasets $\{(\bx^d,\by^d),\ d=1,2,\ldots,D\}$.
\State Sample $\tau^2$ from $\text{InvGamma}\left(\frac{Dp+1}{2}, \frac{1}{\eta}+\sum_{j=1}^p\frac{\bb_j^\top\bb_j}{2\sigma^2\lambda_j^2}\right)$.
\State Sample $\eta$ from $\text{InvGamma}\left(1, 1 + \frac{1}{\tau^2}\right)$.
\State Sample $\kappa_j$ from $\text{InvGamma}\left(1, 1 + \frac{1}{\lambda_j^2}\right)$, $j=1,2,\ldots,p$.
\State Sample $\lambda_j^2$ from its marginal full conditional in Web Formula~\eqref{eq:hs_lambda_marg} by slice sampling, then $\beta_j^d$ from $\mathcal{N}\left(\frac{{\bx^d_{[j]}}^\top(\by^d-\bx^d_{[-j]}\bb^d_{[-j]})}{{\bx^d_{[j]}}^\top\bx^d_{[j]}+1/(\tau^2\lambda_j^2)},\ \frac{\sigma^2}{{\bx^d_{[j]}}^\top\bx^d_{[j]}+1/(\tau^2\lambda_j^2)}\right)$, $d=1,\ldots,D$, $j=1,2,\ldots,p$.
\State Set $\boldsymbol{\Lambda}=diag(\lambda_1^2,\lambda_2^2,\ldots,\lambda_p^2)$ and calculate $SSE = \sum_{d=1}^D\sum_{i=1}^n\left(y_i^d-{\bx_i^d}\bb^d\right)^2$.
\State Sample $\sigma^2$ from $\text{InvGamma}\left(\frac{D(n+p)}{2}, \frac{1}{2}(SSE + \sum_{j=1}^p\frac{\bb_j^\top\bb_j}{\tau^2\lambda_j^2})\right)$.
\end{algorithmic}
\end{algorithm}

\subsection{Gibbs Sampler of the ARD Bayesian MI-LASSO model}
Compared to the Multi-Laplace and Horseshoe priors, the ARD prior behaves differently at the origin $\bb = 0$. As shown in Section 2.3, the marginal prior takes the form $\pi(\beta_j^d) \propto \frac{1}{|\beta_j^d|}$, which is undefined at zero. To address this, we redefine the support of each $\beta_j^d$ as $\mathbb{R} \setminus \{0\}$. Since $\{0\}$ is a null set, this modification has negligible impact on the ARD model. The collapsed Gibbs sampler for the ARD model is presented in Web Algorithm~\ref{alg:ARD}. As in the Horseshoe sampler, we update $\psi_j^2$ and the coefficients $\beta_j^1,\ldots,\beta_j^D$ one covariate at a time, sampling $\psi_j^2$ from its marginal full conditional in Web Formula~\eqref{eq:ard_psi_marg}, with $\beta_j^1,\ldots,\beta_j^D$ integrated out, by slice sampling \citep{Neal2003}, and then the coefficients from their closed-form full conditional. To improve numerical stability, we can set a small threshold $\upsilon$ on $\frac{1}{\psi_j^{2}}$ to prevent over/underflow, i.e. $\psi_j^{2}=\text{min}\left(\psi_j^2, 1/\upsilon\right)$, when $\bb_j$ takes very small values. The default value of $\upsilon$ is $10^{-6}$. Finally, the initial value of $\beta_j^d$ should not be 0.

\begin{equation}\label{eq:ard_psi_marg}
\pi(\psi_j^2\mid\cdot)\ \propto\ (\psi_j^2)^{\frac{D}{2}-1}\prod_{d=1}^D\left({\bx^d_{[j]}}^\top\bx^d_{[j]}+\psi_j^2\right)^{-\frac{1}{2}}\exp\!\left(\frac{\left({\bx^d_{[j]}}^\top(\by^d-\bx^d_{[-j]}\bb^d_{[-j]})\right)^2}{2\sigma^2\left({\bx^d_{[j]}}^\top\bx^d_{[j]}+\psi_j^2\right)}\right).
\end{equation}

\begin{algorithm}[ht]
\caption{The collapsed Gibbs sampler to update ARD Bayesian MI-LASSO}
\label{alg:ARD}
\begin{algorithmic}[1]
\Require MI datasets $\{(\bx^d,\by^d),\ d=1,2,\ldots,D\}$.
\State Sample $\psi_j^2$ from its marginal full conditional in Web Formula~\eqref{eq:ard_psi_marg} by slice sampling, then $\beta_j^d$ from $\mathcal{N}\left(\frac{{\bx^d_{[j]}}^\top(\by^d-\bx^d_{[-j]}\bb^d_{[-j]})}{{\bx^d_{[j]}}^\top\bx^d_{[j]}+\psi_j^2},\ \frac{\sigma^2}{{\bx^d_{[j]}}^\top\bx^d_{[j]}+\psi_j^2}\right)$, $d=1,\ldots,D$, $j=1,2,\ldots,p$.
\State Set $\boldsymbol{\Psi}=diag(\psi_1^2,\psi_2^2,\ldots,\psi_p^2)$ and calculate $SSE = \sum_{d=1}^D\sum_{i=1}^n\left(y_i^d-{\bx_i^d}\bb^d\right)^2$.
\State Sample $\sigma^2$ from $\text{InvGamma}\left(\frac{D(n+p)}{2}, \frac{1}{2}(SSE + \sum_{j=1}^p\bb_j^\top\bb_j\psi_j^2)\right)$.
\end{algorithmic}
\end{algorithm}

\subsection{Gibbs Sampler of the Spike-and-Laplace Bayesian MI-LASSO model}

We define $f(\mathbf{x}; \mu,\boldsymbol{\Sigma})$ as the probability density function of a multivariate normal distribution with mean vector ${\mu}$ and covariance matrix $\boldsymbol{\Sigma}$. For the Spike-and-Laplace model, the standard Gibbs sampler is generally not feasible due to the strong dependence between the regression coefficients $\bb$ and the binary inclusion indicators $\boldsymbol{\gamma}$. Specifically, each inclusion indicator $\gamma_j$ partitions the posterior support of $(\bb_j, \gamma_j)$ into two regions, $\{\bb_j=0, \gamma_j = 0\}$ and $\{\bb_j\in\mathbb{R}^D, \gamma_j = 1\}$. Since $\{\bb_j=0, \gamma_j = 1\}$ is a null set, the two regions $\{\bb_j=0, \gamma_j = 0\}$ and $\{\bb_j\in\mathbb{R}^D, \gamma_j = 1\}$ are disjoint in the posterior support, almost surely. This sharp separation can cause the standard sampler to become trapped within one region, severely hindering efficient mixing and posterior exploration. 

\begin{algorithm}[ht]
\caption{The partially collapsed Gibbs sampler to update Spike-and-Laplace Bayesian MI-LASSO}
\label{alg:collapse_spike_laplace}
\begin{algorithmic}[1]
\Require MI datasets $\{(\bx^d,\by^d),\ d=1,2,\ldots,D\}$, hyperparameters $a$ and $b$.
\State Sample $\rho$ from $\text{Gamma}(a + \frac{p(D+1)}{2}, \frac{1}{\frac{1}{b} + \frac{D}{2}\sum_{j=1}^p\lambda_j^2} )$.
\State Sample $\gamma_j$ from $\text{Bernoulli}\left(\text{prob}_j\right)$, where $\text{prob}_j$ is defined in Web Formula~\eqref{eq:prob}, $j=1,2,\ldots,p$.
\State Sample $\bb_{[\bZ]}^d$ from $\mathcal{MVN}\left([\boldsymbol{\Lambda}^{-1}_{[\bZ]}+{\bx^d}^\top_{[\bZ]}\bx^d_{[\bZ]}]^{-1}{\bx^d}_{[\bZ]}^\top\by^d, \sigma^2[\boldsymbol{\Lambda}^{-1}_{[\bZ]}+{\bx^d}^\top_{[\bZ]}\bx^d_{[\bZ]}]^{-1}\right)$, $d=1,2,\ldots,D$.
\State Set $\bb_j=0$ if $\gamma_j=0$, $j=1,2,\ldots,p$.
\State Sample $\theta_j$ from $\text{Beta}(1 + \gamma_j, 2 - \gamma_j)$, $j=1,2,\ldots, p$. 
\State Sample $\lambda_j^2$ from $\text{GIG}\left(0.5, D\rho,  \frac{\bb_j^\top\bb_j}{\sigma^2}\right)$, if $\gamma_j=1$, $j=1,2,\ldots,p$. 
\State Sample $\lambda_j^2$ from $\text{Gamma}(\frac{D+1}{2}, \frac{2}{D\rho})$, if $\gamma_j=0$, $j=1,2,\ldots,p$.
\State Set $\boldsymbol{\Lambda}=diag(\lambda_1^2, \lambda_2^2,\ldots, \lambda_p^2)$.
\State Calculate $SSE = \sum_{d=1}^D\sum_{i=1}^n\left(y_i^d-{\bx_i^d}\bb^d\right)^2$.
\State Sample $\sigma^2$ from $\text{InvGamma}\left(\frac{D(n+\sum_{j=1}^p\mathbbm{1}(\gamma_j=1))}{2}, \frac{1}{2}(SSE + \sum_{j=1}^p\frac{\bb_j^\top\bb_j\mathbbm{1}(\gamma_j=1)}{\lambda_j^2})\right)$.
\end{algorithmic}
\end{algorithm}

To address this issue, we adopt a partially collapsed Gibbs sampler to improve mixing \citep{Dyk2008}. Specifically, we sample the latent indicator vector $\boldsymbol{\gamma}$ from a collapsed model by integrating out $\bb$. Conditional on $\boldsymbol{\gamma}$, we then sample the coefficients $\bb$ from their uncollapsed full conditional distribution. The remaining parameters are also sampled from the uncollapsed model. We define $\boldsymbol{\Lambda}_{[\bZ]} = \mathrm{diag}(\lambda_j^2; \gamma_j = 1)$ as the submatrix of $\boldsymbol{\Lambda}$ containing the diagonal elements $\lambda_j^2$ corresponding to $\gamma_j = 1$. We further define $\boldsymbol{\gamma}_{-j}$ as the vector of inclusion indicators for all covariates except the $j$-th, $\bx^d_{[\boldsymbol{\gamma}_{-j}, \gamma_j = 1]}$ and $\bx^d_{[\boldsymbol{\gamma}_{-j}, \gamma_j = 0]}$ as the design matrices obtained by holding $\boldsymbol{\gamma}_{-j}$ fixed and setting $\gamma_j$ to 1 or 0, respectively. Similarly, $\bb^d_{[\boldsymbol{\gamma}]}$ denotes the coefficient vector corresponding to selected variables in the $d$-th imputed dataset. We use Woodbury matrix identity to avoid calculation of inverse matrices, and the collapsed likelihood is \[
\by^d \cond \bZ, \boldsymbol{\Lambda}, \bx^d, \sigma^2\sim \mathcal{MVN}(0, \sigma^2(\mathbf{I}+\bx_{[\bZ]}^d\boldsymbol{\Lambda}_{[\bZ]}{\bx^d}_{[\bZ]}^\top)).
\]
Based on this, the full conditional distribution of $\gamma_j$ is a Bernoulli distribution with the probability parameter: 
\begin{equation}
\label{eq:prob}
\resizebox{\textwidth}{!}{$
\frac{\theta_j\prod_{d=1}^Df(\by^d; 0,\sigma^2(\mathbf{I}+\bx_{[\bZ_{-j},\gamma_j=1]}^d\boldsymbol{\Lambda}_{[\bZ_{-j},\gamma_j=1]}{\bx^d}_{[\bZ_{-j},\gamma_j=1]}^\top))}{\theta_j\prod_{d=1}^Df(\by^d; 0,\sigma^2(\mathbf{I}+\bx_{[\bZ_{-j},\gamma_j=1]}^d\boldsymbol{\Lambda}_{[\bZ_{-j},\gamma_j=1]}{\bx^d}_{[\bZ_{-j},\gamma_j=1]}^\top))+(1-\theta_j)\prod_{d=1}^Df(\by^d; 0,\sigma^2(\mathbf{I}+\bx_{[\bZ_{-j},\gamma_j=0]}^d\boldsymbol{\Lambda}_{[\bZ_{-j},\gamma_j=0]}{\bx^d}_{[\bZ_{-j},\gamma_j=0]}^\top))}$} 
\end{equation}
The full algorithm is provided in Web Algorithm~\ref{alg:collapse_spike_laplace}. In steps 2--4, we sample each binary inclusion indicator $\gamma_j$ from its collapsed posterior, followed by steps 3 and 4 for sampling the corresponding coefficients $\bb$. Step 2 is computed on the log scale to avoid numerical overflow or underflow. The remaining parameters are updated in steps 5--10.

\section{Derivation of Modified BIC and Estimated Degrees of Freedom}

\subsection{Modified BIC}

The modified BIC adapts the classical BIC to the multiply-imputed grouped likelihood with a \emph{shared} error variance. We derive it here and then extend it to the projected models used in the main text.

The four Bayesian MI-LASSO models place imputation-specific coefficients $\bb^d$ but a single error variance $\sigma^2$ on the $D$ imputed datasets, $\by^d \sim \mathcal{N}(\bx^d\bb^d, \sigma^2\mathbf{I})$, $d=1,\ldots,D$. Conditioning on the parameters, the $D$ datasets contribute a single Gaussian likelihood over $Dn$ observations,
\[ L\!\left(\{\bb^d\},\sigma^2\right) = (2\pi\sigma^2)^{-\frac{Dn}{2}}\exp\!\left(-\frac{1}{2\sigma^2}\sum_{d=1}^D\|\by^d-\bx^d\bb^d\|_2^2\right), \]
in which the single $\sigma^2$ multiplies all $Dn$ residuals. Because $\sigma^2$ is shared, its maximum-likelihood value pools all $Dn$ residuals,
\[ \hat\sigma^2 = \frac{1}{Dn}\sum_{d=1}^D\|\by^d-\bx^d\hat\bb^d\|_2^2 \;\equiv\; \frac{\mathrm{SSE}}{Dn}, \]
and the profile log-likelihood becomes
\[ -2\hat\ell = Dn\log\!\left(\frac{\mathrm{SSE}}{Dn}\right) + Dn\big[1+\log(2\pi)\big]. \]
The BIC adds to $-2\hat\ell$ a complexity penalty $df\log N$, where $N$ is the number of terms in the likelihood being penalized. For the grouped likelihood this number is $Dn$. With $df$ effective parameters,
\[ \mathrm{BIC} = -2\hat\ell + df\log(Dn) = Dn\log\!\left(\frac{\mathrm{SSE}}{Dn}\right) + df\log(Dn) + Dn\big[1+\log(2\pi)\big]. \]
Dividing by $Dn$ and dropping the model-independent constant $1+\log(2\pi)$ gives the modified criterion
\begin{equation}\label{eq:bic_m_supp}
\mathrm{BIC}_m = \log\!\left(\frac{\mathrm{SSE}}{Dn}\right) + df\,\frac{\log(Dn)}{Dn}.
\end{equation}
The normalization $Dn$ is simply the number of observations sharing $\sigma^2$; it is the same $Dn$ that forms the residual contribution to the shape $D(n+p)/2$ of the $\sigma^2$ full conditional in the Gibbs samplers (Web Algorithms~\ref{alg:ml}--\ref{alg:collapse_spike_laplace}). 

For the projected Bayesian MI-LASSO models we retain \eqref{eq:bic_m_supp}, replacing the fitted values with the projected posterior-mean fit onto the candidate subset $s$ and $df$ with the model-specific $\hat{df}(s)$ of Section~\ref{supsec:df}:
\begin{equation}\label{eq:bic_m_proj_supp}
\mathrm{BIC}_m(s) = \log\!\left(\frac{1}{Dn}\sum_{d=1}^D\big\|\by^d-\bx^d_{[s]}\e[\bb_s^d\cond\text{data}]\big\|_2^2\right) + \hat{df}(s)\,\frac{\log(Dn)}{Dn},
\end{equation}
which is the criterion $\mathrm{BIC}_m(s)$ minimized in Step~3 of the main text.

\subsection{Estimated Degrees of Freedom}
\label{supsec:df}
In this part, we assume $T$ and $|s|$ are finite, and design matrices $\bx^d$ are bounded, for $d=1,\ldots,D$. If we use posterior mean $\e[\bx^d\bb^d \cond \text{data}]$ as predictive values for $\by^d$, we can estimate $df$ by:
\begin{align}
\label{eq:df_general}
    \hat{df}= \sum_{d=1}^D \Tr\left(\frac{\partial \e[\bx^d\bb^d \cond \text{data}]}{\partial \by^d}\right),
\end{align}
where $\Tr(\cdot)$ is the trace of a matrix. When projecting onto selected covariates $\bx^d_{[s]}$, we assume $s\neq \emptyset$ and $\bx_{[s]}^d$ is full rank; we replace~\eqref{eq:df_general} with:
\begin{equation}
    \begin{split}
        \label{eq:df_proj}
    \hat{df}(s) & = \sum_{d=1}^D \Tr\left(\frac{\partial \bx^d_{[s]}\e[\bb_s^d \cond \text{data}]}{\partial \by^d}\right)\\
    & = \sum_{d=1}^D \Tr\left(\bx^d_{[s]}({\bx_{[s]}^d}^\top\bx_{[s]}^d)^{-1}{\bx_{[s]}^d}^\top\frac{\partial \e[\bx^d\bb^d \cond \text{data}]}{\partial \by^d}\right) \\
    \end{split}
\end{equation}

For the Multi-Laplace model, we can exploit MCMC draws of ${\boldsymbol{\Lambda}}^{(t)}$ to calculate $\frac{\partial \e[\bx^d\bb^d \cond \text{data}]}{\partial \by^d}$ based on Monte Carlo integral: 
\begin{equation}
    \begin{split}
        \label{eq:e_beta_ML}
        \frac{\partial \e[\bx^d\bb^d \cond \text{data}]}{\partial \by^d} & =  \frac{\partial \int\e[\bx^d\bb^d \cond \boldsymbol{\Lambda}, \text{data}]\pi(\boldsymbol{\Lambda} \cond \text{data}) d\boldsymbol{\Lambda}}{\partial \by^d}  \\
& = \int \frac{\partial \e[\bx^d\bb^d \cond \boldsymbol{\Lambda}, \text{data}]}{\partial \by^d} \pi(\boldsymbol{\Lambda} \cond \text{data}) d\boldsymbol{\Lambda} \\
& = \int  \left(\bx^d[{\boldsymbol{\Lambda}}^{-1} + {\bx^d}^\top\bx^d]^{-1}{\bx^d}^\top\right)\pi(\boldsymbol{\Lambda} \cond \text{data}) d\boldsymbol{\Lambda} \\
    & \approx \frac{1}{T} \sum_{t=1}^T \bx^d[{\boldsymbol{\Lambda}^{(t)}}^{-1} + {\bx^d}^\top\bx^d]^{-1}{\bx^d}^\top 
    \end{split}
\end{equation}
where $\frac{\partial \e[\bx^d\bb^d \cond \boldsymbol{\Lambda}, \text{data}]}{\partial \by^d}=\bx^d[{\boldsymbol{\Lambda}}^{-1} + {\bx^d}^\top\bx^d]^{-1}{\bx^d}^\top$ comes from Step 3 in Web Algorithm~\ref{alg:ml}. The second equality holds by the dominated convergence theorem. Specifically, the derivative \[\frac{\partial \e[\bx^d\bb^d \cond \boldsymbol{\Lambda}, \text{data}]}{\partial \by^d} =\bx^d[{\boldsymbol{\Lambda}}^{-1} + {\bx^d}^\top\bx^d]^{-1}{\bx^d}^\top\preceq\bx^d[{\bx^d}^\top\bx^d]^{+}{\bx^d}^\top, \forall \boldsymbol{\Lambda}\succ0\]
where $[{\bx^d}^\top\bx^d]^{+}$ is the Moore–Penrose inverse of ${\bx^d}^\top\bx^d$, $A\preceq B$ means $B-A$ is positive semidefinite and $A\succ0$ means $A$ is positive definite. As all $\lambda^2_j>0$, $\boldsymbol{\Lambda}$ is always positive definite. Since $\int \bx^d[{\bx^d}^\top\bx^d]^{+}{\bx^d}^\top \pi(\boldsymbol{\Lambda} \cond \text{data}) d\boldsymbol{\Lambda}=\bx^d[{\bx^d}^\top\bx^d]^{+}{\bx^d}^\top$ and the dominating matrix $\bx^d[{\bx^d}^\top\bx^d]^{+}{\bx^d}^\top$ is idempotent and thus bounded, the derivative and expectation can be interchanged by the dominated convergence theorem. Plug~\eqref{eq:e_beta_ML} in~\eqref{eq:df_proj}, we can derive the estimated degree of freedom for the Multi-Laplace model:
\begin{equation}
    \begin{split}
        \label{eq:df_ML}
    \hat{df}(s)_{Multi-Laplace} & = \sum_{d=1}^D \Tr\left(\bx^d_{[s]}({\bx_{[s]}^d}^\top\bx_{[s]}^d)^{-1}{\bx_{[s]}^d}^\top\frac{\partial \e[\bx^d\bb^d \cond \text{data}]}{\partial \by^d}\right) \\
    & = \sum_{d=1}^D \Tr\left(\frac{1}{T} \sum_{t=1}^T\bx^d_{[s]}({\bx_{[s]}^d}^\top\bx_{[s]}^d)^{-1}{\bx_{[s]}^d}^\top\bx^d [{\boldsymbol{\Lambda}^{(t)}}^{-1} + {\bx^d}^\top\bx^d]^{-1}{\bx^d}^\top\right) \\
    & = \frac{1}{T} \sum_{t=1}^T\sum_{d=1}^D \Tr\left(\bx^d_{[s]}({\bx_{[s]}^d}^\top\bx_{[s]}^d)^{-1}{\bx_{[s]}^d}^\top\bx^d [{\boldsymbol{\Lambda}^{(t)}}^{-1} + {\bx^d}^\top\bx^d]^{-1}{\bx^d}^\top\right) 
    \end{split}
\end{equation}

For the Horseshoe model, similarly, we employ MCMC draws of both ${\boldsymbol{\Lambda}}^{(t)}$ and ${\tau^2}^{(t)}$ to approximate:
\begin{equation}
    \begin{split}
        \label{eq:e_beta_H}
    \frac{\partial \e[\bx^d\bb^d \cond \text{data}]}{\partial \by^d} & = \int \frac{\partial \e[\bx^d\bb^d \cond \boldsymbol{\Lambda}, \tau^2, \text{data}]}{\partial \by^d} \pi(\boldsymbol{\Lambda}, \tau^2 \cond \text{data}) d\boldsymbol{\Lambda} d\tau^2\\
    & \approx \frac{1}{T} \sum_{t=1}^T \bx^d[({\tau^2}^{(t)}\boldsymbol{\Lambda}^{(t)})^{-1} + {\bx^d}^\top\bx^d]^{-1}{\bx^d}^\top
    \end{split}
\end{equation}
Similarly, the derivative is bounded: \[\frac{\partial \e[\bx^d\bb^d \cond \boldsymbol{\Lambda}, \text{data}]}{\partial \by^d} =\bx^d[{(\tau^2\boldsymbol{\Lambda})}^{-1} + {\bx^d}^\top\bx^d]^{-1}{\bx^d}^\top\preceq\bx^d[{\bx^d}^\top\bx^d]^{+}{\bx^d}^\top, \forall \tau^2\boldsymbol{\Lambda}\succ0.\]By using the dominant convergence theorem and
plug~\eqref{eq:e_beta_H} in~\eqref{eq:df_proj}, we have the estimated degree of freedom for Horseshoe model:
\begin{equation}
\label{eq:df_H}
\hat{df}(s)_{Horseshoe} = \frac{1}{T} \sum_{t=1}^T\sum_{d=1}^D \Tr\left(\bx^d_{[s]}({\bx_{[s]}^d}^\top\bx_{[s]}^d)^{-1}{\bx_{[s]}^d}^\top\bx^d [({\tau^2}^{(t)}\boldsymbol{\Lambda}^{(t)})^{-1} + {\bx^d}^\top\bx^d]^{-1}{\bx^d}^\top\right)
\end{equation}

Similarly, we can calculate the estimated degree of freedom for ARD model:
\begin{equation}
\label{eq:df_ARD}
\hat{df}(s)_{ARD} = \frac{1}{T} \sum_{t=1}^T\sum_{d=1}^D \Tr\left(\bx^d_{[s]}({\bx_{[s]}^d}^\top\bx_{[s]}^d)^{-1}{\bx_{[s]}^d}^\top\bx^d [{\boldsymbol{\Psi}}^{(t)} + {\bx^d}^\top\bx^d]^{-1}{\bx^d}^\top\right)
\end{equation}

For the Spike-and-Laplace model, the active covariates set is indexed by $\boldsymbol{\gamma}^{(t)}$ at $t$-th draw. So we need to calculate $\e[\bx^d\bb^d \cond \text{data}]$ in~\eqref{eq:df_general} by accounting for the selection variability by $\boldsymbol{\gamma}$ across draws. Using same Monte Carlo integral, we have 
\begin{equation*}
    \begin{split}
        \label{eq:df_general_ss}
    & \frac{\partial \e[\bx^d\bb^d \cond \text{data}]}{\partial \by^d}  \\
    & = \int \sum_{\bZ} \frac{\partial \e[\bx^d\bb^d \cond \bZ, \boldsymbol{\Lambda}, \text{data}]}{\partial \by^d} \pi(\bZ, \boldsymbol{\Lambda} \cond \text{data}) d\boldsymbol{\Lambda}\\
    &\approx \frac{1}{T} \sum_{t=1}^T \bx^d_{[\bZ^{(t)}]}\left[({\boldsymbol{\Lambda}}^{(t)}_{[\bZ^{(t)}]})^{-1}+{\bx^d}^\top_{[\bZ^{(t)}]}\bx^d_{[\bZ^{(t)}]}\right]^{-1}{\bx^d}_{[\bZ^{(t)}]}^\top
    \end{split}
\end{equation*}
The first equality holds by the dominated convergence theorem. $\frac{\partial \e[\bx^d\bb^d \cond \boldsymbol{\Lambda}, \text{data}]}{\partial \by^d}=\bx^d_{[\bZ]}\left[({\boldsymbol{\Lambda}}_{[\bZ]})^{-1}+{\bx^d}^\top_{[\bZ]}\bx^d_{[\bZ]}\right]^{-1}{\bx^d}_{[\bZ]}^\top$ comes from Step 3 in Web Algorithm~\ref{alg:collapse_spike_laplace}. If $\bZ^{(t)} = 0$, we simply define $\bx^d_{[\bZ^{(t)}]}\left[({\boldsymbol{\Lambda}}^{(t)}_{[\bZ^{(t)}]})^{-1}+{\bx^d}^\top_{[\bZ^{(t)}]}\bx^d_{[\bZ^{(t)}]}\right]^{-1}{\bx^d}_{[\bZ^{(t)}]}^\top\equiv0$, because $\e[\bx^d\bb^d \cond \bZ=0, \boldsymbol{\Lambda}, \text{data}]=0$. Similar to the Multi-Laplace model, when $\bZ^{(t)}\neq 0$, we have 
\[\bx^d_{[\bZ]}\left[({\boldsymbol{\Lambda}}_{[\bZ]})^{-1}+{\bx^d}^\top_{[\bZ]}\bx^d_{[\bZ]}\right]^{-1}{\bx^d}_{[\bZ]}^\top\preceq\bx_{[\bZ]}^d[{\bx_{[\bZ]}^d}^\top\bx_{[\bZ]}^d]^{+}{\bx_{[\bZ]}^d}^\top, \forall \boldsymbol{\Lambda}_{[\bZ]}\succ0.\]Here, the dominant matrix $\bx_{[\bZ]}^d[{\bx_{[\bZ]}^d}^\top\bx_{[\bZ]}^d]^{+}{\bx_{[\bZ]}^d}^\top$ is bounded, so we can interchange differentiation, integration, and even infinite summation over $\boldsymbol{\gamma}$.

Finally, we have the estimated degree of freedom for the Spike-and-Laplace model:
\begin{equation}
\label{eq:df_sl}
\resizebox{0.9\textwidth}{!}{$
\begin{aligned}
\hat{df}(s)_{Spike-and-Laplace} =\frac{1}{T} \sum_{t=1}^T \sum_{d=1}^D \Tr\left(\bx^d_{[s]}({\bx_{[s]}^d}^\top\bx_{[s]}^d)^{-1}{\bx_{[s]}^d}^\top\bx^d_{[\bZ^{(t)}]}\left[({\boldsymbol{\Lambda}}^{(t)}_{[\bZ^{(t)}]})^{-1}+{\bx^d}^\top_{[\bZ^{(t)}]}\bx^d_{[\bZ^{(t)}]}\right]^{-1}{\bx^d}_{[\bZ^{(t)}]}^\top\right)
\end{aligned}
$}
\end{equation}

\section{Calibration of the Projected Posterior}
\label{supsec:calibration}

Fix an imputation $d$ and the selected set $s_m$. The projected coefficients $\bb_{s_m}^d$ have posterior covariance $\mathbf{U}_{s_m}^d=\mathrm{Cov}(\bb_{s_m}^d\cond\text{data})$, which the calibration rescales to the target $\mathbf{V}_{s_m}^d=\E[\sigma_{s_m}^2\cond\text{data}]\,({\bx^d_{[s_m]}}^\top\bx^d_{[s_m]})^{-1}$. This target is the posterior covariance of the submodel coefficients $\check\bb_{s_m}^d$ obtained by fitting the submodel directly under a flat prior. By the law of total variance,
\[ \begin{aligned} \mathrm{Cov}(\check\bb_{s_m}^d\cond\text{data}) &= \E\big[\mathrm{Cov}(\check\bb_{s_m}^d\mid\sigma_{s_m}^2,\text{data})\cond\text{data}\big] + \mathrm{Cov}\big(\E[\check\bb_{s_m}^d\mid\sigma_{s_m}^2,\text{data}]\cond\text{data}\big) \\ &= \E[\sigma_{s_m}^2\cond\text{data}]\,({\bx^d_{[s_m]}}^\top\bx^d_{[s_m]})^{-1} = \mathbf{V}_{s_m}^d, \end{aligned} \]
where the first term uses $\mathrm{Cov}(\check\bb_{s_m}^d\mid\sigma_{s_m}^2,\text{data})=\sigma_{s_m}^2({\bx^d_{[s_m]}}^\top\bx^d_{[s_m]})^{-1}$ and the second vanishes because $\E[\check\bb_{s_m}^d\mid\sigma_{s_m}^2,\text{data}]$ does not depend on $\sigma_{s_m}^2$. Both $\mathbf{U}_{s_m}^d$ and $\mathbf{V}_{s_m}^d$ are symmetric and positive definite, so the symmetric square root $(\mathbf{V}_{s_m}^d)^{1/2}$ and the symmetric inverse square root $(\mathbf{U}_{s_m}^d)^{-1/2}$ used below are well defined, and the calibration in Step~4 of the main text is the affine map
\[ \bb_{s_m,\text{calibrated}}^{d,(t)} = (\mathbf{V}_{s_m}^d)^{1/2}(\mathbf{U}_{s_m}^d)^{-1/2}\big(\bb_{s_m}^{d,(t)}-\E[\bb_{s_m}^d\cond\text{data}]\big)+\E[\bb_{s_m}^d\cond\text{data}]. \]

The matrix $(\mathbf{V}_{s_m}^d)^{1/2}(\mathbf{U}_{s_m}^d)^{-1/2}$ and the vector $\E[\bb_{s_m}^d\cond\text{data}]$ are fixed given the data, so taking the posterior expectation of this map leaves the mean unchanged, $\E[\bb_{s_m,\text{calibrated}}^d\cond\text{data}] = \E[\bb_{s_m}^d\cond\text{data}]$, while for the covariance the additive constant drops out and, using $\mathrm{Cov}(\mathbf{A}\bb_{s_m}^d\cond\text{data})=\mathbf{A}\,\mathbf{U}_{s_m}^d\,\mathbf{A}^\top$ with $\mathbf{A}=(\mathbf{V}_{s_m}^d)^{1/2}(\mathbf{U}_{s_m}^d)^{-1/2}$ together with $(\mathbf{U}_{s_m}^d)^{-1/2}\mathbf{U}_{s_m}^d(\mathbf{U}_{s_m}^d)^{-1/2}=\mathbf{I}$,
\[ \mathrm{Cov}(\bb_{s_m,\text{calibrated}}^d\cond\text{data}) = (\mathbf{V}_{s_m}^d)^{1/2}(\mathbf{U}_{s_m}^d)^{-1/2}\,\mathbf{U}_{s_m}^d\,(\mathbf{U}_{s_m}^d)^{-1/2}(\mathbf{V}_{s_m}^d)^{1/2} = \mathbf{V}_{s_m}^d. \]
The calibrated draws therefore have posterior mean $\E[\bb_{s_m}^d\cond\text{data}]$ and posterior covariance $\mathbf{V}_{s_m}^d$, and the argument holds separately for each imputation $d$.

\clearpage
\newpage

\section{Additional Results of Simulation Scenario 1}

\subsection{Settings of Weaker Signals and High Dimension}
We report the results of weaker signals (SNR=0.2), relatively high dimensionality ($p=80$), and higher dimensionality ($p=200$) in Web Table~\ref{tab:sim_A_additional}. The data-generating processes of weaker signals and relatively high dimensionality were the same as other settings in Simulation Scenario 1. For high dimensional settings ($n<p$), we used classification and regression trees (CART; \citealp{Doove2014}) as the imputer, as CART remains effective in these settings. We generated $D=5$ imputations with five iterations each, using the \texttt{mice} defaults for the CART imputer, a minimum leaf size of five observations and a complexity parameter of $10^{-4}$.

\begin{table}[!htbp]
\centering
\resizebox{\textwidth}{!}{
\begin{tabular}{cccccccccccccccccccccccc}
\toprule
 & \multicolumn{10}{c}{{\normalsize \textbf{n=100, p=20, SNR=0.2, moderate missing}}} & & & & \multicolumn{10}{c}{\normalsize \textbf{n=100, p=200, SNR=1.5, moderate missing}} \\
 \midrule
\textbf{Full Data} &  \multicolumn{2}{c}{\textbf{SEN}} & \multicolumn{2}{c}{\textbf{SPE}} & \multicolumn{2}{c}{\textbf{F$_1$}} & \multicolumn{2}{c}{\textbf{MSE}} & \multicolumn{2}{c}{\textbf{SIGN}} & & & & \multicolumn{2}{c}{\textbf{SEN}} & \multicolumn{2}{c}{\textbf{SPE}} & \multicolumn{2}{c}{\textbf{F$_1$}} & \multicolumn{2}{c}{\textbf{MSE}} & \multicolumn{2}{c}{\textbf{SIGN}} \\
LASSO & \multicolumn{2}{c}{15} & \multicolumn{2}{c}{98} & \multicolumn{2}{c}{19} & \multicolumn{2}{c}{530} & \multicolumn{2}{c}{73} & & & & \multicolumn{2}{c}{92} & \multicolumn{2}{c}{53} & \multicolumn{2}{c}{20} & \multicolumn{2}{c}{1403} & \multicolumn{2}{c}{55}\\
BLASSO & \multicolumn{2}{c}{22} & \multicolumn{2}{c}{98} & \multicolumn{2}{c}{30} & \multicolumn{2}{c}{552} & \multicolumn{2}{c}{75} & & & & \multicolumn{2}{c}{61} & \multicolumn{2}{c}{97} & \multicolumn{2}{c}{56} & \multicolumn{2}{c}{865} & \multicolumn{2}{c}{95}\\
 \midrule
\multirow{2}*{\textbf{Models}} & \multicolumn{5}{c}{MCAR} & & \multicolumn{5}{c}{MAR} & & \multicolumn{5}{c}{MCAR} & & \multicolumn{5}{c}{MAR} \\
\cmidrule{2-6} \cmidrule{8-12} \cmidrule{14-18} \cmidrule{20-24}
 & \textbf{SEN} & \textbf{SPE} & \textbf{F}$_1$ & \textbf{MSE} & \textbf{SIGN} & & \textbf{SEN} & \textbf{SPE} & \textbf{F}$_1$ & \textbf{MSE} & \textbf{SIGN} & & \textbf{SEN} & \textbf{SPE} & \textbf{F}$_1$ & \textbf{MSE} & \textbf{SIGN} & & \textbf{SEN} & \textbf{SPE} & \textbf{F}$_1$ & \textbf{MSE} & \textbf{SIGN} \\
  CC-LASSO & 8 & \textbf{99} & 10 & 575 & 71 & & 6 & 98 & 8 & \textbf{584} & 70 & & 71 & 72 & 23 & 1444 & 72 & & 57 & 70 & 18 & 1432 & 69\\ 
  CC-BLASSO & 8 & \textbf{99} & 12 & 587 & 72 & & 3 & \textbf{99} & 5 & 597 & 70 & & 6 & \textbf{100} & 9 & 1169 & \textbf{94} & & 1 & \textbf{100} & 2 & 1197 & \textbf{94}\\ 
  SaENET & \textbf{55} & 81 & \textbf{53} & \textbf{521} & 73 & & \textbf{56} & 73 & \textbf{51} & 630 & 67 & & 43 & 97 & \textbf{47} & \textbf{921} & \textbf{94} & & 39 & 97 & \textbf{42} & \textbf{989} & 93\\ 
  GaLASSO & 47 & 87 & 49 & 601 & 75 & & 50 & 81 & 49 & 732 & 71 & & 1 & \textbf{100} & 1 & 1198 & \textbf{94} & & 3 & 99 & 3 & 1189 & \textbf{94}\\ 
  MI-LASSO & 32 & 94 & 37 & 606 & 75 & & 35 & 90 & 38 & 687 & 74 & & --- & --- & --- & --- & --- & & --- & --- & --- & --- & ---\\ 
    \cmidrule{1-1}
  Multi-Laplace & 30 & 96 & 40 & 546 & \textbf{77} & & 29 & 94 & 38 & 625 & 74 & & \textbf{91} & 56 & 21 & 4611 & 57 & & \textbf{84} & 58 & 21 & 5919 & 58\\ 
  Horseshoe & 35 & 96 & 45 & 536 & \textbf{77} & & 35 & 92 & 43 & 623 & \textbf{75} & & 78 & 60 & 20 & 2793 & 61 & & 71 & 60 & 18 & 3371 & 60\\ 
  ARD & 32 & 96 & 42 & 559 & \textbf{77} & & 33 & 93 & 41 & 645 & \textbf{75} & & 83 & 56 & 19 & 2656 & 57 & & 76 & 55 & 17 & 3770 & 56\\ 
  Spike-and-Laplace & 28 & 97 & 38 & 595 & 76 & & 29 & 94 & 37 & 674 & 74 & & 78 & 61 & 20 & 2600 & 62 & & 71 & 60 & 18 & 2997 & 61\\ 
\bottomrule
\toprule
\multicolumn{12}{c}{{\normalsize \textbf{n=100, p=80, SNR=1.5, moderate missing}}} \\
 \midrule
\textbf{Full Data} & \multicolumn{2}{c}{\textbf{SEN}} & \multicolumn{2}{c}{\textbf{SPE}} & \multicolumn{2}{c}{\textbf{F$_1$}} & \multicolumn{2}{c}{\textbf{MSE}} & \multicolumn{2}{c}{\textbf{SIGN}} \\
LASSO & \multicolumn{2}{c}{52} & \multicolumn{2}{c}{94} & \multicolumn{2}{c}{48} & \multicolumn{2}{c}{765} & \multicolumn{2}{c}{88}\\
BLASSO & \multicolumn{2}{c}{79} & \multicolumn{2}{c}{93} & \multicolumn{2}{c}{72} & \multicolumn{2}{c}{596} & \multicolumn{2}{c}{91}\\
 \midrule
\multirow{2}*{\textbf{Models}} & \multicolumn{5}{c}{MCAR} & & \multicolumn{5}{c}{MAR} \\
\cmidrule{2-6} \cmidrule{8-12}
 & \textbf{SEN} & \textbf{SPE} & \textbf{F}$_1$ & \textbf{MSE} & \textbf{SIGN} & & \textbf{SEN} & \textbf{SPE} & \textbf{F}$_1$ & \textbf{MSE} & \textbf{SIGN} \\
\midrule
  CC-LASSO & \textbf{86} & 34 & 31 & 2131 & 41 & & 81 & 32 & 28 & 1894 & 39\\ 
  CC-BLASSO & 33 & 97 & 40 & 1056 & 88 & & 10 & \textbf{99} & 15 & 1177 & 86\\ 
  SaENET & 69 & 88 & 62 & 611 & 85 & & 65 & 85 & 56 & 721 & 82\\ 
  GaLASSO & 20 & \textbf{100} & 32 & 1022 & 88 & & 21 & \textbf{99} & 31 & 1025 & 88\\ 
  MI-LASSO & 76 & 88 & 61 & 620 & 86 & & \textbf{85} & 69 & 51 & 2141 & 71\\ 
    \cmidrule{1-1}
  Multi-Laplace & 72 & 94 & \textbf{71} & \textbf{574} & \textbf{91} & & 62 & 95 & 64 & 688 & \textbf{90}\\ 
  Horseshoe & 80 & 90 & 69 & 601 & 89 & & 73 & 91 & \textbf{66} & \textbf{676} & 89\\ 
  ARD & 79 & 90 & 68 & 656 & 88 & & 73 & 90 & 65 & 729 & 88\\ 
  Spike-and-Laplace & 78 & 89 & 67 & 696 & 88 & & 71 & 90 & 64 & 757 & 88\\ 
\bottomrule
\end{tabular}
}
\caption{Results of sensitivity (\texttimes100), specificity (\texttimes100), $\text{F}_1$ score (\texttimes100), MSE (\texttimes100) and sign accuracy (\texttimes100) in Scenario 1. The best value per metric within each setting is in bold. MI-LASSO is omitted in the high dimensional setting ($n=100$, $p=200$), as it is not formulated for $p>n$.}
\label{tab:sim_A_additional}
\end{table}

\subsection{Running Time}
We report the running time of various settings across different $n$ and $p$ in Web Table~\ref{tab:running}.

\begin{table}[!htbp]
\centering
\resizebox{\textwidth}{!}{
\begin{tabular}{cccccccccccccccccccccccccccccc}
\toprule
\multicolumn{30}{c}{{\normalsize \textbf{Median running time in minutes}}}\\
 \midrule
 & \multicolumn{5}{c}{\textbf{$n=100$, $p=20$}} & & \multicolumn{5}{c}{\textbf{$n=100$, $p=40$}} & & \multicolumn{5}{c}{\textbf{$n=100$, $p=80$}} & & \multicolumn{5}{c}{\textbf{$n=100$, $p=200$}} & & \multicolumn{5}{c}{\textbf{$n=200$, $p=20$}} \\
\cmidrule{2-6} \cmidrule{8-12} \cmidrule{14-18} \cmidrule{20-24} \cmidrule{26-30}
  SaENET & \multicolumn{5}{c}{0.23} & & \multicolumn{5}{c}{0.60} & & \multicolumn{5}{c}{3.90} & & \multicolumn{5}{c}{13.54} & & \multicolumn{5}{c}{0.22}\\ 
  GaLASSO & \multicolumn{5}{c}{0.17} & & \multicolumn{5}{c}{0.47} & & \multicolumn{5}{c}{8.21} & & \multicolumn{5}{c}{21.94} & & \multicolumn{5}{c}{0.13}\\ 
  MI-LASSO & \multicolumn{5}{c}{0.13} & & \multicolumn{5}{c}{0.34} & & \multicolumn{5}{c}{1.80} & & \multicolumn{5}{c}{--} & & \multicolumn{5}{c}{0.20}\\ 
  \cmidrule{1-1}
  Multi-Laplace & \multicolumn{5}{c}{0.12 (0.10)} & & \multicolumn{5}{c}{0.14 (0.10)} & & \multicolumn{5}{c}{0.39 (0.24)} & & \multicolumn{5}{c}{1.52 (0.56)} & & \multicolumn{5}{c}{0.12 (0.05)}\\ 
  Horseshoe & \multicolumn{5}{c}{0.06 (0.04)} & & \multicolumn{5}{c}{0.11 (0.08)} & & \multicolumn{5}{c}{0.24 (0.16)} & & \multicolumn{5}{c}{0.68 (0.24)} & & \multicolumn{5}{c}{0.09 (0.05)}\\ 
  ARD & \multicolumn{5}{c}{0.05 (0.04)} & & \multicolumn{5}{c}{0.10 (0.07)} & & \multicolumn{5}{c}{0.23 (0.15)} & & \multicolumn{5}{c}{0.69 (0.24)} & & \multicolumn{5}{c}{0.09 (0.04)}\\ 
  Spike-and-Laplace & \multicolumn{5}{c}{0.19 (0.02)} & & \multicolumn{5}{c}{0.57 (0.02)} & & \multicolumn{5}{c}{1.09 (0.04)} & & \multicolumn{5}{c}{12.38 (0.10)} & & \multicolumn{5}{c}{0.23 (0.02)}\\ 
\bottomrule
\end{tabular}
}
\caption{This table summaries the median total running time in minutes, with the four-step selection time in parentheses for the Bayesian MI-LASSO models. MI-LASSO is omitted in the high dimensional setting ($n=100$, $p=200$), as it is not formulated for $p>n$. }
\label{tab:running}
\end{table}

\section{Sensitivity Analysis}
\subsection{Hyperparameters}
We conducted a sensitivity analysis to evaluate the robustness of coefficient estimates under different prior specifications for the Multi-Laplace and Spike-and-Laplace models. Our analysis was based on two parts: i) the fitted model before variable selection procedure, i.e. the full-model; ii) the best model selected by four-step projection predictive variable selection. Specifically, we assessed both MSE of full-model estimates and different criteria of the selected models. Following the prior specification strategy outlined in Section 2.3 in the main paper, we varied the hyperparameters $(h, v)$ and $(a, b)$ to make the marginal prior variances of the coefficients equal to $0.1\sigma^2$, $\sigma^2$, and $10\sigma^2$. The sensitivity analysis was based on two design settings: independent continuous covariates and correlated continuous covariates with AR(1) structure ($\varrho = 0.5$), using the setting of $n = 100$, $p = 20$, and $\text{SNR} = 1.5$. For each scenario, we considered both moderate and high missingness rates and conducted 200 replicates of simulation. 

To evaluate the stability of full-model fitting before selection, we calculated the MSE of the posterior means of regression coefficients. Results are summarized in Web Table~\ref{tab:full_MSE}, including those for the default prior settings. The results were generally robust to the choice of hyperparameter values, except when $v$ in the Multi-Laplace model and $b$ in the Spike-and-Laplace model were very small. In these cases, the shrinkage effect on the regression coefficients was substantially reduced. Under such conditions, the Multi-Laplace model tended to exhibit higher MSE, whereas the Spike-and-Laplace model more often achieved lower MSE. To help explain this difference, we calculated the posterior means of the pooled coefficients for both models as well as the posterior mean of the selection probabilities in the Spike-Laplace model, as shown in Web Table~\ref{tab:estimates_small}. The results indicate that the Spike-and-Laplace model effectively identified the true variables through binary indicators $\boldsymbol{\gamma}$, with selection probabilities close to 1 for important covariates and near 0 for unimportant ones. This strong variable selection performance contributed to the lower MSE observed in the Spike-and-Laplace model, even though the Multi-Laplace model tended to perform less favorably under the same conditions.

To assess the sensitivity of variable selection results to prior specifications, we evaluated differences in performance criteria between the compared and default priors under the four-step projection predictive selection procedure. Specifically, we computed the average difference for each criterion and its corresponding Monte Carlo standard error (MCSE) across 200 replicates:
\begin{equation*}
  \resizebox{0.9\textwidth}{!}{$
    \begin{aligned}
      \mathrm{Diff}(\mathrm{compared},\mathrm{default})
        &= \frac{1}{200} \sum_{r=1}^{200}
             \Bigl(\mathrm{criterion}_r^{(\mathrm{compared})}
                   - \mathrm{criterion}_r^{(\mathrm{default})}\Bigr), \\[6pt]
      \mathrm{MCSE}(\mathrm{compared},\mathrm{default})
        &= \sqrt{\frac{1}{200\,(200-1)}\,
                    \sum_{r=1}^{200}
                      \Bigl(\mathrm{criterion}_r^{(\mathrm{compared})}
                            - \mathrm{criterion}_r^{(\mathrm{default})}\Bigr)^{2}},
    \end{aligned}
  $}
\end{equation*}
where $criterion_r$ refers to sensitivity, specificity, F$_1$ score, MSE, sign accuracy in the $r$-th replicate. Results are summarized in Web Tables~\ref{tab:sen_independent_mid} and~\ref{tab:sen_independent_high}. We considered the comparison priors to yield different results from the default priors only when both of the following conditions were met: (i) the average difference exceeded the corresponding MCSE, and (ii) the absolute value of the average difference was large enough relative to the default model’s value for that criterion. After selection, robustness improved substantially with most settings produced similar selections and MSE, except when $b$ in the Spike-and-Laplace model were very small under which the model yielded lower specificity and sign accuracy.

\begin{table}[!htbp]
\centering
 \resizebox{\textwidth}{!}{
\begin{tabular}{cclccccccc}
\toprule
\multicolumn{3}{c}{{\textbf{n=100, p=20, SNR=1.5, independent}}} & \multicolumn{7}{c}{{\textbf{MSE of the fitted full-model}}}\\ 
 \midrule
\multirow{2}*{\textbf{Models}} & \multirow{2}*{\makecell[c]{\textbf{Prior}\\\textbf{variance}}} & \multirow{2}*{\makecell[c]{\textbf{Hyper-}\\\textbf{parameters}}} & \multicolumn{3}{c}{\textbf{Moderate missing}} & & \multicolumn{3}{c}{\textbf{High missing}}\\ 
\cmidrule{4-6} \cmidrule{8-10}
& & & MCAR & & MAR & & MCAR & & MAR\\
\midrule
\multirow{8}*{\makecell[c]{Multi-\\Laplace}} & \multirow{3}*{$0.1\sigma^2$} & $h=1.1$, $v=120$ & 82 & & 90 & & 93 & & 113\\
& & $h=2$, $v=12$ & 82 & & 90 & & 92 & & 111\\
& & $h=11$, $v=1.2$ & 83 & & 91 & & 94 & & 114\\ \cmidrule{2-3}
& \multirow{3}*{$1\sigma^2$} & $h=1.1$, $v=12$ & 83 & & 90 & & 93 & & 113\\
& & $h=2$, $v=1.2$ (default) &86 & & 93 & & 97 & & 118\\
& & $h=11$, $v=0.12$ &  94 & & 102 & & 108 & & 130\\ \cmidrule{2-3}
& \multirow{3}*{$10\sigma^2$} & $h=1.1$, $v=1.2$ & 86  & & 93 & & 98 & & 118\\
& & $h=2$, $v=0.12$ & 96 & & 104 & & 110 & & 131\\
& & $h=11$, $v=0.012$ & 105 & & 114 & & 122 & & 146\\
\midrule
\multirow{8}*{\makecell[c]{Spike-and-\\Laplace}} & \multirow{3}*{$0.1\sigma^2$} & $a=1.1$, $b=60$ & 62& & 69 & & 73 & & 93\\
& & $a=2$, $b=6$ & 62 & & 69 & & 73 & & 92\\
& & $a=11$, $b=0.6$ & 62& & 69 & & 73 & & 93\\ \cmidrule{2-3}
& \multirow{3}*{$1\sigma^2$} & $a=1.1$, $b=6$ &62 & & 69 & & 73 & & 93\\
& & $a=2$, $b=0.6$ (default) & 60 & & 68  & & 72 & &  92\\
& & $a=11$, $b=0.06$ & 56 & & 64 & & 68 & & 87\\ \cmidrule{2-3}
& \multirow{3}*{$10\sigma^2$} & $a=1.1$, $b=0.6$ & 60 & & 68 & & 71 & & 92\\
& & $a=2$, $b=0.06$ & 54 & & 61 & & 66 & & 83\\
& & $a=11$, $b=0.006$ & 48 & & 54 & & 59 & & 77\\
\bottomrule
\end{tabular}
}
\caption{Sensitivity analysis of MSE (×100) of fitted full-models.}
\label{tab:full_MSE}
\end{table}

\begin{sidewaystable}[htbp]
\centering
\captionsetup{type=table}
\resizebox{\textwidth}{!}{
\begin{tabular}{ccccccccccccccccccccccc}
\toprule
\multicolumn{23}{c}{{\textbf{n=100, p=20, SNR=1.5, independent, moderate missing, MCAR}}}\\ 
 \midrule
\multirow{2}*{\textbf{Models}} & \multirow{2}*{\makecell[c]{\textbf{Hyper-}\\\textbf{parameters}}} & \multirow{2}*{\textbf{Estimates}} & \multicolumn{20}{c}{\textbf{Variables}} \\
\cmidrule{4-23}
& & & $\bX_1$ & $\bX_2$ & $\bX_3 $ & $\bX_4$ & $\bX_5$ & $\bX_6$ & $\bX_7$ & $\bX_8$ & $\bX_9$ & $\bX_{10}$ & $\bX_{11}$ & $\bX_{12}$ & $\bX_{13}$ & $\bX_{14}$ & $\bX_{15}$ & $\bX_{16}$ & $\bX_{17}$ & $\bX_{18}$ & $\bX_{19}$ & $\bX_{20}$\\
\midrule
\makecell[c]{True\\model} & --- & True $\beta$ & 1 & 1 & 0 & 0 & 1 & 0 & 0 & 0 & 0 & 0 & 1 & 1 & 0 & 0 & 1 & 0 & 0 & 0 & 0 & 0\\
\midrule
\makecell[c]{Multi-\\Laplace} & \makecell[c]{$h=11$\\ $v=0.012$} & $\bb^{\text{pool}}_{full}$ & 0.96 & 0.97 & -0.01 & 0.02 & 0.96 & -0.02 & 0.01 & 0 & -0.03 & 0.03 & 0.99 & 1.02 & 0 & 0.01 & 0.98 & -0.01 & -0.01 & 0.02 & 0.01 & 0.01 \\
\midrule
\multirow{2}*{\makecell[c]{Spike-\\Laplace}} & \multirow{2}*{\makecell[c]{$a=11$\\ $b=0.006$}} & $\bb^{\text{pool}}_{full}$ & 0.97 & 0.99 & 0 & 0 & 0.98 & 0 & 0 & 0 & 0 & 0.01 & 0.99 & 1.01 & 0.01 & 0.01 & 0.97 & 0 & 0 & 0.01 & -0.01 & 0.01 \\
& & $\boldsymbol{\gamma}$ & 0.99 & 0.99 & 0.04 & 0.04 & 1 & 0.03 & 0.03 & 0.05 & 0.02 & 0.03 & 0.98 & 0.98 & 0.05 & 0.05 & 0.99 & 0.05 & 0.06 & 0.05 & 0.06 & 0.06 \\
\bottomrule
\toprule
\multicolumn{23}{c}{{\textbf{n=100, p=20, SNR=1.5, independent, moderate missingness, MAR}}}\\ 
 \midrule
\multirow{2}*{\textbf{Models}} & \multirow{2}*{\makecell[c]{\textbf{Hyper-}\\\textbf{parameters}}} & \multirow{2}*{\textbf{Estimates}} & \multicolumn{20}{c}{\textbf{Variables}} \\
\cmidrule{4-23}
& & & $\bX_1$ & $\bX_2$ & $\bX_3 $ & $\bX_4$ & $\bX_5$ & $\bX_6$ & $\bX_7$ & $\bX_8$ & $\bX_9$ & $\bX_{10}$ & $\bX_{11}$ & $\bX_{12}$ & $\bX_{13}$ & $\bX_{14}$ & $\bX_{15}$ & $\bX_{16}$ & $\bX_{17}$ & $\bX_{18}$ & $\bX_{19}$ & $\bX_{20}$\\
\midrule
\makecell[c]{True\\model} & --- & True $\beta$ & 1 & 1 & 0 & 0 & 1 & 0 & 0 & 0 & 0 & 0 & 1 & 1 & 0 & 0 & 1 & 0 & 0 & 0 & 0 & 0\\
\midrule
\makecell[c]{Multi-\\Laplace} & \makecell[c]{$h=11$\\ $v=0.012$} & $\bb^{\text{pool}}_{full}$ & 0.95 & 0.96 & -0.02 & 0.02 & 0.95 & -0.02 & 0.01 & 0 & -0.04 & 0.03 & 0.98 & 1.01 & 0 & -0.01 & 0.95 & -0.01 & -0.02 & 0.03 & 0.02 & 0 \\
\midrule
\multirow{2}*{\makecell[c]{Spike-\\Laplace}} & \multirow{2}*{\makecell[c]{$a=11$\\ $b=0.006$}} & $\bb^{\text{pool}}_{full}$ & 0.97 & 0.98 & 0 & 0.01 & 0.98 & 0 & 0 & 0 & -0.01 & 0.01 & 0.97 & 1 & -0.01 & 0.01 & 0.94 & 0.01 & -0.01 & 0 & 0 & 0.01 \\
& & $\boldsymbol{\gamma}$ & 0.99 & 0.99 & 0.05 & 0.03 & 0.99 & 0.03 & 0.03 & 0.05 & 0.04 & 0.03 & 0.98 & 0.98 & 0.06 & 0.05 & 0.97 & 0.06 & 0.07 & 0.06 & 0.07 & 0.06\\
\bottomrule
\toprule
\multicolumn{23}{c}{{\textbf{n=100, p=20, SNR=1.5, independent, high missing, MCAR}}}\\ 
 \midrule
\multirow{2}*{\textbf{Models}} & \multirow{2}*{\makecell[c]{\textbf{Hyper-}\\\textbf{parameters}}} & \multirow{2}*{\textbf{Estimates}} & \multicolumn{20}{c}{\textbf{Variables}} \\
\cmidrule{4-23}
& & & $\bX_1$ & $\bX_2$ & $\bX_3 $ & $\bX_4$ & $\bX_5$ & $\bX_6$ & $\bX_7$ & $\bX_8$ & $\bX_9$ & $\bX_{10}$ & $\bX_{11}$ & $\bX_{12}$ & $\bX_{13}$ & $\bX_{14}$ & $\bX_{15}$ & $\bX_{16}$ & $\bX_{17}$ & $\bX_{18}$ & $\bX_{19}$ & $\bX_{20}$\\
\midrule
\makecell[c]{True\\model} & --- & True $\beta$ & 1 & 1 & 0 & 0 & 1 & 0 & 0 & 0 & 0 & 0 & 1 & 1 & 0 & 0 & 1 & 0 & 0 & 0 & 0 & 0\\
\midrule
\makecell[c]{Multi-\\Laplace} & \makecell[c]{$h=11$\\ $v=0.012$} & $\bb^{\text{pool}}_{full}$ & 0.97 & 0.97 & -0.02 & 0.01 & 0.96 & -0.02 & 0.01 & 0 & -0.04 & 0.03 & 1 & 1.02 & 0 & 0 & 0.99 & -0.01 & -0.01 & 0.03 & 0.01 & 0\\
\midrule
\multirow{2}*{\makecell[c]{Spike-\\Laplace}} & \multirow{2}*{\makecell[c]{$a=11$\\ $b=0.006$}} & $\bb^{\text{pool}}_{full}$ & 0.98 & 0.99 & 0 & 0 & 0.98 & -0.01 & 0 & -0.01 & -0.01 & 0 & 0.98 & 1 & -0.01 & 0.01 & 0.97 & 0 & -0.01 & 0.01 & 0 & 0.01 \\
& & $\boldsymbol{\gamma}$ & 0.99 & 0.99 & 0.05 & 0.04 & 1 & 0.04 & 0.05 & 0.06 & 0.03 & 0.03 & 0.97 & 0.97 & 0.07 & 0.07 & 0.98 & 0.05 & 0.08 & 0.08 & 0.07 & 0.08\\
\bottomrule
\toprule
\multicolumn{23}{c}{{\textbf{n=100, p=20, SNR=1.5, independent, high missingness, MAR}}}\\ 
 \midrule
\multirow{2}*{\textbf{Models}} & \multirow{2}*{\makecell[c]{\textbf{Hyper-}\\\textbf{parameters}}} & \multirow{2}*{\textbf{Estimates}} & \multicolumn{20}{c}{\textbf{Variables}} \\
\cmidrule{4-23}
& & & $\bX_1$ & $\bX_2$ & $\bX_3 $ & $\bX_4$ & $\bX_5$ & $\bX_6$ & $\bX_7$ & $\bX_8$ & $\bX_9$ & $\bX_{10}$ & $\bX_{11}$ & $\bX_{12}$ & $\bX_{13}$ & $\bX_{14}$ & $\bX_{15}$ & $\bX_{16}$ & $\bX_{17}$ & $\bX_{18}$ & $\bX_{19}$ & $\bX_{20}$\\
\midrule
\makecell[c]{True\\model} & --- & True $\beta$ & 1 & 1 & 0 & 0 & 1 & 0 & 0 & 0 & 0 & 0 & 1 & 1 & 0 & 0 & 1 & 0 & 0 & 0 & 0 & 0\\
\midrule
\makecell[c]{Multi-\\Laplace} & \makecell[c]{$h=11$\\ $v=0.012$} & $\bb^{\text{pool}}_{full}$ &   0.94 & 0.96 & -0.02 & 0.01 & 0.94 & -0.02 & 0.02 & 0.01 & -0.04 & 0.02 & 0.98 & 1.03 & -0.01 & -0.04 & 0.96 & -0.02 & -0.04 & 0.04 & 0.03 & 0.01\\
\midrule
\multirow{2}*{\makecell[c]{Spike-\\Laplace}} & \multirow{2}*{\makecell[c]{$a=11$\\ $b=0.006$}} & $\bb^{\text{pool}}_{full}$ & 0.95 & 0.97 & -0.01 & -0.01 & 0.97 & -0.01 & 0 & 0 & -0.02 & 0.01 & 0.96 & 0.99 & -0.01 & -0.01 & 0.94 & -0.02 & -0.01 & 0.03 & 0.01 & 0.02 \\
& & $\boldsymbol{\gamma}$ & 0.98 & 0.98 & 0.07 & 0.06 & 0.99 & 0.06 & 0.06 & 0.07 & 0.07 & 0.05 & 0.97 & 0.98 & 0.12 & 0.08 & 0.96 & 0.12 & 0.07 & 0.14 & 0.12 & 0.11\\
\bottomrule
\end{tabular}
}
\caption{Posterior mean of estimates of coefficients and binary inclusion indicators when $v$ and $b$ were very small.}
\label{tab:estimates_small}
\end{sidewaystable}

\begin{table}[!htbp]
\centering
\resizebox{\textwidth}{!}{
\begin{tabular}{cclccccccccccccccccc}
\toprule
 \multicolumn{20}{c}{{\normalsize \textbf{n=100, p=20, SNR=1.5, independent, moderate missingness}}}\\ 
 \midrule
 \multicolumn{20}{c}{{\normalsize \textbf{MCAR}}}\\ 
  \midrule
\multirow{2}*{\textbf{Models}} & \multirow{2}*{\makecell[c]{\textbf{Prior}\\\textbf{variance}}} & \multirow{2}*{\makecell[c]{\textbf{Hyper-}\\\textbf{parameters}}} & \multicolumn{2}{c}{\textbf{SEN}} & & \multicolumn{2}{c}{\textbf{SPE}} & & \multicolumn{2}{c}{\textbf{F}$_1$} & & \multicolumn{2}{c}{\textbf{MSE}} & & \multicolumn{2}{c}{\textbf{SIGN}} & & \multicolumn{2}{c}{\textbf{MSE full}}\\
\cmidrule{4-5} \cmidrule{7-8} \cmidrule{10-11} \cmidrule{13-14} \cmidrule{16-17} \cmidrule{19-20}
& & & Diff & MCSE & & Diff & MCSE & & Diff & MCSE & & Diff & MCSE & & Diff & MCSE & & Diff & MCSE\\
\midrule
\multirow{8}*{\makecell[c]{Multi-\\Laplace}} & \multirow{3}*{$0.1\sigma^2$} & $h=1.1$, $v=120$ & 0 & 0 & & 0 & 0 & & 0 & 0 & & 2 & 1 & & 0 & 0 & & -3 & 0\\
 &  & $h=2$, $v=12$ & 0 & 0 & & 0 & 0 & & 0 & 0 & & 1 & 1 & & 0 & 0 & & -3 & 0\\
 &  & $h=11$, $v=1.2$ & 0 & 0 & & 0 & 0 & & 0 & 0 & & 0 & 0 & & 0 & 0 & & -3 & 0\\
\cmidrule{2-3}
 & \multirow{2}*{$1\sigma^2$} & $h=1.1$, $v=12$ & 0 & 0 & & 0 & 0 & & 0 & 0 & & 1 & 1 & & 0 & 0 & & -3 & 0\\
 &  & $h=11$, $v=0.12$ & 0 & 0 & & 0 & 0 & & 0 & 0 & & 0 & 1 & & 0 & 0 & & 9 & 1\\
\cmidrule{2-3}
 & \multirow{3}*{$10\sigma^2$} & $h=1.1$, $v=1.2$ & 0 & 0 & & 0 & 0 & & 0 & 0 & & 0 & 0 & & 0 & 0 & & 0 & 0\\
 &  & $h=2$, $v=0.12$ & 0 & 0 & & 0 & 0 & & 0 & 0 & & -1 & 1 & & 0 & 0 & & 10 & 1\\
 &  & $h=11$, $v=0.012$ & 0 & 0 & & 0 & 0 & & 0 & 0 & & 2 & 1 & & 0 & 0 & & 20 & 1\\
\midrule
\multirow{8}*{\makecell[c]{Spike-and-\\Laplace}} & \multirow{3}*{$0.1\sigma^2$} & $a=1.1$, $b=60$ & 0 & 0 & & -1 & 0 & & 0 & 0 & & 2 & 1 & & 0 & 0 & & 2 & 0\\
 &  & $a=2$, $b=6$ & 0 & 0 & & -1 & 0 & & -1 & 0 & & 2 & 1 & & 0 & 0 & & 2 & 0\\
 &  & $a=11$, $b=0.6$ & 0 & 0 & & -1 & 0 & & -1 & 0 & & 2 & 1 & & 0 & 0 & & 2 & 0\\
\cmidrule{2-3}
 & \multirow{2}*{$1\sigma^2$} & $a=1.1$, $b=6$ & 0 & 0 & & -1 & 0 & & 0 & 0 & & 2 & 1 & & 0 & 0 & & 2 & 0\\
 &  & $a=11$, $b=0.06$ & 0 & 0 & & -3 & 0 & & -2 & 0 & & 1 & 1 & & -2 & 0 & & -4 & 1\\
\cmidrule{2-3}
 & \multirow{3}*{$10\sigma^2$} & $a=1.1$, $b=0.6$ & 0 & 0 & & 0 & 0 & & 0 & 0 & & 0 & 1 & & 0 & 0 & & 0 & 0\\
 &  & $a=2$, $b=0.06$ & 1 & 0 & & -7 & 1 & & -7 & 0 & & 3 & 1 & & -5 & 0 & & -7 & 1\\
 &  & $a=11$, $b=0.006$ & 1 & 0 & & -10 & 1 & & -8 & 1 & & -1 & 1 & & -6 & 1 & & -10 & 2\\
 \midrule
 \multicolumn{20}{c}{{\normalsize \textbf{MAR}}}\\ 
  \midrule
\multirow{2}*{\textbf{Models}} & \multirow{2}*{\makecell[c]{\textbf{Prior}\\\textbf{variance}}} & \multirow{2}*{\makecell[c]{\textbf{Hyper-}\\\textbf{parameters}}} & \multicolumn{2}{c}{\textbf{SEN}} & & \multicolumn{2}{c}{\textbf{SPE}} & & \multicolumn{2}{c}{\textbf{F}$_1$} & & \multicolumn{2}{c}{\textbf{MSE}} & & \multicolumn{2}{c}{\textbf{SIGN}} & & \multicolumn{2}{c}{\textbf{MSE full}}\\
\cmidrule{4-5} \cmidrule{7-8} \cmidrule{10-11} \cmidrule{13-14} \cmidrule{16-17} \cmidrule{19-20}
& & & Diff & MCSE & & Diff & MCSE & & Diff & MCSE & & Diff & MCSE & & Diff & MCSE & & Diff & MCSE\\
\midrule
\multirow{8}*{\makecell[c]{Multi-\\Laplace}} & \multirow{3}*{$0.1\sigma^2$} & $h=1.1$, $v=120$ & 0 & 0 & & 0 & 0 & & 0 & 0 & & 2 & 1 & & 0 & 0 & & -4 & 0\\
 &  & $h=2$, $v=12$ & 0 & 0 & & 0 & 0 & & 0 & 0 & & 2 & 0 & & 0 & 0 & & -4 & 0\\
 &  & $h=11$, $v=1.2$ & 0 & 0 & & 0 & 0 & & 0 & 0 & & 1 & 0 & & 0 & 0 & & -3 & 0\\
\cmidrule{2-3}
 & \multirow{2}*{$1\sigma^2$} & $h=1.1$, $v=12$ & 0 & 0 & & 0 & 0 & & 0 & 0 & & 2 & 0 & & 0 & 0 & & -3 & 0\\
 &  & $h=11$, $v=0.12$ & 0 & 0 & & 0 & 0 & & 0 & 0 & & -1 & 1 & & 0 & 0 & & 9 & 1\\
\cmidrule{2-3}
 & \multirow{3}*{$10\sigma^2$} & $h=1.1$, $v=1.2$ & 0 & 0 & & 0 & 0 & & 0 & 0 & & 0 & 0 & & 0 & 0 & & 0 & 0\\
 &  & $h=2$, $v=0.12$ & 0 & 0 & & 0 & 0 & & 0 & 0 & & -1 & 1 & & 0 & 0 & & 11 & 1\\
 &  & $h=11$, $v=0.012$ & 0 & 0 & & 0 & 0 & & 0 & 0 & & 0 & 1 & & 0 & 0 & & 21 & 1\\
\midrule
\multirow{8}*{\makecell[c]{Spike-and-\\Laplace}} & \multirow{3}*{$0.1\sigma^2$} & $a=1.1$, $b=60$ & 0 & 0 & & 0 & 0 & & 0 & 0 & & 0 & 0 & & 0 & 0 & & 2 & 0\\
 &  & $a=2$, $b=6$ & 0 & 0 & & 0 & 0 & & 0 & 0 & & 0 & 1 & & 0 & 0 & & 2 & 0\\
 &  & $a=11$, $b=0.6$ & 0 & 0 & & 0 & 0 & & 0 & 0 & & 0 & 0 & & 0 & 0 & & 2 & 0\\
\cmidrule{2-3}
 & \multirow{2}*{$1\sigma^2$} & $a=1.1$, $b=6$ & 0 & 0 & & 0 & 0 & & 0 & 0 & & 0 & 1 & & 0 & 0 & & 2 & 0\\
 &  & $a=11$, $b=0.06$ & 0 & 0 & & -2 & 0 & & -2 & 0 & & -1 & 1 & & -2 & 0 & & -5 & 1\\
\cmidrule{2-3}
 & \multirow{3}*{$10\sigma^2$} & $a=1.1$, $b=0.6$ & 0 & 0 & & 0 & 0 & & 0 & 0 & & 0 & 1 & & 0 & 0 & & 0 & 0\\
 &  & $a=2$, $b=0.06$ & 1 & 0 & & -7 & 1 & & -6 & 1 & & -1 & 1 & & -5 & 0 & & -7 & 1\\
 &  & $a=11$, $b=0.006$ & 1 & 0 & & -12 & 1 & & -10 & 1 & & -3 & 1 & & -8 & 1 & & -13 & 2\\
\bottomrule
\end{tabular}
}
\caption{Sensitivity analysis of mean difference (x 100) and corresponding MCSE (x 100) between different and default hyperparameters specifications, in the scenario of independent design and moderate missingness. The ``MSE full'' criterion refers to the MSE of the fitted full-models before the four-step variable selection procedure, and averaged across 200 replicates.}
\label{tab:sen_independent_mid}
\end{table}

\begin{table}[!htbp]
\centering
\resizebox{\textwidth}{!}{
\begin{tabular}{cclccccccccccccccccc}
\toprule
 \multicolumn{20}{c}{{\normalsize \textbf{n=100, p=20, SNR=1.5, independent, high missingness}}}\\ 
 \midrule
 \multicolumn{20}{c}{{\normalsize \textbf{MCAR}}}\\ 
  \midrule
\multirow{2}*{\textbf{Models}} & \multirow{2}*{\makecell[c]{\textbf{Prior}\\\textbf{variance}}} & \multirow{2}*{\makecell[c]{\textbf{Hyper-}\\\textbf{parameters}}} & \multicolumn{2}{c}{\textbf{SEN}} & & \multicolumn{2}{c}{\textbf{SPE}} & & \multicolumn{2}{c}{\textbf{F}$_1$} & & \multicolumn{2}{c}{\textbf{MSE}} & & \multicolumn{2}{c}{\textbf{SIGN}} & & \multicolumn{2}{c}{\textbf{MSE full}}\\
\cmidrule{4-5} \cmidrule{7-8} \cmidrule{10-11} \cmidrule{13-14} \cmidrule{16-17} \cmidrule{19-20}
& & & Diff & MCSE & & Diff & MCSE & & Diff & MCSE & & Diff & MCSE & & Diff & MCSE & & Diff & MCSE\\
\midrule
\multirow{8}*{\makecell[c]{Multi-\\Laplace}} & \multirow{3}*{$0.1\sigma^2$} & $h=1.1$, $v=120$ & 0 & 0 & & 0 & 0 & & 0 & 0 & & 1 & 1 & & 0 & 0 & & -4 & 0\\
 &  & $h=2$, $v=12$ & 0 & 0 & & 0 & 0 & & 0 & 0 & & 1 & 1 & & 0 & 0 & & -4 & 0\\
 &  & $h=11$, $v=1.2$ & 0 & 0 & & 0 & 0 & & 0 & 0 & & 0 & 1 & & 0 & 0 & & -3 & 0\\
\cmidrule{2-3}
 & \multirow{2}*{$1\sigma^2$} & $h=1.1$, $v=12$ & 0 & 0 & & 0 & 0 & & 0 & 0 & & 1 & 1 & & 0 & 0 & & -4 & 0\\
 &  & $h=11$, $v=0.12$ & 0 & 0 & & 0 & 0 & & 0 & 0 & & 0 & 1 & & 0 & 0 & & 11 & 1\\
\cmidrule{2-3}
 & \multirow{3}*{$10\sigma^2$} & $h=1.1$, $v=1.2$ & 0 & 0 & & 0 & 0 & & 0 & 0 & & 0 & 0 & & 0 & 0 & & 0 & 0\\
 &  & $h=2$, $v=0.12$ & 0 & 0 & & 0 & 0 & & 0 & 0 & & 0 & 1 & & 0 & 0 & & 13 & 1\\
 &  & $h=11$, $v=0.012$ & -1 & 0 & & 0 & 0 & & 0 & 0 & & 3 & 1 & & 0 & 0 & & 25 & 1\\
\midrule
\multirow{8}*{\makecell[c]{Spike-and-\\Laplace}} & \multirow{3}*{$0.1\sigma^2$} & $a=1.1$, $b=60$ & 0 & 0 & & 0 & 0 & & 0 & 0 & & 1 & 1 & & 0 & 0 & & 2 & 0\\
 &  & $a=2$, $b=6$ & 0 & 0 & & 0 & 0 & & 0 & 0 & & 0 & 1 & & 0 & 0 & & 2 & 0\\
 &  & $a=11$, $b=0.6$ & 0 & 0 & & 0 & 0 & & 0 & 0 & & -1 & 1 & & 0 & 0 & & 2 & 0\\
\cmidrule{2-3}
 & \multirow{2}*{$1\sigma^2$} & $a=1.1$, $b=6$ & 0 & 0 & & 0 & 0 & & 0 & 0 & & 0 & 1 & & 0 & 0 & & 2 & 0\\
 &  & $a=11$, $b=0.06$ & 0 & 0 & & -3 & 0 & & -2 & 0 & & 1 & 1 & & -2 & 0 & & -4 & 1\\
\cmidrule{2-3}
 & \multirow{3}*{$10\sigma^2$} & $a=1.1$, $b=0.6$ & 0 & 0 & & 0 & 0 & & 0 & 0 & & 0 & 0 & & 0 & 0 & & 0 & 0\\
 &  & $a=2$, $b=0.06$ & 1 & 0 & & -7 & 1 & & -6 & 0 & & 1 & 1 & & -5 & 0 & & -6 & 1\\
 &  & $a=11$, $b=0.006$ & 1 & 0 & & -12 & 1 & & -9 & 1 & & -2 & 1 & & -8 & 1 & & -12 & 2\\
 \midrule
 \multicolumn{20}{c}{{\normalsize \textbf{MAR}}}\\ 
  \midrule
\multirow{2}*{\textbf{Models}} & \multirow{2}*{\makecell[c]{\textbf{Prior}\\\textbf{variance}}} & \multirow{2}*{\makecell[c]{\textbf{Hyper-}\\\textbf{parameters}}} & \multicolumn{2}{c}{\textbf{SEN}} & & \multicolumn{2}{c}{\textbf{SPE}} & & \multicolumn{2}{c}{\textbf{F}$_1$} & & \multicolumn{2}{c}{\textbf{MSE}} & & \multicolumn{2}{c}{\textbf{SIGN}} & & \multicolumn{2}{c}{\textbf{MSE full}}\\
\cmidrule{4-5} \cmidrule{7-8} \cmidrule{10-11} \cmidrule{13-14} \cmidrule{16-17} \cmidrule{19-20}
& & & Diff & MCSE & & Diff & MCSE & & Diff & MCSE & & Diff & MCSE & & Diff & MCSE & & Diff & MCSE\\
\midrule
\multirow{8}*{\makecell[c]{Multi-\\Laplace}} & \multirow{3}*{$0.1\sigma^2$} & $h=1.1$, $v=120$ & 0 & 0 & & -1 & 0 & & 0 & 0 & & 1 & 1 & & 0 & 0 & & -5 & 0\\
 &  & $h=2$, $v=12$ & 0 & 0 & & 0 & 0 & & 0 & 0 & & 1 & 1 & & 0 & 0 & & -5 & 0\\
 &  & $h=11$, $v=1.2$ & 0 & 0 & & 0 & 0 & & 0 & 0 & & 2 & 1 & & 0 & 0 & & -4 & 0\\
\cmidrule{2-3}
 & \multirow{2}*{$1\sigma^2$} & $h=1.1$, $v=12$ & 0 & 0 & & -1 & 0 & & 0 & 0 & & 1 & 1 & & 0 & 0 & & -4 & 0\\
 &  & $h=11$, $v=0.12$ & 0 & 0 & & 1 & 0 & & 1 & 0 & & -1 & 1 & & 0 & 0 & & 12 & 1\\
\cmidrule{2-3}
 & \multirow{3}*{$10\sigma^2$} & $h=1.1$, $v=1.2$ & 0 & 0 & & 0 & 0 & & 0 & 0 & & 0 & 0 & & 0 & 0 & & 0 & 0\\
 &  & $h=2$, $v=0.12$ & 0 & 0 & & 1 & 0 & & 0 & 0 & & 0 & 1 & & 0 & 0 & & 15 & 1\\
 &  & $h=11$, $v=0.012$ & -1 & 0 & & 1 & 0 & & 0 & 0 & & 4 & 2 & & 0 & 0 & & 28 & 1\\
\midrule
\multirow{8}*{\makecell[c]{Spike-and-\\Laplace}} & \multirow{3}*{$0.1\sigma^2$} & $a=1.1$, $b=60$ & 0 & 0 & & 0 & 0 & & 0 & 0 & & 1 & 1 & & 0 & 0 & & 2 & 0\\
 &  & $a=2$, $b=6$ & 0 & 0 & & 0 & 0 & & 0 & 0 & & 1 & 1 & & 0 & 0 & & 2 & 0\\
 &  & $a=11$, $b=0.6$ & 0 & 0 & & 0 & 0 & & 0 & 0 & & 0 & 1 & & 0 & 0 & & 2 & 0\\
\cmidrule{2-3}
 & \multirow{2}*{$1\sigma^2$} & $a=1.1$, $b=6$ & 0 & 0 & & 0 & 0 & & 0 & 0 & & 1 & 1 & & 0 & 0 & & 2 & 0\\
 &  & $a=11$, $b=0.06$ & 0 & 0 & & -2 & 0 & & -2 & 0 & & 3 & 1 & & -2 & 0 & & -5 & 1\\
\cmidrule{2-3}
 & \multirow{3}*{$10\sigma^2$} & $a=1.1$, $b=0.6$ & 0 & 0 & & 0 & 0 & & 0 & 0 & & 0 & 1 & & 0 & 0 & & 0 & 0\\
 &  & $a=2$, $b=0.06$ & 0 & 0 & & -7 & 1 & & -6 & 1 & & 2 & 1 & & -5 & 0 & & -7 & 1\\
 &  & $a=11$, $b=0.006$ & 2 & 0 & & -15 & 1 & & -11 & 1 & & -2 & 1 & & -10 & 1 & & -13 & 2\\
\bottomrule
\end{tabular}
}
\caption{Sensitivity analysis of mean difference (x 100) and corresponding MCSE (x 100) between different and default hyperparameters specifications, in the scenario of independent design and high missingness. The ``MSE full'' criterion refers to the MSE of the fitted full-models before the four-step variable selection procedure, and averaged across 200 replicates.}
\label{tab:sen_independent_high}
\end{table}

\clearpage

\subsection{Resolution of the Threshold Grid}

The first step of the four-step procedure builds the candidate collection $S$ from a grid of thresholds $\{\phi_k \in [0,1]\}$, so its increment is a tuning choice. We repeated the selection with increments $0.1$, $0.02$, $0.01$, $0.002$ and $0.001$ (the paper default is $0.01$) for the Horseshoe and Spike-and-Laplace models under the baseline ($p=20$) and the $p=40$ settings across 200 replicates, with results in Web Table~\ref{tab:sup_grid}. For increments of $0.02$ or smaller, all five metrics were essentially unchanged, and only the coarsest $0.1$ increment gave a small drop for the Horseshoe model. Because many thresholds map to the same inclusion pattern, the candidate collection stays a small and sparse subset of the $2^p$ possible submodels, which is what makes the selection insensitive to the increment. A finer grid only raises the selection time, which grows as the increment shrinks, without changing the selected model.

As $p$ grows, more inclusion patterns become possible, so we recommend a smaller increment for larger $p$ as a precaution. Because of the sparsity of $S$, however, the selection stays robust to this choice and it need not be tuned precisely. In the real-data application, where $p$ is larger than in the simulations, we therefore used the default $0.01$ increment.

\begin{table}[!htbp]
\centering
\resizebox{\textwidth}{!}{
\begin{tabular}{ll ccccc c ccccc c c}
\toprule
\multicolumn{15}{c}{\normalsize \textbf{Scenario 1, n=100, p=20, SNR=1.5, moderate missing}}\\
\midrule
 & & \multicolumn{5}{c}{MCAR} & & \multicolumn{5}{c}{MAR} & & selection time(s)\\
\cmidrule{3-7}\cmidrule{9-13}
 & \textbf{Increment} & \textbf{SEN} & \textbf{SPE} & \textbf{F}$_1$ & \textbf{MSE} & \textbf{SIGN} & & \textbf{SEN} & \textbf{SPE} & \textbf{F}$_1$ & \textbf{MSE} & \textbf{SIGN} & & \\
\midrule
\multirow{5}{*}{\textbf{Horseshoe}}
  & 0.1 & 99 & 94 & 93 & 53 & 95 & & 99 & 93 & 92 & 58 & 95 & & 1.4\\
  & 0.02 & 99 & 95 & 95 & 50 & 96 & & 99 & 95 & 94 & 56 & 96 & & 1.9\\
  & 0.01 (default) & 99 & 95 & 95 & 50 & 96 & & 99 & 95 & 94 & 56 & 96 & & 2.4\\
  & 0.002 & 99 & 95 & 95 & 50 & 96 & & 99 & 95 & 94 & 56 & 96 & & 8.2\\
  & 0.001 & 99 & 95 & 95 & 50 & 96 & & 99 & 95 & 94 & 56 & 96 & & 15.1\\
\midrule
\multirow{5}{*}{\textbf{Spike-and-Laplace}}
  & 0.1 & 99 & 97 & 96 & 46 & 97 & & 98 & 96 & 95 & 54 & 97 & & 0.9\\
  & 0.02 & 99 & 97 & 96 & 47 & 97 & & 98 & 96 & 95 & 54 & 97 & & 0.9\\
  & 0.01 (default) & 99 & 97 & 96 & 47 & 97 & & 98 & 96 & 95 & 55 & 97 & & 1.2\\
  & 0.002 & 99 & 97 & 96 & 47 & 97 & & 98 & 96 & 95 & 54 & 97 & & 1.4\\
  & 0.001 & 99 & 97 & 96 & 47 & 97 & & 98 & 96 & 95 & 54 & 97 & & 1.9\\
\midrule
\multicolumn{15}{c}{\normalsize \textbf{Scenario 1, n=100, p=40, SNR=1.5, moderate missing}}\\
\midrule
 & & \multicolumn{5}{c}{MCAR} & & \multicolumn{5}{c}{MAR} & & selection time(s)\\
\cmidrule{3-7}\cmidrule{9-13}
 & \textbf{Increment} & \textbf{SEN} & \textbf{SPE} & \textbf{F}$_1$ & \textbf{MSE} & \textbf{SIGN} & & \textbf{SEN} & \textbf{SPE} & \textbf{F}$_1$ & \textbf{MSE} & \textbf{SIGN} & & \\
\midrule
\multirow{5}{*}{\textbf{Horseshoe}}
  & 0.1 & 85 & 93 & 84 & 348 & 91 & & 83 & 90 & 81 & 406 & 88 & & 2.2\\
  & 0.02 & 82 & 94 & 83 & 372 & 90 & & 80 & 92 & 80 & 428 & 88 & & 3.2\\
  & 0.01 (default) & 82 & 94 & 83 & 373 & 90 & & 80 & 92 & 80 & 427 & 88 & & 4.8\\
  & 0.002 & 82 & 94 & 83 & 374 & 90 & & 80 & 92 & 80 & 428 & 88 & & 15.7\\
  & 0.001 & 82 & 94 & 83 & 374 & 90 & & 80 & 92 & 80 & 429 & 88 & & 29.1\\
\midrule
\multirow{5}{*}{\textbf{Spike-and-Laplace}}
  & 0.1 & 80 & 94 & 82 & 394 & 90 & & 78 & 91 & 78 & 458 & 87 & & 1.2\\
  & 0.02 & 80 & 94 & 83 & 387 & 90 & & 79 & 93 & 80 & 441 & 89 & & 1.3\\
  & 0.01 (default) & 80 & 95 & 83 & 386 & 90 & & 79 & 93 & 80 & 441 & 89 & & 1.2\\
  & 0.002 & 80 & 95 & 83 & 389 & 90 & & 78 & 93 & 80 & 449 & 88 & & 2.2\\
  & 0.001 & 80 & 95 & 83 & 390 & 90 & & 78 & 93 & 80 & 449 & 89 & & 3.0\\
\bottomrule
\end{tabular}
}
\caption{Sensitivity of the four-step selection to the threshold-grid resolution, for the Horseshoe and Spike-and-Laplace models over 200 replicates. The candidate-set grid $\{\phi_k\}$ is an evenly spaced sequence on $[0,1]$ with the stated increment ($0.01$ is the default used throughout the paper). SEN, SPE, F$_1$, MSE and SIGN are $\times100$. Time in seconds is the median wall-clock of the selection step.}
\label{tab:sup_grid}
\end{table}

\section{Additional Simulations}

\subsection{Fixed Thresholds}
Table~\ref{tab:sup_nothreshold} reports fixed-threshold selection, in which a shrinkage-prior covariate is retained when its $95\%$ pooled credible interval excludes zero and a Spike-and-Laplace covariate is retained when its posterior inclusion probability is at least $0.5$. For the shrinkage priors, the fixed threshold performed similarly to the four-step procedure when the covariates were independent or weakly correlated. As the correlation increased, the signals became harder to separate, and the $95\%$ credible-interval rule retained fewer variables, attaining higher specificity at the expense of sensitivity, consistent with \citet{Li2010}. This behavior became increasingly pronounced under strong correlation, and the four-step procedure substantially improved the balance between sensitivity and specificity by selecting among a range of candidate thresholds. The Spike-and-Laplace model showed a different pattern. Its fixed posterior inclusion probability threshold of $0.5$, corresponding to the median probability model, generally yielded higher sensitivity and lower specificity than the four-step procedure. This tradeoff led to worse performance when the covariates were independent, but the fixed threshold became competitive as the correlation increased, with similar or slightly higher $F_1$ scores and lower MSE. Thus, the four-step procedure performs better for the Spike-and-Laplace prior when the covariates are independent, whereas the median probability model performs better under strong correlation. For the shrinkage priors, in contrast, the advantage of the four-step procedure becomes more pronounced as the correlation increases.

\begin{table}[!htbp]
\centering
\resizebox{\textwidth}{!}{
\begin{tabular}{cccccccccccccccccccccccc}
\toprule
 & \multicolumn{10}{c}{{\normalsize \textbf{Scenario 1, n=100, p=20, SNR=1.5, moderate missing}}} & & & & \multicolumn{10}{c}{\normalsize \textbf{Scenario 2, n=100, p=20, SNR=1.5, moderate missing, $\boldsymbol{\varrho=0.5}$}} \\ 
 \midrule
\multirow{2}*{\textbf{Models}} & \multicolumn{5}{c}{MCAR} & & \multicolumn{5}{c}{MAR} & & \multicolumn{5}{c}{MCAR} & & \multicolumn{5}{c}{MAR} \\ 
\cmidrule{2-6} \cmidrule{8-12} \cmidrule{14-18} \cmidrule{20-24}
 & \textbf{SEN} & \textbf{SPE} & \textbf{F}$_1$ & \textbf{MSE} & \textbf{SIGN} & & \textbf{SEN} & \textbf{SPE} & \textbf{F}$_1$ & \textbf{MSE} & \textbf{SIGN} & & \textbf{SEN} & \textbf{SPE} & \textbf{F}$_1$ & \textbf{MSE} & \textbf{SIGN} & & \textbf{SEN} & \textbf{SPE} & \textbf{F}$_1$ & \textbf{MSE} & \textbf{SIGN} \\
\midrule
  Multi-Laplace & 98 & 96 & 95 & 57 & 97 & & 97 & 96 & \textbf{95} & 66 & 96 & & 83 & 97 & 87 & 135 & 93 & & 77 & \textbf{98} & 84 & 171 & \textbf{92}\\
  Horseshoe & 98 & \textbf{97} & \textbf{96} & 53 & \textbf{98} & & 97 & \textbf{97} & \textbf{95} & 63 & \textbf{97} & & 84 & \textbf{98} & \textbf{89} & 119 & \textbf{94} & & 79 & \textbf{98} & \textbf{85} & 153 & \textbf{92}\\
  ARD & 98 & \textbf{97} & \textbf{96} & \textbf{51} & \textbf{98} & & 97 & \textbf{97} & \textbf{95} & \textbf{61} & \textbf{97} & & 84 & \textbf{98} & 88 & 117 & \textbf{94} & & 79 & 97 & \textbf{85} & 146 & \textbf{92}\\
  Spike-and-Laplace & \textbf{100} & 89 & 89 & 57 & 92 & & \textbf{99} & 89 & 88 & 65 & 92 & & \textbf{96} & 89 & 87 & \textbf{85} & 91 & & \textbf{92} & 88 & 84 & \textbf{103} & 89\\
\bottomrule
\toprule
 & \multicolumn{10}{c}{{\normalsize \textbf{Scenario 2, n=100, p=20, SNR=1.5, moderate missing, $\boldsymbol{\varrho=0.7}$}}} & & & & \multicolumn{10}{c}{\normalsize \textbf{Scenario 2, n=100, p=20, SNR=1.5, moderate missing, $\boldsymbol{\varrho=0.9}$}} \\ 
 \midrule
\multirow{2}*{\textbf{Models}} & \multicolumn{5}{c}{MCAR} & & \multicolumn{5}{c}{MAR} & & \multicolumn{5}{c}{MCAR} & & \multicolumn{5}{c}{MAR} \\ 
\cmidrule{2-6} \cmidrule{8-12} \cmidrule{14-18} \cmidrule{20-24}
 & \textbf{SEN} & \textbf{SPE} & \textbf{F}$_1$ & \textbf{MSE} & \textbf{SIGN} & & \textbf{SEN} & \textbf{SPE} & \textbf{F}$_1$ & \textbf{MSE} & \textbf{SIGN} & & \textbf{SEN} & \textbf{SPE} & \textbf{F}$_1$ & \textbf{MSE} & \textbf{SIGN} & & \textbf{SEN} & \textbf{SPE} & \textbf{F}$_1$ & \textbf{MSE} & \textbf{SIGN} \\
\midrule
  Multi-Laplace & 54 & \textbf{97} & 65 & 343 & 84 & & 48 & \textbf{97} & 60 & 411 & 82 & & 11 & \textbf{99} & 18 & 1495 & 72 & & 9 & \textbf{99} & 14 & 1596 & 72\\
  Horseshoe & 57 & \textbf{97} & 68 & 266 & \textbf{85} & & 52 & \textbf{97} & 64 & 306 & \textbf{83} & & 21 & 97 & 32 & 801 & \textbf{74} & & 17 & 97 & 26 & 901 & \textbf{73}\\
  ARD & 59 & 96 & 69 & 243 & \textbf{85} & & 54 & 96 & 65 & 280 & \textbf{83} & & 23 & 96 & 34 & 715 & \textbf{74} & & 19 & 96 & 28 & 791 & \textbf{73}\\
  Spike-and-Laplace & \textbf{79} & 86 & \textbf{75} & \textbf{141} & 84 & & \textbf{76} & 85 & \textbf{72} & \textbf{151} & 82 & & \textbf{43} & 84 & \textbf{48} & \textbf{207} & 72 & & \textbf{39} & 83 & \textbf{44} & \textbf{223} & 70\\
\bottomrule
\end{tabular}
}
\caption{Selection performance (\texttimes100) of the four Bayesian MI-LASSO models under fixed-threshold selection (shrinkage priors: $95\%$ credible interval excluding zero; Spike-and-Laplace: posterior inclusion probability $\ge 0.5$), without the four-step procedure, as the AR(1) correlation increases from the independent baseline ($\varrho=0$) to $\varrho=0.9$. The MSE is calculated using the pooled posterior mean of the calibrated coefficients for the variables selected by the fixed-threshold rule. The best value per metric within each setting is in bold.}
\label{tab:sup_nothreshold}
\end{table}

\subsection{Scaled Neighborhood Criterion}
Web Table~\ref{tab:snc_bml} reports the four-step procedure using the scaled neighborhood criterion \citep{Li2010} to generate the candidate path, the alternative to the threshold-grid search used in the main text (Step 1). The results were nearly identical for all three shrinkage-based methods across the four settings in Web Table~\ref{tab:snc_bml}. For the Spike-and-Laplace model, the results were similar when signal was strong, and threshold-grid search had smaller MSE than scaled neighborhood criterion when $p=40$ or $\varrho=0.5$.

\begin{table}[!htbp]
\centering
\resizebox{\textwidth}{!}{
\begin{tabular}{cccccccccccccccccccccccc}
\toprule
\multicolumn{24}{c}{\textbf{Scaled Neighborhood Criterion}} \\
\midrule
 & \multicolumn{10}{c}{{\normalsize \textbf{Scenario 1, n=100, p=20, SNR=1.5, moderate missing}}} & & & & \multicolumn{10}{c}{{\normalsize \textbf{Scenario 1, n=200, p=20, SNR=1.5, moderate missing}}} \\
\midrule
\multirow{2}*{\textbf{Models}} & \multicolumn{5}{c}{MCAR} & & \multicolumn{5}{c}{MAR} & & \multicolumn{5}{c}{MCAR} & & \multicolumn{5}{c}{MAR} \\
\cmidrule{2-6} \cmidrule{8-12} \cmidrule{14-18} \cmidrule{20-24}
 & \textbf{SEN} & \textbf{SPE} & \textbf{F}$_1$ & \textbf{MSE} & \textbf{SIGN} & & \textbf{SEN} & \textbf{SPE} & \textbf{F}$_1$ & \textbf{MSE} & \textbf{SIGN} & & \textbf{SEN} & \textbf{SPE} & \textbf{F}$_1$ & \textbf{MSE} & \textbf{SIGN} & & \textbf{SEN} & \textbf{SPE} & \textbf{F}$_1$ & \textbf{MSE} & \textbf{SIGN} \\
\midrule
  Multi-Laplace & 99 & 97 & 96 & 48 & 97 & & 98 & 96 & 95 & 56 & 97 & & 100 & 98 & 98 & 18 & 99 & & 100 & 98 & 98 & 21 & 99 \\
  Horseshoe & 99 & 95 & 95 & 50 & 96 & & 98 & 95 & 94 & 57 & 96 & & 100 & 98 & 98 & 18 & 98 & & 100 & 97 & 97 & 21 & 98 \\
  ARD & 99 & 95 & 95 & 50 & 96 & & 98 & 95 & 94 & 57 & 96 & & 100 & 98 & 98 & 18 & 98 & & 100 & 97 & 97 & 21 & 98 \\
  Spike-and-Laplace & 98 & 97 & 96 & 47 & 98 & & 98 & 97 & 95 & 53 & 97 & & 100 & 99 & 98 & 18 & 99 & & 100 & 98 & 98 & 20 & 99 \\
\bottomrule
\toprule
 & \multicolumn{10}{c}{{\normalsize \textbf{Scenario 1, n=100, p=40, SNR=1.5, moderate missing}}} & & & & \multicolumn{10}{c}{{\normalsize \textbf{Scenario 2, n=100, p=20, SNR=1.5, moderate missing, $\boldsymbol{\varrho=0.5}$}}} \\
\midrule
\multirow{2}*{\textbf{Models}} & \multicolumn{5}{c}{MCAR} & & \multicolumn{5}{c}{MAR} & & \multicolumn{5}{c}{MCAR} & & \multicolumn{5}{c}{MAR} \\
\cmidrule{2-6} \cmidrule{8-12} \cmidrule{14-18} \cmidrule{20-24}
 & \textbf{SEN} & \textbf{SPE} & \textbf{F}$_1$ & \textbf{MSE} & \textbf{SIGN} & & \textbf{SEN} & \textbf{SPE} & \textbf{F}$_1$ & \textbf{MSE} & \textbf{SIGN} & & \textbf{SEN} & \textbf{SPE} & \textbf{F}$_1$ & \textbf{MSE} & \textbf{SIGN} & & \textbf{SEN} & \textbf{SPE} & \textbf{F}$_1$ & \textbf{MSE} & \textbf{SIGN} \\
\midrule
  Multi-Laplace & 78 & 95 & 82 & 391 & 90 & & 76 & 94 & 79 & 449 & 88 & & 86 & 96 & 88 & 105 & 93 & & 86 & 94 & 86 & 114 & 92 \\
  Horseshoe & 82 & 94 & 83 & 376 & 90 & & 80 & 92 & 80 & 426 & 88 & & 88 & 95 & 88 & 105 & 93 & & 87 & 93 & 85 & 118 & 91 \\
  ARD & 82 & 94 & 83 & 378 & 90 & & 80 & 92 & 80 & 438 & 88 & & 88 & 95 & 88 & 105 & 93 & & 86 & 93 & 85 & 123 & 91 \\
  Spike-and-Laplace & \textbf{78} & 96 & 82 & \textbf{403} & 90 & & \textbf{75} & 94 & 79 & \textbf{478} & 88 & & \textbf{85} & 96 & 87 & \textbf{110} & 93 & & 84 & 94 & 85 & \textbf{128} & 91 \\
\bottomrule
\end{tabular}
}
\caption{Selection and estimation performance ($\times100$) of the four Bayesian MI-LASSO models under the \emph{scaled-neighborhood} candidate-set criterion. The major differences with threshold-grid search (Step 1) are in bold.}
\label{tab:snc_bml}
\end{table}

\subsection{PSIS-LOO and Forward Search}
Table~\ref{tab:sup_criterion} compares the modified BIC and PSIS-LOO \citep{Vehtari2017} selection criteria, with an additional comparison to forward search for candidate set generation. For PSIS-LOO, we consider both the model with the highest ELPD and the 1-SE rule \citep{Piironen2020}, which selects the smallest candidate model whose ELPD is within one standard error of the maximum. The standard error is calculated from the paired pointwise ELPD differences between the two models. We evaluate PSIS-LOO at the unit level, since the $D$ imputed datasets represent the same $n$ units. Holding out unit $i$ removes its observations $(\bx_i^d, y_i^d)$ from all $D$ imputed datasets at once, so the same unit is left out across every imputation, and the pointwise log predictive density of unit $i$ sums its per-imputation log-likelihoods over $d = 1, \ldots, D$. In addition to the selection evaluation, we also report the median running time.

The results show that the forward and grid candidate paths produced nearly the same selection and estimation across all settings, so the way the candidate set is generated has little effect on the final model. Their computational cost diverged as $p$ grew. The forward path was far slower than the grid path in high dimensions, because each forward step projects the full model onto every remaining variable to score its inclusion, whereas the grid path obtains its candidates directly from the pooled posterior. The model-selection criteria showed a clearer difference. Selecting the model with the maximum PSIS-LOO ELPD tended to favor larger models and performed worse than the modified BIC in the $p=20$ settings. The 1-SE rule substantially reduced this difference and produced selection performance broadly comparable to the modified BIC across settings. The modified BIC, however, remained computationally more efficient than PSIS-LOO.

\begin{table}[!htbp]
\centering
\resizebox{0.72\textwidth}{!}{
\begin{tabular}{ll ccccc c ccccc c c}
\toprule
\multicolumn{15}{c}{\normalsize \textbf{Scenario 1, n=100, p=20, SNR=1.5, moderate missing}}\\
\midrule
 & & \multicolumn{5}{c}{MCAR} & & \multicolumn{5}{c}{MAR} & & running time(s)\\
\cmidrule{3-7}\cmidrule{9-13}
 & & \textbf{SEN} & \textbf{SPE} & \textbf{F}$_1$ & \textbf{MSE} & \textbf{SIGN} & & \textbf{SEN} & \textbf{SPE} & \textbf{F}$_1$ & \textbf{MSE} & \textbf{SIGN} & & \\
\midrule
\multirow{6}{*}{\textbf{Multi-Laplace}}
  & modified BIC $+$ grid (default) & 99 & 97 & \textbf{96} & \textbf{49} & \textbf{97} & & 98 & 96 & \textbf{95} & 56 & \textbf{97} & & \textbf{6.5}\\
  & modified BIC $+$ forward & 98 & 97 & \textbf{96} & 50 & \textbf{97} & & 98 & 96 & \textbf{95} & \textbf{54} & \textbf{97} & & 6.6\\
  & PSIS-LOO (max) $+$ grid & 100 & 77 & 80 & 71 & 84 & & 100 & 75 & 78 & 78 & 83 & & 8.8\\
  & PSIS-LOO (max) $+$ forward & 100 & 77 & 80 & 71 & 84 & & 100 & 76 & 79 & 78 & 83 & & 9.5\\
  & PSIS-LOO (1-SE) $+$ grid & 99 & 94 & 94 & 53 & 96 & & 98 & 93 & 92 & 61 & 94 & & 8.8\\
  & PSIS-LOO (1-SE) $+$ forward & 99 & 95 & 94 & 51 & 96 & & 98 & 94 & 93 & 58 & 95 & & 9.5\\
\midrule
\multirow{6}{*}{\textbf{Horseshoe}}
  & modified BIC $+$ grid (default) & 99 & 95 & 95 & \textbf{50} & 96 & & 99 & 95 & 94 & \textbf{56} & 96 & & \textbf{5.9}\\
  & modified BIC $+$ forward & 99 & 95 & 95 & 51 & 96 & & 99 & 95 & 94 & \textbf{56} & 96 & & 6.0\\
  & PSIS-LOO (max) $+$ grid & 100 & 87 & 87 & 57 & 91 & & 99 & 85 & 86 & 63 & 89 & & 7.9\\
  & PSIS-LOO (max) $+$ forward & 100 & 87 & 87 & 57 & 91 & & 99 & 85 & 86 & 64 & 89 & & 8.6\\
  & PSIS-LOO (1-SE) $+$ grid & 98 & 97 & \textbf{96} & 51 & \textbf{97} & & 98 & 96 & 94 & 58 & 96 & & 7.9\\
  & PSIS-LOO (1-SE) $+$ forward & 98 & 97 & \textbf{96} & 53 & \textbf{97} & & 98 & 96 & \textbf{95} & \textbf{56} & \textbf{97} & & 8.6\\
\midrule
\multirow{6}{*}{\textbf{ARD}}
  & modified BIC $+$ grid (default) & 99 & 95 & 95 & \textbf{50} & 96 & & 99 & 95 & 94 & 56 & 96 & & \textbf{5.9}\\
  & modified BIC $+$ forward & 99 & 95 & 95 & \textbf{50} & 96 & & 99 & 95 & 94 & \textbf{55} & 96 & & 6.0\\
  & PSIS-LOO (max) $+$ grid & 99 & 91 & 91 & 54 & 94 & & 99 & 89 & 89 & 60 & 92 & & 7.8\\
  & PSIS-LOO (max) $+$ forward & 99 & 91 & 91 & 54 & 94 & & 99 & 90 & 89 & 60 & 92 & & 8.7\\
  & PSIS-LOO (1-SE) $+$ grid & 98 & 98 & \textbf{97} & \textbf{50} & \textbf{98} & & 97 & 97 & 95 & 57 & \textbf{97} & & 7.8\\
  & PSIS-LOO (1-SE) $+$ forward & 97 & 98 & 96 & 53 & \textbf{98} & & 97 & 97 & \textbf{96} & 57 & \textbf{97} & & 8.7\\
\midrule
\multirow{6}{*}{\textbf{Spike-and-Laplace}}
  & modified BIC $+$ grid (default) & 99 & 97 & 96 & \textbf{47} & 97 & & 98 & 96 & 95 & 55 & 97 & & \textbf{9.3}\\
  & modified BIC $+$ forward & 98 & 97 & 96 & 49 & 97 & & 98 & 96 & 95 & \textbf{53} & 97 & & 9.4\\
  & PSIS-LOO (max) $+$ grid & 99 & 94 & 93 & 52 & 95 & & 99 & 92 & 92 & 60 & 94 & & 10.3\\
  & PSIS-LOO (max) $+$ forward & 99 & 94 & 93 & 52 & 95 & & 99 & 92 & 92 & 60 & 94 & & 11.9\\
  & PSIS-LOO (1-SE) $+$ grid & 98 & 98 & \textbf{97} & \textbf{47} & \textbf{98} & & 98 & 97 & 95 & 55 & 97 & & 10.3\\
  & PSIS-LOO (1-SE) $+$ forward & 97 & 98 & 96 & 53 & \textbf{98} & & 97 & 98 & \textbf{96} & 55 & \textbf{98} & & 11.9\\
\bottomrule
\toprule
\multicolumn{15}{c}{\normalsize \textbf{Scenario 1, n=100, p=40, SNR=1.5, moderate missing}}\\
\midrule
 & & \multicolumn{5}{c}{MCAR} & & \multicolumn{5}{c}{MAR} & & running time(s)\\
\cmidrule{3-7}\cmidrule{9-13}
 & & \textbf{SEN} & \textbf{SPE} & \textbf{F}$_1$ & \textbf{MSE} & \textbf{SIGN} & & \textbf{SEN} & \textbf{SPE} & \textbf{F}$_1$ & \textbf{MSE} & \textbf{SIGN} & & \\
\midrule
\multirow{6}{*}{\textbf{Multi-Laplace}}
  & modified BIC $+$ grid (default) & 78 & 95 & 82 & 391 & \textbf{90} &  & 77 & 93 & \textbf{79} & 443 & \textbf{88} & & \textbf{14.3}\\
  & modified BIC $+$ forward & 77 & 95 & 81 & 405 & \textbf{90} &  & 72 & 94 & 76 & 493 & 87 & & 14.8\\
  & PSIS-LOO (max) $+$ grid & 95 & 76 & 76 & \textbf{325} & 82 &  & 94 & 71 & 72 & 390 & 78 & & 20.8\\
  & PSIS-LOO (max) $+$ forward & 95 & 76 & 76 & 329 & 82 &  & 94 & 72 & 73 & \textbf{388} & 79 & & 23.1\\
  & PSIS-LOO (1-SE) $+$ grid & 89 & 89 & 83 & 327 & 89 &  & 87 & 84 & 78 & 396 & 85 & & 20.8\\
  & PSIS-LOO (1-SE) $+$ forward & 88 & 90 & \textbf{84} & 326 & \textbf{90} &  & 86 & 86 & \textbf{79} & 392 & 86 & & 23.1\\
\midrule
\multirow{6}{*}{\textbf{Horseshoe}}
  & modified BIC $+$ grid (default) & 82 & 94 & \textbf{83} & 373 & 90 &  & 80 & 92 & 80 & 427 & 88 & & \textbf{11.8}\\
  & modified BIC $+$ forward & 80 & 94 & 82 & 387 & 90 &  & 77 & 92 & 79 & 453 & 88 & & 12.3\\
  & PSIS-LOO (max) $+$ grid & 89 & 88 & \textbf{83} & \textbf{333} & 89 &  & 88 & 85 & 79 & 392 & 86 & & 16.2\\
  & PSIS-LOO (max) $+$ forward & 89 & 88 & 82 & 335 & 88 &  & 88 & 84 & 79 & \textbf{389} & 86 & & 18.8\\
  & PSIS-LOO (1-SE) $+$ grid & 80 & 95 & \textbf{83} & 385 & \textbf{91} &  & 80 & 93 & \textbf{81} & 429 & \textbf{89} & & 16.2\\
  & PSIS-LOO (1-SE) $+$ forward & 79 & 96 & \textbf{83} & 395 & \textbf{91} &  & 78 & 93 & 80 & 446 & 88 & & 18.8\\
\midrule
\multirow{6}{*}{\textbf{ARD}}
  & modified BIC $+$ grid (default) & 82 & 94 & 83 & 375 & 90 &  & 80 & 92 & 80 & 438 & 88 & & \textbf{11.1}\\
  & modified BIC $+$ forward & 80 & 94 & 82 & 393 & 90 &  & 77 & 93 & 78 & 463 & 88 & & 11.5\\
  & PSIS-LOO (max) $+$ grid & 87 & 92 & \textbf{85} & \textbf{337} & 90 &  & 85 & 89 & \textbf{81} & 404 & 88 & & 14.7\\
  & PSIS-LOO (max) $+$ forward & 88 & 91 & 84 & 338 & 90 &  & 85 & 89 & \textbf{81} & \textbf{400} & 88 & & 17.4\\
  & PSIS-LOO (1-SE) $+$ grid & 78 & 96 & 83 & 395 & \textbf{91} &  & 76 & 94 & 80 & 454 & \textbf{89} & & 14.7\\
  & PSIS-LOO (1-SE) $+$ forward & 76 & 97 & 82 & 416 & 90 &  & 74 & 94 & 79 & 470 & 88 & & 17.4\\
\midrule
\multirow{6}{*}{\textbf{Spike-and-Laplace}}
  & modified BIC $+$ grid (default) & 80 & 95 & 83 & 386 & 90 &  & 79 & 93 & 80 & 441 & \textbf{89} & & \textbf{25.8}\\
  & modified BIC $+$ forward & 76 & 96 & 81 & 422 & 90 &  & 73 & 94 & 77 & 492 & 88 & & 26.2\\
  & PSIS-LOO (max) $+$ grid & 87 & 92 & \textbf{85} & 337 & 90 &  & 85 & 89 & \textbf{81} & 407 & 88 & & 27.7\\
  & PSIS-LOO (max) $+$ forward & 88 & 92 & \textbf{85} & \textbf{336} & 90 &  & 85 & 89 & \textbf{81} & \textbf{404} & 88 & & 30.8\\
  & PSIS-LOO (1-SE) $+$ grid & 81 & 95 & 84 & 370 & \textbf{91} &  & 79 & 93 & \textbf{81} & 436 & \textbf{89} & & 27.7\\
  & PSIS-LOO (1-SE) $+$ forward & 77 & 96 & 82 & 408 & 90 &  & 75 & 94 & 79 & 462 & 88 & & 30.8\\
\bottomrule
\toprule
\multicolumn{15}{c}{\normalsize \textbf{Scenario 1, n=100, p=200, SNR=1.5, moderate missing}}\\
\midrule
 & & \multicolumn{5}{c}{MCAR} & & \multicolumn{5}{c}{MAR} & & running time(s)\\
\cmidrule{3-7}\cmidrule{9-13}
 & & \textbf{SEN} & \textbf{SPE} & \textbf{F}$_1$ & \textbf{MSE} & \textbf{SIGN} & & \textbf{SEN} & \textbf{SPE} & \textbf{F}$_1$ & \textbf{MSE} & \textbf{SIGN} & & \\
\midrule
\multirow{6}{*}{\textbf{Multi-Laplace}}
  & modified BIC $+$ grid (default) & 91 & 56 & \textbf{21} & 4616 & 57 &  & 84 & 57 & \textbf{21} & 5850 & 58 & & \textbf{125.7}\\
  & modified BIC $+$ forward & 82 & 54 & 18 & 3519 & 55 &  & 76 & 54 & 17 & 4174 & 55 & & 155.8\\
  & PSIS-LOO (max) $+$ grid & 92 & 55 & \textbf{21} & 4022 & 57 &  & 88 & 55 & 20 & 4764 & 57 & & 150.7\\
  & PSIS-LOO (max) $+$ forward & 80 & 58 & 19 & 3483 & 59 &  & 75 & 58 & 18 & 4100 & 58 & & 207.7\\
  & PSIS-LOO (1-SE) $+$ grid & 92 & 56 & \textbf{21} & 3636 & 57 &  & 88 & 56 & 20 & 4564 & 57 & & 150.7\\
  & PSIS-LOO (1-SE) $+$ forward & 80 & 59 & 20 & \textbf{3470} & \textbf{60} &  & 74 & 59 & 18 & \textbf{4070} & \textbf{59} & & 207.7\\
\midrule
\multirow{6}{*}{\textbf{Horseshoe}}
  & modified BIC $+$ grid (default) & 78 & 60 & \textbf{20} & \textbf{2780} & \textbf{61} &  & 71 & 60 & \textbf{18} & \textbf{3381} & \textbf{60} & & \textbf{72.9}\\
  & modified BIC $+$ forward & 82 & 53 & 18 & 3555 & 55 &  & 77 & 53 & 17 & 4252 & 54 & & 101.2\\
  & PSIS-LOO (max) $+$ grid & 79 & 60 & 19 & 2783 & 60 &  & 72 & 59 & \textbf{18} & \textbf{3381} & 59 & & 75.5\\
  & PSIS-LOO (max) $+$ forward & 82 & 53 & 18 & 3555 & 55 &  & 77 & 53 & 17 & 4246 & 54 & & 130.5\\
  & PSIS-LOO (1-SE) $+$ grid & 79 & 60 & 19 & 2783 & 60 &  & 71 & 59 & \textbf{18} & 3383 & 59 & & 75.5\\
  & PSIS-LOO (1-SE) $+$ forward & 82 & 53 & 18 & 3555 & 55 &  & 77 & 53 & 17 & 4243 & 54 & & 130.5\\
\midrule
\multirow{6}{*}{\textbf{ARD}}
  & modified BIC $+$ grid (default) & 83 & 56 & \textbf{19} & \textbf{2647} & \textbf{57} &  & 77 & 55 & \textbf{17} & \textbf{3775} & \textbf{56} & & \textbf{71.6}\\
  & modified BIC $+$ forward & 82 & 53 & 18 & 3555 & 55 &  & 77 & 53 & \textbf{17} & 4273 & 54 & & 99.8\\
  & PSIS-LOO (max) $+$ grid & 83 & 55 & \textbf{19} & \textbf{2647} & \textbf{57} &  & 77 & 55 & \textbf{17} & \textbf{3775} & \textbf{56} & & 73.3\\
  & PSIS-LOO (max) $+$ forward & 82 & 53 & 18 & 3555 & 55 &  & 77 & 53 & \textbf{17} & 4273 & 54 & & 126.8\\
  & PSIS-LOO (1-SE) $+$ grid & 83 & 55 & \textbf{19} & \textbf{2647} & \textbf{57} &  & 77 & 55 & \textbf{17} & \textbf{3775} & \textbf{56} & & 73.3\\
  & PSIS-LOO (1-SE) $+$ forward & 82 & 53 & 18 & 3555 & 55 &  & 77 & 53 & \textbf{17} & 4273 & 54 & & 126.8\\
\midrule
\multirow{6}{*}{\textbf{Spike-and-Laplace}}
  & modified BIC $+$ grid (default) & 78 & 61 & 20 & \textbf{2587} & 62 &  & 71 & 61 & 18 & \textbf{2999} & 61 & & \textbf{725.8}\\
  & modified BIC $+$ forward & 79 & 61 & 20 & 3336 & 62 &  & 74 & 59 & 18 & 3785 & 60 & & 749.0\\
  & PSIS-LOO (max) $+$ grid & 74 & 71 & \textbf{24} & 2603 & 71 &  & 65 & 70 & \textbf{21} & 3019 & 70 & & 729.4\\
  & PSIS-LOO (max) $+$ forward & 79 & 64 & 21 & 3319 & 64 &  & 72 & 63 & 19 & 3735 & 63 & & 767.0\\
  & PSIS-LOO (1-SE) $+$ grid & 74 & 72 & \textbf{24} & 2597 & \textbf{72} &  & 64 & 72 & \textbf{21} & 3015 & \textbf{71} & & 729.4\\
  & PSIS-LOO (1-SE) $+$ forward & 79 & 65 & 22 & 3294 & 66 &  & 72 & 65 & 20 & 3692 & 65 & & 767.0\\
\bottomrule
\end{tabular}
}
\caption{Selection and estimation performance ($\times100$) of the four Bayesian MI-LASSO models under the two candidate-path generators (threshold-grid and forward selection) crossed with the three model-selection criteria (modified BIC, PSIS-LOO (max) and PSIS-LOO (1-SE)), over 200 replicates. SEN, SPE, F$_1$, MSE and SIGN are all reported $\times100$. Time in seconds is the median total wall-clock of the model fit and the selection procedure. Within each model the best F$_1$, MSE and SIGN under each missingness mechanism, together with the shortest selection time, are in bold.}
\label{tab:sup_criterion}
\end{table}

\subsection{Stacking Methods}
Web Table~\ref{tab:sup_stack} reports stacking fits by Bayesian MI-LASSO, which concatenates the $D$ imputed datasets into a single dataset of $Dn$ rows. This approach attained high sensitivity but much lower specificity than the grouped-likelihood models, and its $95\%$ intervals covered the true signals only about half of the time, even under the oracle selection. Stacking treats the $D$ imputations as independent observations and understates the between-imputation uncertainty, which inflates selection and produces intervals that are too narrow. These results support modeling the imputations through the grouped likelihood rather than by stacking.

\begin{table}[!htbp]
\centering
\resizebox{\textwidth}{!}{
\begin{tabular}{l ccccc c ccccc}
\multicolumn{12}{c}{\textbf{(a) Selection performance ($\times100$)}}\\
\toprule
 & \multicolumn{5}{c}{MCAR} & & \multicolumn{5}{c}{MAR} \\
\cmidrule(lr){2-6}\cmidrule(lr){8-12}
 & SEN & SPE & F$_1$ & MSE & SIGN & & SEN & SPE & F$_1$ & MSE & SIGN \\
\midrule
Multi-Laplace (stacking) & \textbf{100} & \textbf{64} & \textbf{71} & 92 & \textbf{75} & & \textbf{100} & \textbf{62} & \textbf{70} & 101 & \textbf{73} \\
Horseshoe (stacking) & \textbf{100} & 61 & 70 & 89 & 73 & & \textbf{100} & 60 & 69 & 97 & 72 \\
ARD (stacking) & \textbf{100} & 57 & 67 & \textbf{86} & 70 & & \textbf{100} & 55 & 66 & \textbf{94} & 69 \\
Spike-and-Laplace (stacking) & \textbf{100} & 55 & 66 & 90 & 68 & & \textbf{100} & 53 & 65 & 98 & 67 \\
\bottomrule
\end{tabular}
}

\vspace{8pt}

\resizebox{\textwidth}{!}{
\begin{tabular}{l cccc c cccc}
\multicolumn{10}{c}{\textbf{(b) 95\% interval coverage ($\times100$) and width ($\times100$)}}\\
\toprule
 & \multicolumn{4}{c}{MCAR} & & \multicolumn{4}{c}{MAR} \\
\cmidrule(lr){2-5}\cmidrule(lr){7-10}
 & Cov$_{\ne0}$ & Cov$_{\ne0}^{\text{sel}}$ & Cov$_{=0}$ & Width & & Cov$_{\ne0}$ & Cov$_{\ne0}^{\text{sel}}$ & Cov$_{=0}$ & Width \\
\midrule
\multicolumn{10}{l}{\textit{Best select}}\\
\quad Multi-Laplace (stacking) & 54 & 54 & 64 & 34 & & 52 & 53 & 62 & 34 \\
\quad Horseshoe (stacking) & 53 & 53 & 63 & 34 & & 52 & 52 & 62 & 34 \\
\quad ARD (stacking) & 53 & 53 & 63 & 34 & & 53 & 53 & 62 & 34 \\
\quad Spike-and-Laplace (stacking) & 53 & 53 & 59 & 33 & & 52 & 52 & 58 & 33 \\
\addlinespace
\multicolumn{10}{l}{\textit{True select (oracle)}}\\
\quad Multi-Laplace (stacking) & 58 & 58 & 100 & 36 & & 57 & 57 & 100 & 36 \\
\quad Horseshoe (stacking) & 57 & 57 & 100 & 35 & & 56 & 56 & 100 & 36 \\
\quad ARD (stacking) & 57 & 57 & 100 & 35 & & 56 & 56 & 100 & 36 \\
\quad Spike-and-Laplace (stacking) & 57 & 57 & 100 & 35 & & 57 & 57 & 100 & 36 \\
\bottomrule
\end{tabular}
}
\caption{Stacking fits of the four Bayesian MI-LASSO models at the Scenario 1 baseline, over 200 replicates. Panel (a): selection performance, best value per column in bold. Panel (b): empirical coverage and mean width of the $95\%$ credible intervals, for the four-step selection and for the oracle that fixes the true signal set.}
\label{tab:sup_stack}
\end{table}

\clearpage
\bibliography{bibliography}